\documentclass[lettersize,journal]{IEEEtran}
\usepackage{amsmath,amsfonts}
\usepackage{algorithmic}
\usepackage{algorithm}
\usepackage{array}
\usepackage[caption=false,font=normalsize,labelfont=sf,textfont=sf]{subfig}
\usepackage{textcomp}
\usepackage{stfloats}
\usepackage{url}
\usepackage{verbatim}
\usepackage{graphicx}
\usepackage{cite}
\usepackage{hyperref}

\usepackage{caption}
\usepackage{soul, color, xcolor}
\usepackage{multirow}
\usepackage{threeparttable}

\sethlcolor{yellow}

\newcommand{\tabincell}[2]{\begin{tabular}{@{}#1@{}}#2\end{tabular}}%

\begin{document}

\title{Efficient Sensor Scheduling Strategy Based on Spatio-temporal Scope Information Model}

\author{Yang Liu, Chen Dong*, Xiaoqi Qin, and Xiaodong Xu
        
\thanks{
  This work was supported in part by Beijing Natural Science Foundation under Grant No. 19L2033.  (Corresponding author: Chen Dong.)  
}
\thanks{Yang Liu, Chen Dong, Xiaoqi Qin, and Xiaodong Xu are with the State Key Laboratory of Networking and Switching Technology, Beijing University of Posts and Telecommunications, Beijing, 100876, China. (e-mail: liu\_young@bupt.edu.cn; dongchen@bupt.edu.cn; xiaoqiqin@bupt.edu.cn; xuxiaodong@bupt.edu.cn;).}
}

\markboth{Journal of \LaTeX\ Class Files,~Vol.~XX, No.~X, XXX~2022}%
{Shell \MakeLowercase{\textit{et al.}}: A Sample Article Using IEEEtran.cls for IEEE Journals}


\maketitle

\begin{abstract}

% 在本文中，基于物联网（IoT）中传感器节点的时空相关性，提出了一种时空范围信息模型（SSIM）来量化传感器数据的范围价值信息，该信息随空间和时间衰减，以指导系统进行感知范围区域的高效决策。考虑了包含三个传感器节点的简易传感器监控系统，并针对优化问题，提出了两种优化调度决策机制，分别为单步最优和长期最优决策机制。对于单步机制，对其调度结果进行了理论分析，得到了部分调度结果之间节点布局的近似数值边界，与仿真结果一致。对于长期机制，利用Q-学习算法获得了具有不同节点布局的长期信息的最优调度结果。通过利用相对湿度数据集进行了实验，验证了两种机制的性能，并讨论了两种机制性能表现的不同，总结了模型的局限性。
In this paper, based on the spatio-temporal correlation of sensor nodes in the Internet of Things (IoT), a Spatio-temporal Scope information model (SSIM) is proposed to quantify the scope valuable information of sensor data, which decays with space and time, to guide the system for efficient decision making in the sensed region. A simple sensor monitoring system containing three sensor nodes is considered, and two optimal scheduling decision mechanisms, single-step optimal and long-term optimal decision mechanisms, are proposed for the optimization problem. For the single-step mechanism, the scheduling results are analyzed theoretically, and approximate numerical bounds on the node layout between some of the scheduling results are obtained, consistent with the simulation results. For the long-term mechanism, the scheduling results with different node layouts are obtained using the Q-learning algorithm. The performance of the two mechanisms is verified by conducting experiments using the relative humidity dataset, and the differences in performance of the two mechanisms are discussed; in addition, the limitations of the model are summarized.
\end{abstract}


\section{Introduction}

 %%--------------------------2022-----------------------------------------:
 
% 物联网通过利用其基础技术，如无处不在和普及的计算、嵌入式设备、通信技术、传感器网络、互联网协议和应用程序，连接物理世界并将物理对象从传统转变为智能\cite{ref0}.
% 在物联网中往往存在大量传感节点，他们的作用则是采集数据，为上层决策系统提供信息，从而对整个系统进行相关调整与决策。典型的应用场景包括城市城市的温度和空气质量监测[2]，智能交通\cite{smart traffic}和智能农业\cite{smart agriculture}等。
% 而在现阶段，由于成本的限制，在大部分物联网应用中，处于网络末端且数量较大的感知节点通常是由不可充电电池供电的低成本传感器，其对功耗非常敏感，很多节点是微功耗型节点，甚至还有一些无源的节点。
% 在这些应用中，提高节点的感知效率，在确保感知准确性和有效性的同时，尽可能延长节点的工作寿命一直是面临的主要研究问题之一。

\IEEEPARstart{T}{he}  Internet of Things connects the physical world and transforms physical objects from being traditional to smart by exploiting its underlying technologies such as ubiquitous and pervasive computing, embedded devices, communication technologies, sensor networks, Internet protocols and applications\cite{ref0}. 
There are a large number of sensing nodes in the IoT, whose role is to collect data and provide information to the upper layers to make relevant adjustments and decisions for the whole system.
Typical application scenarios include temperature and air quality monitoring in urban cities \cite{air pollution}, smart traffic \cite{smart traffic} and smart agriculture \cite{smart agriculture}, etc.
However, due to the cost constraints, most of the sensing nodes at the end of the network are usually low-cost sensors powered by batteries, which are very sensitive to power consumption, including micropower nodes and passive nodes. 
It has been one of the leading research problems to improve nodes sensing efficiency and prolong the nodes working life in these applications.

% 对于一些时间敏感型应用，其对数据即时性的要求较大，即过时的数据对系统没有太大的价值。基于这个理念，AoI的概念\cite{ref1}被提出，其被定义为感知数据产生时刻和被接收端接受时刻的时间差值。一方面，系统趋于让节点频繁的发送更新，从而能够确保数据信息的即时性，但另一方面，频繁的更新会造成系统的拥堵，并过多消耗节点的能量，造成性能不佳。所以，众多学者将AoI结合各种排队和通信模型，对系统的权衡策略进行了研究。……(一些AoI+排队引文)

For some time-sensitive applications, the requirement for data immediacy is stronger, i.e., outdated data is not of much value to the system. Based on this idea, the concept of Age of Information (AoI) \cite{ref1} is proposed, which is defined as the time difference between the moment of sensing data generation and the moment of acceptance by the receiver.
On the one hand, the system expects nodes to send frequent updates so as to be able to ensure the immediacy of data information, but on the other hand, frequent updates can cause congestion in the system and consume too much energy of nodes, resulting in poor performance. 
Therefore, numerous scholars have studied the trade-off strategy of the system by combining AoI with various queuing and communication models.
% \hl{AoI+queueing}
Specifically, the optimal service rates to minimize the average AoI in different queuing models are derived in \cite{ref1}. 
The problem of minimizing the expected weighted sum AoI while simultaneously satisfying timely-throughput constraints is addressed in \cite{ref4}.
Reference \cite{ref5} investigates the problem of minimizing average and peak age of information (PAoI) under general interference constraints. 
A discrete-time queueing model to derive the exact distributions of the AoI and PAoI sequences in a multisource status update system is proposed in \cite{Discrete-Time}.
In \cite{ref6}, it is found that the optimal arrival-independent renewal (AIR) scheduling policy to minimize the time-average AoI is RR-ONE, i.e., scheduling terminals in a round-robin fashion, and each terminal only retains the most up-to-date packet.
% \hl{Discrete-Time Queueing Model of Age of
% Information With Multiple Information Sources}
% \hl{Timely Updates With Priorities: Lexicographic Age Optimality}
Reference \cite{ref25} studies the optimal device scheduling process that jointly minimizes the average AoI and the energy cost with two types of correlated devices. 

% AoI的定义中，信息的新鲜度是线性下降的，但由于监测信源的物理性质，其在时间域上存在一定相关性，这可能会造成信息老化导致的性能下降不是时间的线性函数。所以一些学者提出关于AoI的非线性函数，并进行了相关研究……(一些AoI非线性函数引文)%信息源知识监控的不确定性随时间呈非线性增长。

The freshness of information is linearly decreasing in the definition of AoI, but there is some correlation in the time domain due to the physical nature of monitoring sources, which may cause the performance degradation due to information aging is not a linear function of time. 
Therefore, some scholars have proposed nonlinear functions about AoI and conducted related researches 
% \hl{(some AoI nonlinear function citations)}
Reference \cite{ref2} introduces a general age penalty function to characterize the level of dissatisfaction on data staleness.  
In \cite{ref3}, a general expression of the generating function of AoI and the peak age of information (PAoI) metric is provided, which provides a methodology for analyzing general non-linear age functions. 
Exploiting the temporal correlation between consecutive samples of a Markov source, reference \cite{ref10} considers a generalized incremental update scheme by sending differential updates.

% 此外，在空间上密集部署的传感节点，不可避免的造成节点的观测值在空间域中具有高度相关性。所以，一些学者研究了节点在空间域上的布局设置和空间临界采样率等等。具体的(一些节点空间放置的论文)。

In addition, the dense deployment of sensing nodes on the space inevitably causes the observations of the nodes to be highly correlated in the spatial domain. 
Therefore, some scholars have studied the layout setting of nodes on the spatial domain and the spatial critical sampling rate, etc.
% \hl{(some papers on the spatial placement of nodes)}.
% Reference \cite{ref8} uses 
The correlation between nodes is used to reconstruct the observed physical phenomena based on a fraction of all available sensor nodes in \cite{ref8}, where a framework for the analysis of sensor density based on asymptotic analysis is proposed.
In \cite{ref9}, a theoretical analysis of spatio-temporal correlation characteristics of point and field sources in WSN is performed, and the spatio-temporal characteristics are analytically derived along with the distortion functions.
Based on the confident information coverage (CIC) model, \cite{ref17} and \cite{ref18} study the critical sensor density and find the optimal placement pattern to achieve complete coverage in randomly deployed sensor networks.
In addition, a node deployment scheme to maximize the network lifetime and ensure CIC in a field with obstacles is proposed in \cite{Nature-Inspired}.

% 与此同时，一些学者同时结合物联网中节点在时间和空间上的相关性，对节点之间的合作调度和时空采样率等，进行了研究。(时空联合调度引文)这方面的详细工作将在子章节B详细展开。

Meanwhile, some scholars have simultaneously investigated the cooperative scheduling and spatio-temporal sampling rate among nodes in conjunction with the correlation of nodes in time and space domain. 
% \hl{(Joint spatio-temporal scheduling citation)}
In \cite{ref7}, %\hl{(undetermined)} 
a theoretical framework to capture the spatial and temporal correlations in wireless sensor networks is first introduced, which constitutes a basis for the development of such energy-efficient communication protocols for Wireless Sensor Networks (WSN). More work related to this paper in this regard is described in detail in subsection B.

% 本文的主要思路便是利用待测区域信源的时空相关性质，对节点数据时间域和空间域上的实时有用信息价值进行衡量。例如，节点在时间域上连续采集的数据，第一次数据由于先验知识的缺少，其能带来最多的有用信息量，而在时间域上后几次的数据，由于信源物理性的短时相关性，即第一次数据能为后几次提供一定的先验信息，从而导致后几次数据能带来的有用信息明显比第一次少。同样的，在空间域上离得很近的节点，其同时获取的数据提供的有用信息存在重叠。
% 本文建立实时衡量数据信息价值模型，表示数据信息的有效时空范围，进而研究不同节点布局下的最佳节点调度策略，提高感知信息的效率。
The main idea of this paper is to measure the value of real-time useful information of sensor data in the temporal and spatial domains by using the spatio-temporal correlation of the sources.
For example, the first data of sensors collected continuously in the time domain can bring the most amount of valuable information due to the lack of a priori knowledge, while the later data in the time domain can bring less valuable information than the first data due to the short time correlation of the physical properties of the sources, i.e., the first data can provide a certain amount of priori information for the later ones.
Similarly, there is an overlap in the valuable information provided by the data acquired simultaneously by nodes close to each other in the spatial domain.
In this paper, a model to measure real-time valuable information of sensor data is established to represent the effective spatio-temporal scope of data information, and the optimal node scheduling strategy under different node layouts is investigated to improve the efficiency of sensing information.

\subsection{Contributions and Paper Outline}
% 本文则主要利用待测区域信源的时空相关性质，建立节点感知数据的实时范围信息模型，实时衡量节点数据对整个系统的有效信息时空范围。简单来讲，例如某个节点数据的信息价值在时间域上，随时间流逝逐渐衰减，在空间域上，在节点所在位置附近邻域逐渐衰减。
% 因此，本文主要针对传感节点数据的有效信息时空域覆盖范围，研究三个节点在不同位置布局的情况下，最高效的激活调度方案，确保系统对全局的有效掌控。本文的主要贡献如下：

In this paper, utilizing the spatio-temporal correlation of the sources in the area to be measured, we establish a Spatio-temporal Scope Information Model, which measure the effective information spatio-temporal domian of node data for the whole system in real time. 
In simple terms, the information value of a sensor data gradually decays in the temporal domain with the passage of time, and it also gradually decays in the neighborhood of the node location in the spatial domain.
Moreover, this paper investigates the efficient activation scheduling scheme for three nodes in different location layouts to ensure the effective grasp of the system over the whole area. The main contributions of this paper are as follows.
\begin{enumerate}
\item
  % 利用待测区域信源的时空相关性，提出范围信息模型，量化节点感知数据有用信息
  Utilizing the spatio-temporal correlation of sensor nodes, a Spatio-temporal Scope Information Model is proposed to quantify the valuable information of sensor data, which decays with space and time.
\item
  % 提出了单步最优信息获取决策机制，遍历不同节点布局情形得到具有三个节点的感知系统的的优化调度结果，且理论分析得到的不同调度情况的数值近似边界与仿真相吻合。
  A single-step optimal decision-making mechanism for a three-node sensing system is proposed. The theoretical analysis is made, and a method to solve the boundary node distribution among various scheduling situations is provided.
  
\item
  % 提出了长程平均信息获取决策机制，将三节点感知系统的信息获取建模为Markov决策过程，并利用Q学习算法求解出不同节点布局的最优调度结果。
  A long-term optimal decision-making mechanism for a three-node sensing system is proposed, which is modeled as a Markov decision process, and the Q-learning algorithm is utilized to solve the optimal scheduling results.
\item
  % 比较分析两种机制调度结果的不同性能表现，讨论两种机制的优缺点。
  % 单步机制的仿真结果与理论分析的各调度结果的边界节点布局吻合，长期机制利用Q学习算法得到了不同布局下对应的长期最优节点激活调度结果。并对两种机制的性能表现进行对比分析，分析了其优缺点。
  % 通过实验验证了两种机制的不同性能，对各自的优势和局限性进行了总结。

  With single-step mechanism, the approximate bounds for the node layout between partial scheduling results are obtained from theoretical analysis and numerical calculation, which match with the simulation results. 
  The optimal scheduling results with long-term mechanism corresponding to different node layout are obtained. 
  Finally, the different performances of the two mechanisms are experimentally verified, and the advantages and limitations of each are summarized.
  
\end{enumerate}
% 本文的其余部分结构如下：第二节主要描述了系统模型和优化问题；第三节则针对建立的模型提出两种优化机制，并进行对应部分的理论分析；第四节则主要给出部分理论数值计算相关结果与仿真结果；最后，第五节给出本文结论。
% 第六节进行了实验评估。
% \hl{(to be rewrited)}

The rest of the paper is organized as follows: In Section II, the system model and optimization problem are described. 
In Section III, a optimal decision-making mechanism of single-step information acquisition is proposed for the established model, and the related theoretical analysis is made. 
In Section IV, a optimal decision-making mechanism of long-term information acquisition is proposed and the modeling and solution methods are described.
In Section V, numerical results and simulation results are presented. 
In Section VI, experimental evaluation is performed. Finally, conclusions are drawn and discussions are made in Section VII.

\subsection{Related work}

\begin{table*}[t]
  \caption{Comparison of References.\label{ref compare}}
  \centering
  \begin{threeparttable}[b]
  \begin{tabular}{|c|c|c|c|c|c|}
  \hline 
  Reference & information source & sensors & correlation & metrics & Research results \\
  \hline 
  \cite{ref11} & Single-point  & Two & Spatial-temporal & Estimation Error & \tabincell{c}{optimal time
  shift \\between two sensors }\\
  \hline 
  \cite{ref12}, \cite{ref13} & multi-point  & Multiple & Spatial-temporal & Estimation Error & \tabincell{c}{DRL -based \\scheduling mechanism } \\
  \hline
  % \cite{ref14} &  & single & T & mutual information & \hl{undetermined} \\
  % \hline 
  \cite{ref15}, \cite{ref16} & \tabincell{c}{Single-point \\(noisy Ornstein-\\Uhlenbeck process)} & Single  & source and noise & mutual information & closed-form VoI\ \ expressions \\
  \hline 
  \cite{ref19}, \cite{ref20} & Regional  & Multiple & Spatial-temporal&\tabincell{c}{error-tolerable sensing \\(ETS) coverage }&  \tabincell{c}{AoI violation probability, \\optimal sensors\\ transmission power }  \\
  \hline 
  \cite{ref21} & Regional & Multiple & Spatial-temporal & \tabincell{c}{Node Spatial-temporal\\ coverage radius}& Node activation scheduling\\
  \hline 
  \cite{ref22}, \cite{ref23} & \tabincell{c}{time-varying Gauss-\\Markov Random
  Field\\ (GMRF) }& Multiple  &Spatial-temporal & \tabincell{c}{mean squared \\estimation error} & \tabincell{c}{Closed-form expressions \\for estimation error, \\optimal Spatial-temporal \\sampling
  rates  }\\
  \hline 
  \cite{ref24}, \cite{ref24Sensors} & N Gaussian process & Multiple & Spatial-temporal & mean squared error (MSE) &  optimal scheduling policy\\
  \hline 
  \cite{Satellite STI} & Single-point  & Multiple & Spatial-temporal & \tabincell{c}{Spatial-temporal \\mutual information}  & optimal update interval \\
  \hline 
  \cite{ref29}  & multi-point  &  Multiple& Spatial-temporal &\tabincell{c}{overall utility of\\information update}   & Status update node set\\
  \hline 
  this work & Regional  & Three & Spatial-temporal & \tabincell{c}{Spatial-temporal \\Scope information} & \tabincell{c}{Node activation  \\strategy in any layout }\\
  \hline 
  % &  &  &  &  &  \\
  % \hline
  \end{tabular}
  % \begin{tablenotes}
    % \tnote{1}
  %     \item[1] Spatial-temporal  
  %     \item[2] Deep Reinforcement Learning
  %     \item[3] Value of Information
  
  % \end{tablenotes}
 \end{threeparttable}
\end{table*}

% 在最近的一些研究中，以下几篇研究与本文工作较为相关。具体的，
Among some recent studies, the following ones are more relevant to the work of this paper. 
Specifically, reference \cite{ref11} considers a system consisting of two correlated information sources ,and establishes a minimal time shift between the two sources' updates, for which the system estimation error is minimal. 
An energy-aware scheduling mechanism based on Deep Reinforcement Learning (DRL) is proposed in \cite{ref12} and \cite{ref13}, which is capable of significantly prolonging the lifetime of a network of battery-powered sensors. %without hindering the overall performance of the sensing process.
A measure for the freshness of information is proposed in \cite{ref14}, which uses the mutual information between the real-time source value and the delivered samples at the receiver.
In \cite{ref15} and \cite{ref16}, a mutual-information based Value of Information (VoI) framework is formalised to characterise how valuable the status updates are for Hidden Markov Models.
An error-tolerable sensing (ETS) coverage as the area where the estimated information is with smaller error than the target value is defined in \cite{ref19} and \cite{ref20}, where the $\eta $-coverage probability is presented, and the optimal transmission power which minimizes the average energy consumption while guaranteeing a certain level of the $\eta $-coverage probability is provided.
In \cite{ref21}, a transmission scheduling approach is presented through a geometric approach with individual node coverage model, which is a function of the estimation accuracy in a region near the node.
The performance of state updates is studied in \cite{ref22} and \cite{ref23}, where the status is modeled as a time-varying Gauss-Markov Random Field (GMRF) and the estimation error of status update at the fusion center is analyzed. 
In \cite{ref24} and \cite{ref24Sensors}, an optimal scheduling policy for transmitting observations of spatio-temporally dependent processes over a limited number of communication channels is derived that minimizes the time-average mean squared error (MSE), resulting in a periodic scheduling sequence.
A novel timeliness metric spatially temporally correlative mutual information (STI) is proposed in \cite{Satellite STI}, where an optimal update interval is found by solving an integer optimization problem about slot allocation in satellite.
Assuming that each information can be commonly observed by multiple sensors, two multi-source information update problems, named AoI-aware Multi-Source Information Updating (AoI-MSIU) and AoI-Reduction-aware Multi-Source Information Updating (AoIR-MSIU) problems, are formulated in \cite{ref29}.

% 表1
% 纵观上述研究，尽管不少学者都结合了信源的时空相关性，对节点之间的合作调度和采样策略进行了研究，但并没有针对不同节点布局，在对整个区域有效掌握的前提下，分析所有可能出现的调度情况，并分析结果的变化规律。

% 例如在本文中，不同节点布局情况下的节点依次激活调度问题。

% 极为简单三个节点的依次激活调度问题，如何调度才能确保对整个平面的有效掌控，

% 研究不同节点分布下可能的优化调度结果,并针对所有节点布局的调度情形，分析其优化调度结果的变化规律，则是本文的主要研究动机。

Table \ref{ref compare} shows in detail the comparison between our work and the above references.
Throughout the above studies, although many scholars have studied the cooperative scheduling and sampling strategies among nodes by combining the spatio-temporal correlation of the sources, less attention has been devoted to all possible scheduling results and change rules of the scheduling results for different node layouts with an effective grasp of the whole region.
For detail, this paper discusses the node sequential activation scheduling problem for different node layout cases.
It is the main motivation of this paper to study the possible optimal scheduling results for different node layout cases, and analyze the change law of optimal scheduling results for different node layouts.
% \hl{to be modified}

% 综上所述，目前大部分研究集中在优化感知通信系统的AoI性能表现上，利用信源时空相关性的研究也较偏向于优化数据感知精度和数据AoI方面。由于时空相关性，考虑感知数据在特定的“面”或者“空间”内能够提供的有效范围信息更为合理，而研究感知数据的有效信息的相关研究还较少，且大多针对传感器部署所处的点区域的信息感知。

\section{System model and Problem formulation}

  % 系统构成：
 \subsection{System model}

% 本文主要考虑研究三个感知节点构成的简易感知系统，信息感知区域为整个二维平面，且对区域内每个位置的重视程度几乎相同，不存在感知优先级的设定，即如图\ref{System model}所示。
% 为了初步分析的便捷，假定系统定期决策激活一个感知节点进行信息获取，即每次决策时间间隔不变，并暂定数据传输条件理想，忽略数据发送与传输的影响，即假设节点采集完后监控端可以立刻收到其信息。

\begin{figure}[ht]
\centering
\includegraphics[width=2.5in]{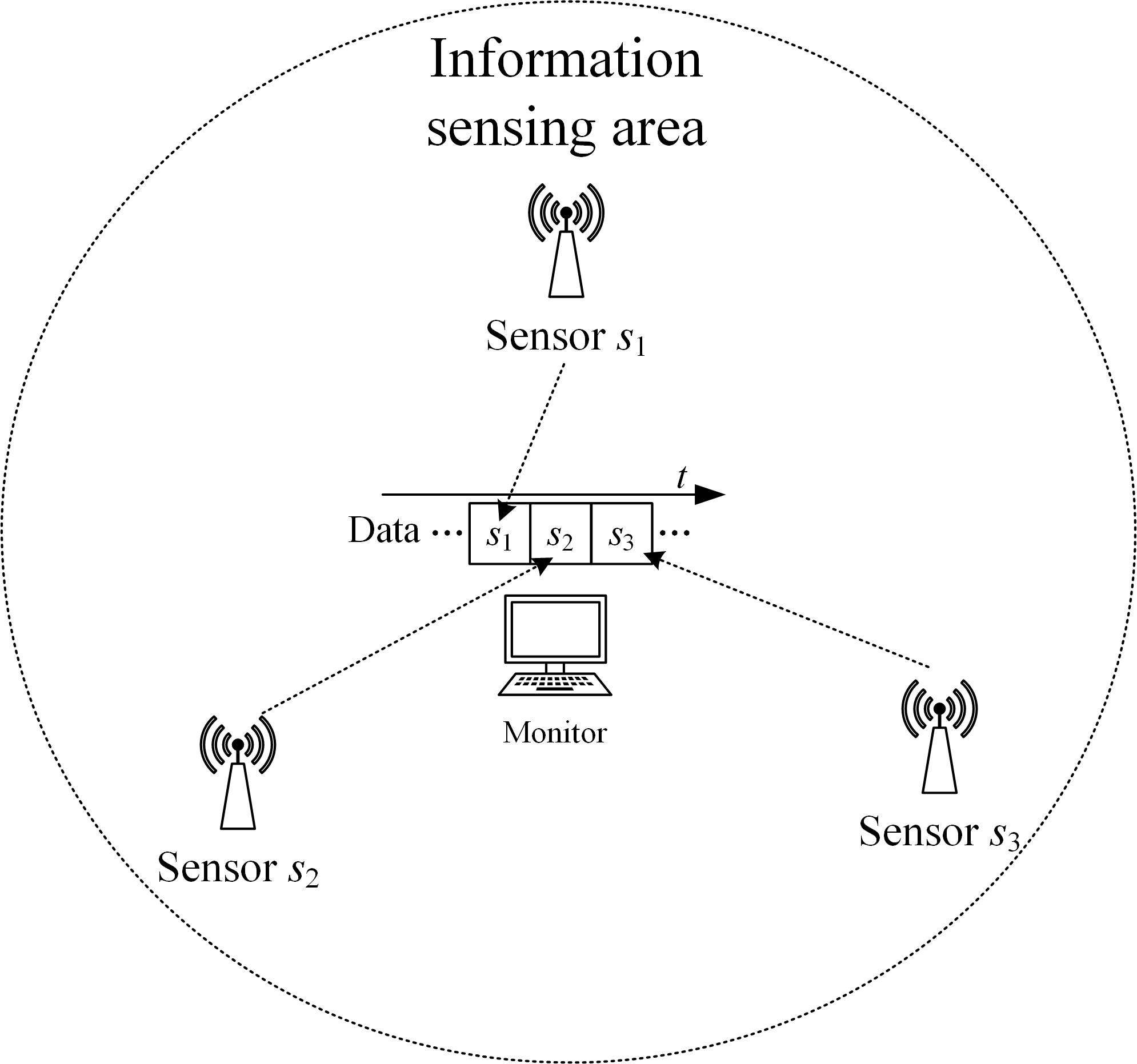}
\caption{System model}
\label{System model}
\end{figure}

In this paper, we mainly consider a simple sensing system composed of three sensing nodes, where the information sensing area is the whole two-dimensional (2D) plane, and the importance attached to each location in the area is the same, that is, there is no setting of sensing priority, as shown in Fig. \ref{System model}. For the convenience of preliminary analysis, it is assumed that the system periodically decides to activate a sensing node for information acquisition, i.e., the interval time between each decision is constant. The data transmission conditions are assumed to be ideal, ignoring the influence of data sending and transmission, i.e., it is assumed that the monitor can receive sensor data immediately after the node is activated.

% \begin{figure} [ht]
% 	\centering
% 	\subfloat[\label{model fig:a}]{
% 		\includegraphics[width=2.5in]{System Model}}
%     \\
% 	\subfloat[\label{model fig:b}]{
% 		\includegraphics[width=2.5in]{info Model}}
% 	\caption{Field Information Model}
% 	\label{model fig} 
% \end{figure}

\subsection{Spatio-temporal Scope Information Model}

\begin{figure}[ht]
  \centering
  \includegraphics[width=3.0in]{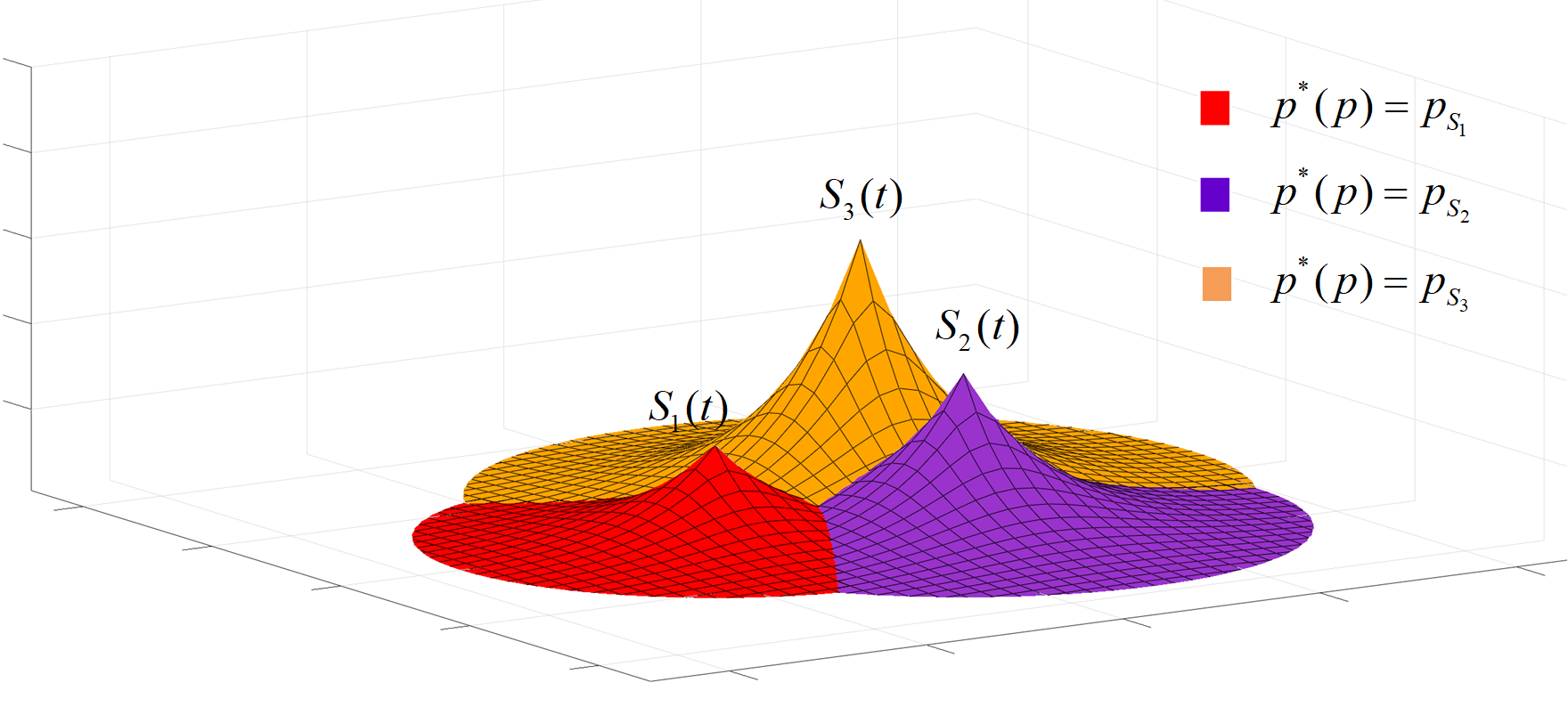}
  \caption{Spatio-temporal scope information map}
  \label{info model}
\end{figure}

% 在本文中，假设待测二维区域内为高斯随机场，即空间内任意点之间待测变量符合联合高斯分布。
% 且不考虑节点硬件采集误差，即认为节点采集的数据足够准确，完全代表了待测随机变量的取值。则针对某一节点来说，设该节点所在空间位置对应的待监测随机变量为X，则第一次激活该节点能为X带来的有价值信息量即为X的熵H(x)。
% 与此同时，针对节点附近邻域任一点p，设其对应的随机变量为Yp，且Yp与X符合联合高斯分布。记相关系数记为$\rho$，则该节点数据能再位置p提供的信息量为
In this paper, a Gaussian random field is assumed in the two-dimensional region to be measured, i.e., the variables to be measured between any points in space conform to the joint Gaussian distribution.
Without considering the node hardware acquisition error, i.e., the data collected by the nodes are considered accurate enough to completely represent the values of the random variables to be measured.
Then, for a certain node, let the random variable to be monitored corresponding to the spatial location of the node be $X$. The amount of valuable information that the first activation of the node can bring to $X$ is the entropy $H(x)$.
Meanwhile, for any point $p$ in the neighborhood near the node, let its corresponding random variable be $Y_p$, and $Y_p$ and $X$ conform to the joint Gaussian distribution. Denote the correlation coefficient as $\rho$, then the amount of information that the node data can provide at position $p$ is
\begin{equation}\label{info_p}
  I(p)=I(Y_p;X)=h(Y_p)-h(Y_p|X)=-\frac{1}{2}\log (1-{{\rho }^{2}})
\end{equation}
% 那么二维平面总信息量为
Then the total amount of information in the two-dimensional plane is
\begin{equation}
  \begin{split}
    {{I}_{D}}& =\int_{0}^{2\pi }{\int_{0}^{1}{-\frac{1}{2}\log (1-{{\rho }^{2}})\cdot \rho \cdot d\theta d}}\rho  \\ 
   & =-\frac{\pi }{2}\int_{0}^{1}{\log (1-{\rho }^{2})d{{\rho }^{2}}}=\frac{\pi }{2}  
  \end{split}  
\end{equation}
% 由上式可知虽然信息量计算式\cite{info_p}在相关系数$\rho$趋于1时该结果趋于无穷大，但二维平面的范围总信息积分收敛，可以计算。
It can be seen from the above equation that although the information calculation result in Eq. (\ref{info_p}) tends to infinity when the correlation coefficient $\rho$ tends to $1$, the total scope information integral in the two-dimensional plane converges and can be calculated.
% 在本文中，简化起见，我们采用时空可分离协方差函数\cite{ref11}，即
In this paper, for simplicity, we consider the spatiotemporally separable covariance function \cite{ref11}
\begin{equation}
  \rho ={{e}^{-{{\lambda }_{d}}\cdot d-{{\lambda }_{t}}\cdot t}}
\end{equation}
% 其中d代表空间距离，t代表时间差。lamda是缩放参数
where $d$ represents the spatial distance, and $t$ represents the time difference, in addition, $\lambda_d$ and $\lambda_t$ are the scaling parameters of information relevance with respect to space and time, respectively.
% 以上叙述都是针对单个节点且第一次激活采集的情况。而当系统有多个节点，并且按照一定顺序激活了相关节点时，二维平面内的每个位置会存在多个节点与其产生时空关联，换句话说，即在空间内的任一点，系统每个节点以往的感知数据都能提供一定量的信息，消除部分的不确定性。但为了简单起见，系统只保留每个节点相对最新的一次感知数据，且在该位置只选取能够消除最多不确定性，提供最多信息的节点感知数据为该位置提供参考，即认为在时空距离上最接近，最相关的节点数据为当前时刻该位置提供信息，暂不考虑多节点以往数据共同提供的联合信息量。数学表达，即为系统t时刻在二维平面内任意一点p掌握的信息量为
The above description is for a single node and the first activation of the information acquisition. 
However, when the system has multiple nodes and the nodes are activated in a certain order, there will be multiple nodes with spatio-temporal association at each location in the two-dimensional plane. In other words, at any point in space, the previous sensor data of each node can provide a certain amount of information and eliminate part of the uncertainty. In this paper, for the sake of simplicity, the system keeps only the latest sensor data of each node, and only the sensor data that can eliminate the most uncertainty and provide the most information at that location is selected to provide a reference for that location. In other words, it is considered that the node data closest to the location in terms of spatio-temporal domian, i.e., the most relevant node data, provides information for that location, and the amount of joint information provided by multiple nodes previous data together is not considered. 
That is, the amount of information available for the system at any point $p$ in the two-dimensional plane at moment $t$ is

\begin{equation}\label{2D_info}
  I(p,t)=-\frac{1}{2}{{\log }}\left( 1-{{e}^{-2{{\lambda }_{d}}\cdot |p-{{p}^{*}}(p)|-2{{\lambda }_{t}}\cdot |t-{{t}^{*}}(p)|}} \right).
\end{equation}
% 其中p*(p)为针对位置p其时空上最相关的感知数据的节点位置坐标，且有
where ${p^*}(p)$ is the node location coordinates of the most relevant sensor data in space-time for location $p$, and 
\begin{equation}
  {{p}^{*}}(p)=\mathop{\arg\min}\limits_{p_{s_i}}{{\lambda }_{d}}|p-p_{s_i}|+{{\lambda }_{t}}|t-t_{s_i}|,\ p_{s_i}\in S,
\end{equation}
% 其中S是所有节点坐标组成的集合。%此外，$AoI^*(p)$是在位置p处与其时空最相关节点数据的AoI值。 
% 此外，\ref{2D_info}式中的$|t-{{t}^{*}}(p)|$为当前时刻与位置p最相关数据的AoI取值
where $S$ is the set of all node coordinates. Furthermore, $|t-{{t}^{*}}(p)|$ in  Eq. (\ref{2D_info}) is the value of AoI for the most relevant data at position $p$ at the current moment.
According to the above equation, the spatio-temporal scope information map of the system at time t for the whole 2D space can be obtained as illustrated in Fig. \ref{info model}, where the vertical height represents the amount of valuable information, and the location of the peak is the sensing nodes location. 
% 则二维空间内总信息量为
% \hl{ Proof of convergence of the integral }
% 根据上式即可得到系统t时刻对于整个二维空间所掌握的范围信息地图如图\ref{info model}示意所示，其中尖峰的高度代表了信息量，尖峰所在的位置即为感知节点的位置。
% 此时若激活节点si，则针对平面内每个点p能够获取的信息增量为

At this moment, if the node $s_i$ is activated, the incremental information that can be obtained for each point $p$ in the 2D plane is 
\begin{equation}
  \begin{split}
    &  inf{{o}_{gain}}(p,s_i)=h({{Y}_{p}}|{{X}_{past}^*})-h({{Y}_{p}}|{{X}_{{{s}_{i}}}}) \\ 
   & =\left[ h({{Y}_{p}})-h({{Y}_{p}}|{{X}_{{{s}_{i}}}}) \right]-\left[ h({{Y}_{p}})-h({{Y}_{p}}|{{X}_{past}^*}) \right] \\ 
   & =-\frac{1}{2}\log \frac{1-{{e}^{-2{{\lambda }_{d}}\cdot |p-{{p}_{{{s}_{i}}}}\text{ }\!\!|\!\!\text{ }}}}{1-{{e}^{-2{{\lambda }_{d}}\cdot |p-{{p}^{*}}(p)|-2{{\lambda }_{t}}\cdot  AoI^*(p)}}}  
  \end{split}
\end{equation}
% 其中${Y}_{p}$为位置p处对应的随机变量，$X_{s_i}$为节点si处对应的随机变量, ${X}_{past}^*$为在位置p处最相关节点数据对应的随机变量。$AoI^*(p)$为与位置$p$最相关数据的AoI值。
where ${Y}_{p}$ is the random variable corresponding to position p, $X_{s_i}$ is the random variable corresponding to position of node $s_i$, and ${X}_{past}^*$ is the random variable corresponding to the most relevant node data at position $p$. In addition, $AoI^*(p)$ is the AoI value of the most spatiotemporally relevant data at position $p$.
% 然而，每次被系统决策激活的感知节点在某些区域内可能获取不到有价值的新信息。换句话说，即存在其他节点的先前数据在这些区域提供的信息要比此刻激活的节点提供的信息更多，在这些区域内上式的取值为负值，但这不符合信息的定义，因为即时不能提供更多的信息，但暂时不会造成损失. 所以在这些区域内能获取的信息增量定义为0,即激活节点si整个二维平面内的范围信息增量为
However, each time a sensor node activated by system may acquire no new valuable information in certain regions. In other words, the previous data of other nodes provide more information in these regions than the node activated at this moment, and the result of the above equation takes a negative value in these regions, but which does not meet the definition of information. Because no more information can be provided, but temporarily no loss is caused. 
Thus, the information increment that can be obtained in these areas is defined as 0, i.e., the scope information increment in the whole two-dimensional plane of the activated node $s_i$ is

\begin{equation}
  \begin{split}
    & {{I}_{gain}}=\iint\limits_{D}{\max \{inf{{o}_{gain}}(p,s_i),0\}}d\sigma  \\ 
   & =\iint\limits_{D}{\max \left\{ -\frac{1}{2}\log \frac{1-{{e}^{-2{{\lambda }_{d}}\cdot |p-{{p}_{{{s}_{i}}}}\text{ }\!\!|\!\!\text{ }}}}{1-{{e}^{-2{{\lambda }_{d}}\cdot |p-{{p}^{*}}(p)|-2{{\lambda }_{t}}\cdot AoI^*(p)}}},0 \right\}}d\sigma  \\ 
  \end{split}
\end{equation}

\subsection{Problem formulation}

% 在本文中，我们的目标是在给定节点位置布局和待测区域时空相关系数的情况下，寻找最高效的节点激活调度方式，从而达到节能的目的。即希望系统对整个二维平面掌握的信息总量在时间上的均值最大，如下式
In this paper, it is the research objective to find the most efficient way of node activation scheduling given the node location layout and the spatio-temporal correlation coefficient of the region to be measured. That is, it is desired that the total amount of information mastered by the system for the entire two-dimensional plane has the maximum mean value in time domian, as in the following equation
\begin{equation}
  \max \overline{I}=\underset{T\to \infty }{\mathop{\lim }}\,\frac{1}{T}\int_{0}^{T}{\big[ \iint\limits_{D}{I(p,t)d\sigma} \big]}dt 
\end{equation}
% 设每次激活节点时刻ti时系统掌握的信息总量为Ii,即
Let the total amount of information held by the system at each activation node moment $t_i$ be $I_i$, that is,
\begin{equation}
  {{I}_{i}}=\iint\limits_{D}{I(p,{{t}_{i}})d\sigma}.
\end{equation}
% 并记系统第i次激活节点后二维信息总量衰减函数为fi(t),即
and let the total two-dimensional information decay function after the $i_{th}$ activation node of the system be $f_i(t)$ as follows
\begin{equation}
  {{f}_{i}}(t)=\frac{\iint\limits_{D}{-\frac{1}{2}{{\log }_{2}}\left( 1-{{e}^{-2{{\lambda }_{d}}\cdot |p-{{p}^{*}}(p)|-2{{\lambda }_{t}}\cdot [Ao{{I}^{*}}(p)+t]}} \right)}d\sigma }{\iint\limits_{D}{-\frac{1}{2}{{\log }_{2}}\left( 1-{{e}^{-2{{\lambda }_{d}}\cdot |p-{{p}^{*}}(p)|-2{{\lambda }_{t}}\cdot Ao{{I}^{*}}(p)}} \right)}d\sigma }
\end{equation}
% 不妨令，则信息均值表达式可进一步写成：
Letting $T=n*\Delta t$, the expression for the information mean can be further written as
\begin{equation}
  \begin{split}
       \bar{I}
      %  &=\underset{n\to \infty }{\mathop{\lim }}\,\frac{1}{n\cdot \Delta t}\sum\limits_{i=1}^{n}{\int_{0}^{\Delta t}{{{I}_{i}}\cdot {{f}_{i}}(t)dt}} \\ 
     =\underset{n\to \infty }{\mathop{\lim }}\,\frac{\sum\limits_{i=1}^{n}{{{I}_{i}}\cdot \int_{0}^{\Delta t}{{{f}_{i}}(t)dt}}}{n\cdot \Delta t}   
  \end{split}
\end{equation}
% 且二维信息总量和激活节点信息增量有以下关系成立：
and the following relationship holds for the total amount of two-dimensional information and the incremental information of the activated nodes
\begin{equation}
  {{I}_{i+1}}={{I}_{i}}\cdot {{f}_{i}}(\Delta t)+{{I}_{gain}}(i),\ i=1,2,3,\cdots 
\end{equation}

% 将上式进行归纳易得以下结果
Recursive induction of the above equation leads to the following result
\begin{equation}
  \begin{split}
    {{I}_{n}} &={{I}_{gain}}(n)+ \sum\limits_{i=1}^{n-1}{\left[{{I}_{gain}}(i)\cdot \prod\limits_{j=i}^{n-1}{{{f}_{j}}(\Delta t)}\right]}, \\  
  \end{split}  
\end{equation}
then,
% 然后便有
\begin{equation}\label{Igain_fi}
  \begin{split}
    & \sum\limits_{i=1}^{n}{{{I}_{i}}\cdot \int_{0}^{\Delta t}{{{f}_{i}}(t)dt}} \\ 
    =& \sum\limits_{i=1}^{n}{\left[ \left( {I}_{gain}(i)+\sum\limits_{j=1}^{i-1}{{{I}_{gain}}(j)\cdot \prod\limits_{k=j}^{i-1}{{{f}_{k}}(\Delta t)}} \right)\cdot \int_{0}^{\Delta t}{{{f}_{i}}(t)dt} \right]} \\ 
     =&\sum\limits_{i=1}^{n}\left\{{{{I}_{gain}}(i)\cdot }\left[ \sum\limits_{j=i}^{n}{\left( \int_{0}^{\Delta t}{{{f}_{j}}(t)dt}\cdot \prod\limits_{k=i}^{j-1}{{{f}_{k}}(\Delta t)} \right)} \right] \right\}. \\ 
  \end{split}
\end{equation}
% 上式难以继续推导，主要原因在于fi(t)难以得到一个清晰的闭式，所以这里拟将所有时刻信息总量衰减式统一写为f(t),即
The above equation is difficult to continue to derive, mainly because $f_i(t)$ is challenging to get a clear closed form, so here it is proposed to unify the total information decay of all moments approximately expressed as $f(t)$, that is, 
\begin{equation}
 {{f}_{i}}(t)\approx f(t), i=1, 2, 3\cdots.
\end{equation}
% 换句话说，认为所有激活节点时刻信息总量衰减趋势统一。
% 这里这样近似假设的原因，主要出于考虑系统趋于获取更多，平稳的信息，从而导致每次激活节点时刻信息总量相差不多，从而造成衰减趋势可能基本相同。(这里要不要说数值计算验证了这一假设)
In other words, it is assumed that the total information decay trend is uniform at all activation node moments.
The reason for such an approximate assumption here is mainly motivated by the consideration that the system tends to acquire more and smoother information, which leads to similar total information at each activation node moment, thus causing the decay trend may be approximately the same.
% \hl{ the numerical calculation verifies this assumption or not? }
% 然后\ref{Igain_fi}式可进一步写为
Then Eq. (\ref{Igain_fi}) can be further written as
\begin{equation}
  \begin{split}
   & \int_{0}^{\Delta t}{f(t)dt}\cdot \sum\limits_{i=1}^{n}{\left( {{I}_{gain}}(i)\cdot \sum\limits_{j=0}^{n-i}{f{{(\Delta t)}^{j}}} \right)} \\ 
    =&\int_{0}^{\Delta t}{f(t)dt}\cdot \sum\limits_{i=1}^{n}{\left( {{I}_{gain}}(i)\cdot \frac{1-f{{(\Delta t)}^{n-i+1}}}{1-f(\Delta t)} \right)} \\ 
    =&\frac{\int_{0}^{\Delta t}{f(t)dt}}{1-f(\Delta t)}\cdot \sum\limits_{i=1}^{n}{\left[ {{I}_{gain}}(i)\cdot \left( 1-f{{(\Delta t)}^{n-i+1}} \right) \right]}  
  \end{split}
\end{equation}
% 所以二维总信息积分平均值进而可以改写成
% \begin{equation}
%   \bar{I}=\frac{\int_{0}^{\Delta t}{f(t)dt}}{\Delta t\left( 1-f(\Delta t) \right)}\cdot \underset{n\to \infty }{\mathop{\lim }}\,\cdot \frac{1}{n}\sum\limits_{i=1}^{n}{\left[ {{I}_{gain}}(i)\cdot \left( 1-f{{(\Delta t)}^{n-i+1}} \right) \right]}
% \end{equation}
% 因为$f(\Delta t)$<1,所以对任一大于0，且足够小的正数$\varepsilon $，
Since $f(\Delta t)$ < 1, for any positive number $\varepsilon $ that is greater than 0 and small enough,
\begin{equation}
  \begin{split}
  & 1-f{{(\Delta t)}^{n-i+1}}<1-\varepsilon  \\ 
 & i>n+1-{{\log }_{f(\Delta t)}}\varepsilon,  \\ 
\end{split}
\end{equation}
% 所以当n趋于无穷大时，信息增量$I_{gain}(i)$系数里小于$1-\varepsilon$的项数占比为
so when $n$ tends to infinity, the percentage of terms with information increment $I_{gain}(i)$ coefficients less than $1-\varepsilon$ is
\begin{equation}
  \underset{n\to \infty }{\mathop{\lim }}\,\frac{{{\log }_{f(\Delta t)}}\varepsilon }{n}=0.
\end{equation}
% 所以系数可以近似直接取1，即优化问题变为以下形式：
Therefore the coefficients can be approximated by 1, i.e., the optimization problem becomes the following form
\begin{equation}
  \begin{split}
    \max \ & \bar{I}=\frac{\int_{0}^{\Delta t}{f(t)dt}}{\Delta t\left( 1-f(\Delta t) \right)}\cdot \underset{n\to \infty }{\mathop{\lim }}\,\cdot \frac{1}{n}\sum\limits_{i=1}^{n}{{{I}_{gain}}(i)} \\ 
    \text{s}\text{.t}\text{.  }&Se({{t}_{i}})={{s}_{i}}\in \left\{ {{s}_{1}},{{s}_{2}},{{s}_{3}} \right\} \\ 
   & {{t}_{i}}=0,\Delta t,2\Delta t,\cdots, n\Delta t, \\ 
  \end{split}
\end{equation}
% 其中$Se(t_i)$为$t_i$时刻激活的节点。由上述表达可知，原问题转换为令每次激活节点获取的信息增量均值最大问题。
where $Se(t_i)$ is the node activated at the moment $t_i$.
From the above expression, the original problem is converted to the problem of maximizing the mean value of the incremental information obtained by each activated node.

\section{Single-step optimal mechanism}%Analysis
% \subsection{Single-step optimal decision mechanism}%单步信息获取最优决策机制
% 针对上述问题，首先容易想到的优化机制是单步决策法，即系统每次根据掌握的已知范围信息残余地图，通过计算求出当前时刻能够获取最多信息增量的感知节点。
% 然后系统使其激活并采集，记录并更新掌握的范围信息地图。
% 以后每个离散的决策时刻系统采取相同的步骤。
% 数学表达为每次激活节点si的准则可以写成：
Regarding the above-mentioned problems, the first optimization mechanism readily comes to mind is the single-step decision method, which means that the system computationally finds out the sensing node that can obtain the most information increment at the current moment based on the known spatio-temporal information residual map, then makes it active and updates the spatio-temporal information map.
The system takes the same actions at each discrete decision moment after that.
The mathematical expression as a rule for each node activation $s_i$ can be written as follows%Eq. (\ref{One step decision principle}), 
% \begin{equation}
%   {{s}_{i}}=\arg \max \iint\limits_{D}{\max \{0,{{e}^{-{{\lambda }_{d}}\cdot |p-{{p}_{{{s}_{i}}}}|}}-{{e}^{-{{\lambda }_{d}}\cdot |p-{{p}^{*}}(p)|-{{\lambda }_{t}}\cdot |t-{{t}^{*}}(p)|}}\}}d\sigma 
% \end{equation}
\begin{equation}\label{One step decision principle}
  Se(t_i)={ \underset{s_i\in \left\{ {{s}_{1}},{{s}_{2}},{{s}_{3}} \right\} }{\arg\max} \, \iint\limits_{D}{\max \{inf{{o}_{gain}}(p,s_i),0\}}d\sigma }.
\end{equation}
% 其中${{p}_{{{s}_{i}}}}\in S$，为激活的感知节点的位置坐标。
% 根据积分式中取最大值函数后一项的不同的取值，可以将积分区域分为后一项取值大于0和小于等于0的D'和D'\,'两部分，则其积分可写成：
where ${{p}_{{{s}_{i}}}}$ is the position coordinate of the activated sensing node.
According to whether the value of the latter term in the integral equation is greater than 0, the integration region can be divided into two parts, $D'$ and $D''$, thus the integral in Eq. (\ref{One step decision principle}) is equivalent to
\begin{equation}
  \iint\limits_{D'}{-\frac{1}{2}\log \frac{1-{{e}^{-2{{\lambda }_{d}}\cdot |p-{{p}_{{{s}_{i}}}}\text{ }\!\!|\!\!\text{ }}}}{1-{{e}^{-2{{\lambda }_{d}}\cdot |p-{{p}^{*}}(p)|-2{{\lambda }_{t}}\cdot AoI^*(p)}}}}d\sigma,
\end{equation}
% 其中D'和D'\,'分别为满足以下条件的点构成的集合：
where $D'$ and $D''$ are the set consisting of points that respectively satisfy the following conditions:
\begin{equation}
\left\{
   \begin{split}
  & p\in D',\ |p-{{p}_{{{s}_{i}}}}|-|p-{{p}^{*}}(p)|<\frac{{{\lambda }_{t}}}{{{\lambda }_{d}}}\cdot |t-{{t}^{*}}(p)| \\ 
  & p\in D'',\ |p-{{p}_{{{s}_{i}}}}|-|p-{{p}^{*}}(p)|\ge \frac{{{\lambda }_{t}}}{{{\lambda }_{d}}}\cdot |t-{{t}^{*}}(p)|.  
   \end{split} 
\right.
\end{equation}
% 对于三个点的系统来说，针对不同位置x的信息残留时空上最相关的节点x*也各不相同，具体存在以下关系：
% For a system of three nodes, 
In addition, the most spatially and temporally correlated node $p^*(p)$ varies for different locations $p$ with the following equations:
\begin{equation}
    \begin{split}
       {{p}^{\text{*}}}(p)\text{=}&\arg \min {{\lambda }_{d}}|p-{{p}^{*}}(p)|+{{\lambda }_{t}}|t-{{t}^{*}}(p)| \\ 
       =&
      \begin{cases}
        {{p}_{{{s}_{1}}}}, & p\in {{S}_{1}}  \\
        {{p}_{{{s}_{2}}}}, & p\in {{S}_{2}}  \\
        {{p}_{{{s}_{3}}}}, & p\in {{S}_{3}} , \\
      \end{cases}
    \end{split}
\end{equation}
% 且$S_1$,$S_2$,$S_3$分别为满足以下条件的点组成的集合：
where $S_1$, $S_2$ and $S_3$ are the sets of points that respectively meet the conditions as Eq. (\ref{remain judgment}), 
% $\Vert x \Vert_2$ %二范数的写法
% \begin{figure*}
%   \begin{equation}\label{remain judgment}
%     \left\{ \begin{split}
%       {{S}_{1}}=\left\{ p\in D\Big||p-{{p}_{{{s}_{1}}}}|-|p-{{p}_{{{s}_{2}}}}|<\frac{{{\lambda }_{t}}}{{{\lambda }_{d}}}\cdot ({{AoI}_{{{s}_{2}}}}-{{AoI}_{{{s}_{1}}}}) \right\} 
%       \cap \left\{ p\in D\Big| |p-{{p}_{{{s}_{1}}}}|-|p-{{p}_{{{s}_{3}}}}|<\frac{{{\lambda }_{t}}}{{{\lambda }_{d}}}\cdot ({{AoI}_{{{s}_{3}}}}-{{AoI}_{{{s}_{1}}}}) \right\} & \\
%       {{S}_{2}}=\left\{ p\in D\Big||p-{{p}_{{{s}_{2}}}}|-|p-{{p}_{{{s}_{1}}}}|<\frac{{{\lambda }_{t}}}{{{\lambda }_{d}}}\cdot ({{AoI}_{{{s}_{1}}}}-{{AoI}_{{{s}_{2}}}}) \right\} 
%       \cap \left\{ p\in D\Big||p-{{p}_{{{s}_{2}}}}|-|p-{{p}_{{{s}_{3}}}}|<\frac{{{\lambda }_{t}}}{{{\lambda }_{d}}}\cdot ({{AoI}_{{{s}_{3}}}}-{{AoI}_{{{s}_{2}}}}) \right\} & \\
%       {{S}_{3}}=\left\{ p\in D\Big||p-{{p}_{{{s}_{3}}}}|-|p-{{p}_{{{s}_{1}}}}|<\frac{{{\lambda }_{t}}}{{{\lambda }_{d}}}\cdot ({{AoI}_{{{s}_{1}}}}-{{AoI}_{{{s}_{3}}}}) \right\} 
%       \cap \left\{ p\in D\Big||p-{{p}_{{{s}_{3}}}}|-|p-{{p}_{{{s}_{2}}}}|<\frac{{{\lambda }_{t}}}{{{\lambda }_{d}}}\cdot ({{AoI}_{{{s}_{2}}}}-{{AoI}_{{{s}_{3}}}}) \right\} & \\ 
%     \end{split} \right.,
%   \end{equation}
% \end{figure*}
\begin{figure*}
  \begin{equation}\label{remain judgment} %写法2
    \left\{ \begin{split}
      {{S}_{1}}=\left\{ p\in D\Big||p-{{p}_{{{s}_{1}}}}|-|p-{{p}_{{{s}_{2}}}}|<\frac{{{\lambda }_{t}}}{{{\lambda }_{d}}}\cdot ({{AoI}_{{{s}_{2}}}}-{{AoI}_{{{s}_{1}}}})
      ,\ |p-{{p}_{{{s}_{1}}}}|-|p-{{p}_{{{s}_{3}}}}|<\frac{{{\lambda }_{t}}}{{{\lambda }_{d}}}\cdot ({{AoI}_{{{s}_{3}}}}-{{AoI}_{{{s}_{1}}}}) \right\} & \\
      {{S}_{2}}=\left\{ p\in D\Big||p-{{p}_{{{s}_{2}}}}|-|p-{{p}_{{{s}_{1}}}}|<\frac{{{\lambda }_{t}}}{{{\lambda }_{d}}}\cdot ({{AoI}_{{{s}_{1}}}}-{{AoI}_{{{s}_{2}}}}) 
      ,\ |p-{{p}_{{{s}_{2}}}}|-|p-{{p}_{{{s}_{3}}}}|<\frac{{{\lambda }_{t}}}{{{\lambda }_{d}}}\cdot ({{AoI}_{{{s}_{3}}}}-{{AoI}_{{{s}_{2}}}}) \right\} & \\
      {{S}_{3}}=\left\{ p\in D\Big||p-{{p}_{{{s}_{3}}}}|-|p-{{p}_{{{s}_{1}}}}|<\frac{{{\lambda }_{t}}}{{{\lambda }_{d}}}\cdot ({{AoI}_{{{s}_{1}}}}-{{AoI}_{{{s}_{3}}}})
      ,\ |p-{{p}_{{{s}_{3}}}}|-|p-{{p}_{{{s}_{2}}}}|<\frac{{{\lambda }_{t}}}{{{\lambda }_{d}}}\cdot ({{AoI}_{{{s}_{2}}}}-{{AoI}_{{{s}_{3}}}}) \right\} & . 
    \end{split} \right.
  \end{equation}
\end{figure*}
% 其中${{AoI}_{{{s}_{1}}}}$,${{AoI}_{{{s}_{2}}}}$,${{AoI}_{{{s}_{3}}}}$分别为每个节点最新一次感知数据的AoI取值。由上述表达式得激活节点si所能够获取的信息增量可以写成：
in which the ${{AoI}_{{{s}_{1}}}}$, ${{AoI}_{{{s}_{2}}}}$ and ${{AoI}_{{{s}_{3}}}}$ are the AoI of the latest sensing data of each node. The increment of information that can be obtained by the activated node $s_i$ from the above expression can be written as
\begin{equation}\label{Node information increment}
  \begin{split}
     I_{gain} 
    =&\sum\limits_{j=1}^{3}{\iint\limits_{{{S}_{j}}}{\max \{inf{{o}_{gain}}(p,s_i),0\}}d\sigma } \\ 
    =&\sum\limits_{j=1}^{3}{\iint\limits_{{{S}_{j}}'}{ {info_{gain}}(p,s_i)}d\sigma },
  \end{split}
\end{equation}
% 其中Sj'为满足条件$|x-{{x}_{{{s}_{i}}}}|-|x-{{x}_{{{s}_{j}}}}|<\frac{{{\lambda }_{t}}}{{{\lambda }_{d}}}\cdot |t-{{t}_{{{s}_{j}}}}|,x\in {{S}_{j}},j=1,2,3$的点组成的集合。
where $S_j'$ is the set of points that meet the condition as 
% \begin{equation}
%   |x-{{x}_{{{s}_{i}}}}|-|x-{{x}_{{{s}_{j}}}}|<\frac{{{\lambda }_{t}}}{{{\lambda }_{d}}}\cdot |t-{{t}_{{{s}_{j}}}}|,x\in {{S}_{j}},\ j=1,\ 2,\ 3.
% \end{equation}
\begin{equation}\label{set S'}
  |p-{{p}_{{{s}_{i}}}}|-|p-{{p}_{{{s}_{j}}}}|<\frac{{{\lambda }_{t}}}{{{\lambda }_{d}}}\cdot{AoI}_{{{s}_{j}}},\ p\in {{S}_{j}},\ j\in \left\{1,2,3\right\}.
\end{equation}

\begin{figure} [ht]
	\centering
	\subfloat[\label{fig:a}]{
		\includegraphics[width=1.60in]{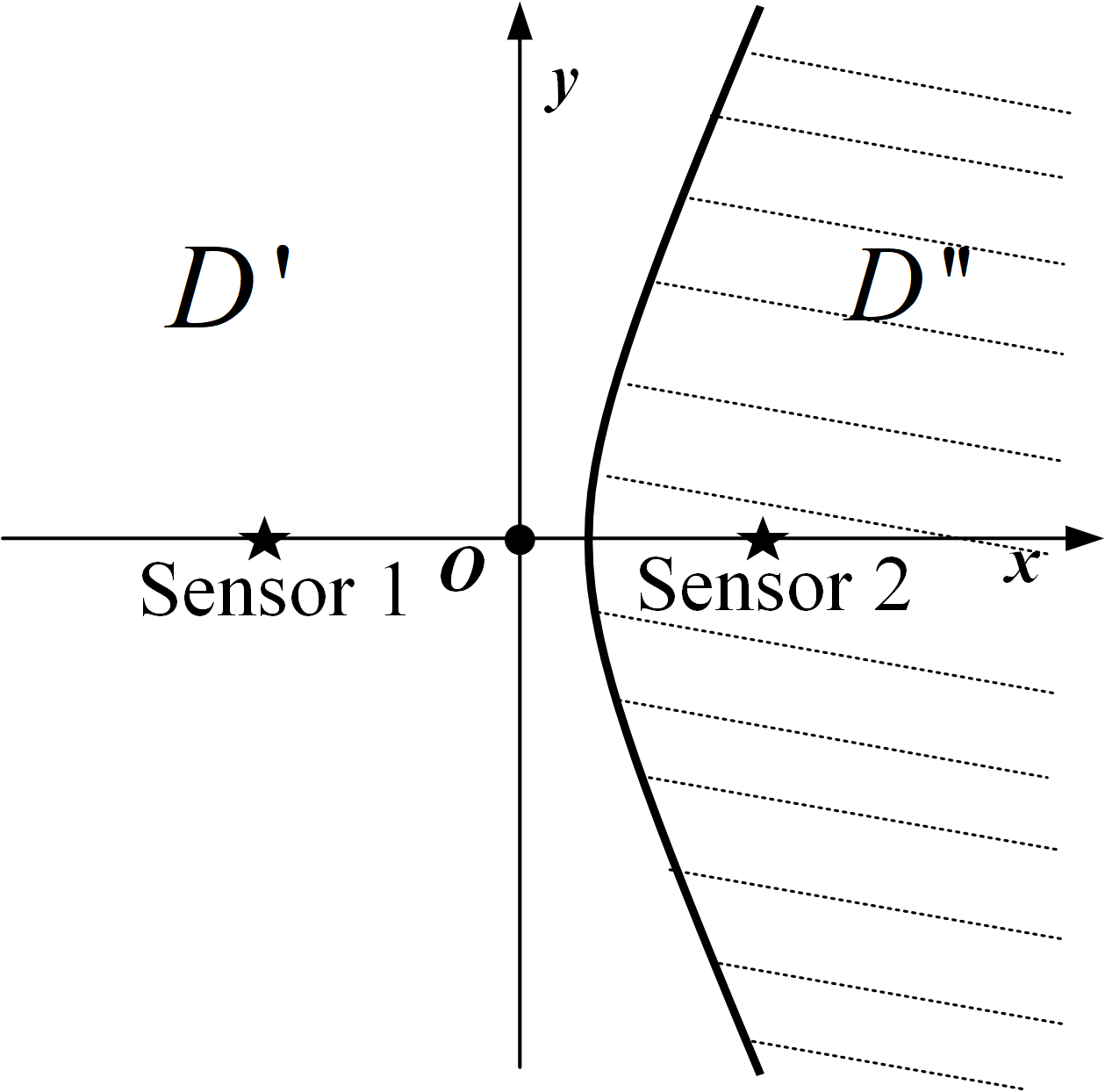}}
	\subfloat[\label{fig:b}]{
		\includegraphics[width=1.7in]{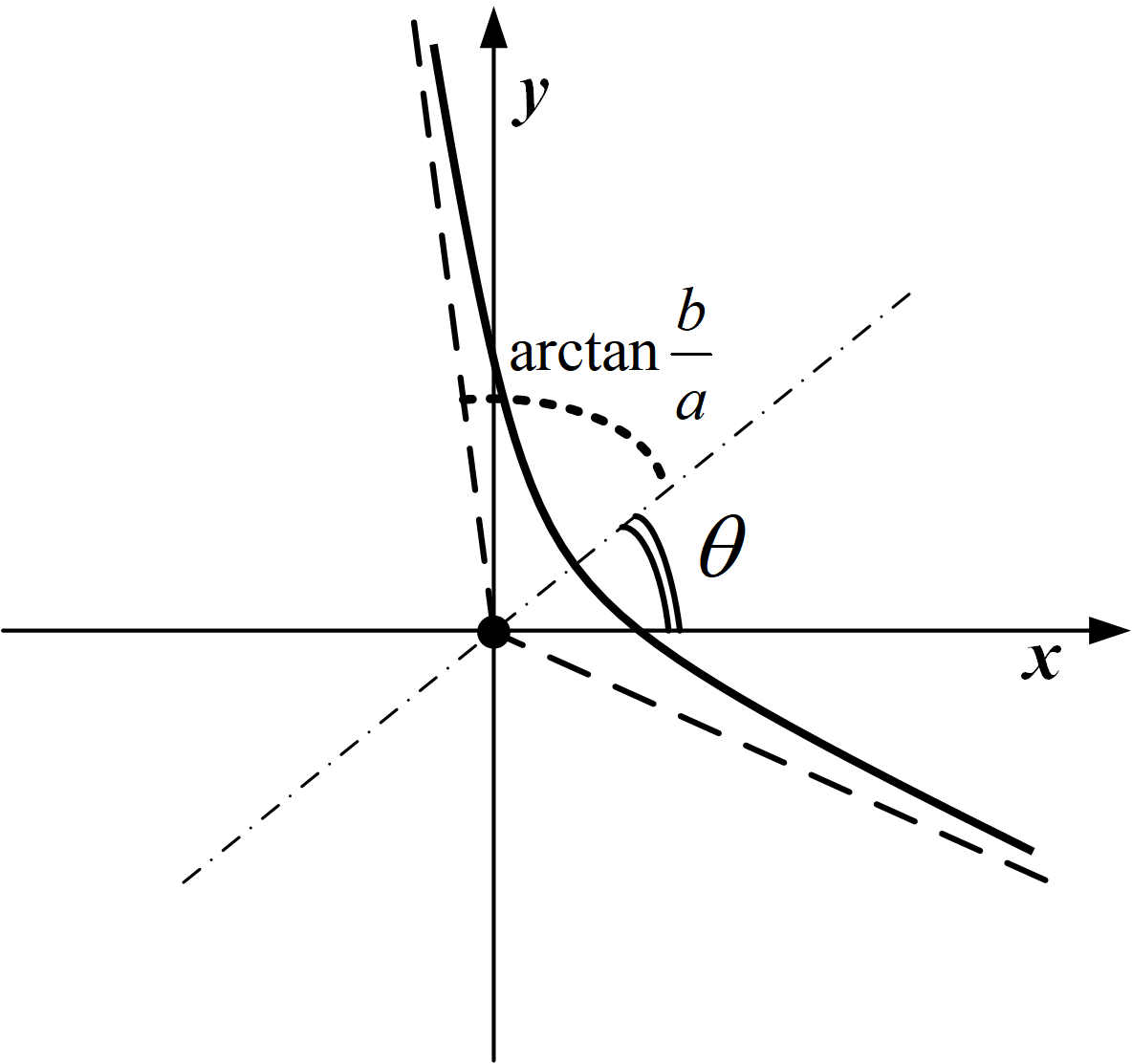}}
	% \caption{Regional distribution}
  \caption{Schematic diagram of the distribution of set elements and boundary curves. (a) Schematic of the set distribution and boundary curve of the second activation node of the two-point system. (b) Schematic of the curve after rotating $\theta$ counterclockwise around the origin, where the dashed line represents the asymptote of the curve. }
	\label{analysis fig} 
\end{figure}

% 由平面几何中到两点距离之差为定值的点的轨迹为双曲线可初步判断上述集合D'的边界可能包含若干条曲线。例如当系统只有两个节点，且${{p}_{{{s}_{i}}}}$，${{p}^{*}}(p)$分别取${{p}_{{{s}_{1}}}}$，${{p}_{{{s}_{2}}}}$，，$|t-{{t}^{*}}(p)|$取$\Delta t $时，设节点1,2在二维空间的坐标分别为（-c，0）,(c，0),由三角形两边之差小于第三边的性质易得：
From the fact that the trajectory of a point in plane geometry where the difference between the distances to two points is fixed is a hyperbola, we can preliminarily determine that the boundary of the above set $D'$ may contain several curves.
For example, when the system only has two nodes, and ${{p}_{{{s}_{i}}}}$, ${{p}^{*}}(p)$ are  taken as ${{p}_{{{s}_{1}}}}$, ${{p}_{{{s}_{2}}}}$, respectively, let the coordinates of node $s_1$ and node $s_2$ in 2D space be $(-c, 0)$, $(c, 0)$ as shown in Fig. \ref{analysis fig}(a). From the property that the difference between two sides of a triangle is less than the third side, it is easy to obtain the inequality as follows:
\begin{equation}
  |p-{{p}_{{{s}_{1}}}}|-|p-{{p}_{{{s}_{2}}}}|< d_{s_1,s_2}=2c .
\end{equation}
% 将$\frac{{{\lambda }_{t}}}{{{\lambda }_{d}}}\cdot \Delta t$简记为2a，则D''对应的不等式可以进一步写为
Abbreviating $\frac{{{\lambda }_{t}}}{{{\lambda }_{d}}}\cdot \Delta t$ to $2a$, the inequality corresponding to the set $D''$ can be written as
\begin{equation}\label{two node example}
  \begin{matrix}
    &2c>|p-{{p}_{{{s}_{1}}}}|-|p-{{p}_{{{s}_{2}}}}|\ge \frac{{{\lambda }_{t}}}{{{\lambda }_{d}}}\cdot \Delta t=2a>0\\ 
   \Leftrightarrow &
    \frac{{x}^{2}}{{a}^{2}}-\frac{{{y}^{2}}}{{{c}^{2}}-{{a}^{2}}}\ge 1,\ x>0,\ c>a>0 ,
   \end{matrix}
\end{equation} 
% 且当$c<a$,即$d_{s_1,s_2}<\frac{{{\lambda }_{t}}}{{{\lambda }_{d}}}\cdot \Delta t$时D''为空集，此时在整个二维平面内皆可获取一定的信息增量。但是需要注意的是，此时激活节点s2和激活节点s1能够获取的信息增量相同，从只考虑信息获取的角度上来讲此时不存在最优的激活方案。所以在之后的分析建模中，各节点之间距离需至少满足大于$\frac{{{\lambda }_{t}}}{{{\lambda }_{d}}}\cdot \Delta t$的条件。
and $D''$ is an empty set when $c<a$, i.e., $d_{s_1,s_2}<\frac{{{\lambda }_{t}}}{{{\lambda }_{d}}}\cdot \Delta t$, in which case, a certain increment of information can be obtained in the whole 2D plane.
However, it should be noted that activating node $s_1$ or node $s_2$ can obtain the same incremental information at this point, and there is no optimal activation scheme from the perspective of information acquisition only. 
Therefore, in the subsequent analytical modelling, the distance between the nodes must be at least greater than $\frac{{{\lambda }_{t}}}{{{\lambda }_{d}}}\cdot \Delta t$.%\hl{understood?}

% 由\ref{two node example}式可得D''的分布区域如图\ref{analysis fig} (a)阴影部分所示，在这片区域激活节点s1获取不到新的信息增量。而其余空白区域则为能够获取信息增量的区域，即集合D'。若要计算此时的信息增量的值，则需写出D''区域的边界曲线，以便信息增量积分限的表达。
The area distribution of $D''$ that can be obtained from Eq. (\ref{two node example}) is shown in the shaded part of Fig. \ref{analysis fig}(a), and no new information increment can be obtained by activating node $s_1$ in this area. The remaining blank areas are where new information increments can be obtained, that is, the set $D'$. The boundary curve formula of the $D''$ is required for the expression of the information increment's integral limit to calculate the information increment's value.
 
% 集合元素分布和分界曲线示意图。(a) 两点系统第二次激活节点的集合分布和边界曲线示意。(b)绕原点逆时针旋转theta后的曲线示意，且其中虚线代表该曲线的渐进线。

% \begin{figure}[ht]
%   \centering
%   \includegraphics[width=3.0in]{two nodes case.png}
%   \caption{two nodes}
%   \label{two nodes case}
% \end{figure}
% 然而，由于节点位置的不同以及坐标系角度的不同，当曲线的横轴不是处在x轴时，则利用二维旋转矩阵可以求出$\frac{{{x}^{2}}}{{{a}^{2}}}-\frac{{{y}^{2}}}{{{b}^{2}}}=1$
% 绕原点逆时针旋转$\theta$变换后的曲线方程为
However, due to different node positions and different angles of the coordinate system, when the horizontal axis of the curve is not parallel to the x-axis, a 2D rotation matrix can be used to derive the equation of the curve $\frac{{{x}^{2}}}{{{a}^{2}}}-\frac{{{y}^{2}}}{{{b}^{2}}}=1$ after rotating $\theta$ counterclockwise around the origin as 
\begin{equation}
  \label{Rotation curve}
  (\frac{{{\cos }^{2}}\theta }{{{a}^{2}}}-\frac{{{\sin }^{2}}\theta }{{{b}^{2}}}){{x}^{2}}+(\frac{{{\sin }^{2}}\theta }{{{a}^{2}}}-\frac{{{\cos }^{2}}\theta }{{{b}^{2}}}){{y}^{2}}+\sin 2\theta (\frac{1}{{{a}^{2}}}+\frac{1}{{{b}^{2}}})xy=1.
\end{equation} 
% 上述方程对应了两条曲线，而信息残余地图中不同区域的分界处仅为双曲线其中的一边，并随着旋转角度的不同，单条曲线的表达方程针对不同情况可能只方便写成$y=f(x)$或$x=g(y)$两种形式中的其中的一种。例如当旋转角为0时，位于y轴右侧的曲线的表达方程只方便写成$x=g(y)$，因为针对每个取值x，其存在两个与其对应的y值，所以无法表示成唯一的$y=f(x)$的表达形式。
% 而主要的判别方法为根据该曲线的两条渐进线与水平和垂直方向的直线最多的交点数进行判别。例如，如图	\ref{analysis fig} (b) 所示，当曲线的两条渐近线和与x轴平行的直线最多只有一个交点时，曲线可表示为x=g(y)的形式，当和与y轴方向平行的直线最多有一个交点时，曲线可表示为y=f(x)的形式. 然后可以得到两种表达形式的具体适用范围如下：
The above equation corresponds to a hyperbola, while the dividing curve of different regions in the information residual map is only one side of the hyperbola. Moreover, the expression equation of a single curve may be conveniently written as only one of two forms $y=f(x)$ or $x=g(y)$ with different rotation angles. For example, when the rotation angle is 0, the expression equation of the curve located on the right side of the y-axis is only conveniently written as $x=g(y)$. Because there are two corresponding $y$ values for each value of $x$, the curve cannot be expressed as a unique expression of $y=f(x) $. And the primary method of discrimination is based on the maximum number of intersections of the two asymptotic lines of the curve with the horizontal and vertical lines. For example, as shown in Fig. \ref{analysis fig}(b), the curve can be expressed as $x=g(y)$ when there is at most one intersection point between the two asymptotes of the curve and the line parallel to the x-axis. Meanwhile, the curve can be expressed as $y=f(x)$ when there is at most one intersection point between the two asymptotes of the curve and the line parallel to the y-axis. The applicability of the two expression forms can then be obtained as follows:
\begin{equation}\label{range}
  \left\{
    \begin{split}
      & x=g(y) \Leftrightarrow \theta \in [0,\arctan \frac{b}{a}]\cup [\pi -\arctan \frac{b}{a},\pi ] \\ 
    & y=f(x) \Leftrightarrow \theta \in [\frac{\pi }{2}-\arctan \frac{b}{a},\frac{\pi }{2}+\arctan \frac{b}{a}], \\ 
    \end{split}
  \right.
\end{equation} 
% \begin{figure}[ht]
%   \centering
%   \includegraphics[width=2.5in]{曲线表达适用范围.png}
%   \caption{适用范围}
%   \label{judge curve equation form}
% \end{figure}
% 其中$\frac{b}{a}$为双曲线渐近线的斜率的绝对值。容易发现，当$\arctan \frac{b}{a}<\frac{\pi }{4}$时，即$c<\sqrt{2}a$时,(12)式中的$\theta$取值范围没有完全覆盖$[0,\pi ]$,即存在某些范围内，单条曲线无法简单表示成以上两种函数形式，在这种情形下一种解决方案是通过调整坐标系角度，以便后续曲线方程表达。
where $\frac{b}{a}$ is the absolute value of the slope of the hyperbola asymptote. It is easy to find that when $\arctan \frac{b}{a}<\frac{\pi }{4}$, i.e., $c<\sqrt{2}a$, the range of values of $\theta$ in Eq. (\ref{range}) does not completely cover $[0,\pi ]$. In other words, there are certain ranges where a single curve cannot be expressed in the above two functional forms, and one solution in this case is to adjust the angle of the coordinate system for subsequent expression of the curve equation.

% 当三个节点呈现等边三角布局时，根据信息随时间的衰减，且个节点之间空间相关性相等，容易分析出此时在单步决策机制下其优化调度顺序为三个节点依次交替激活。但需要注意的是，上述节点交替激活的前提是节点之间距离大于$\frac{{{\lambda }_{t}}}{{{\lambda }_{d}}}\cdot 2\Delta t$。原因是若节点距离若不大于该值，每个节点的感知数据包含的有用信息经过两个时刻就被其余节点的数据覆盖，从而在每次决策时刻存在两个节点能够获取的信息增量相同，所以从信息获取的角度来讲此时不存在最优的决策。
% 下面则主要初步考虑只改变其中两个节点之间距离的取值，其他条件不变，即保持节点呈现等腰三角布局，分析其可能出现的调度情形。且三个节点的位置分布假设如图\label{Nodes location distribution} 所示。

\begin{figure} [ht]
	\centering
	\subfloat[isosceles triangle\label{fig:a}]{
		\includegraphics[width=1.5in]{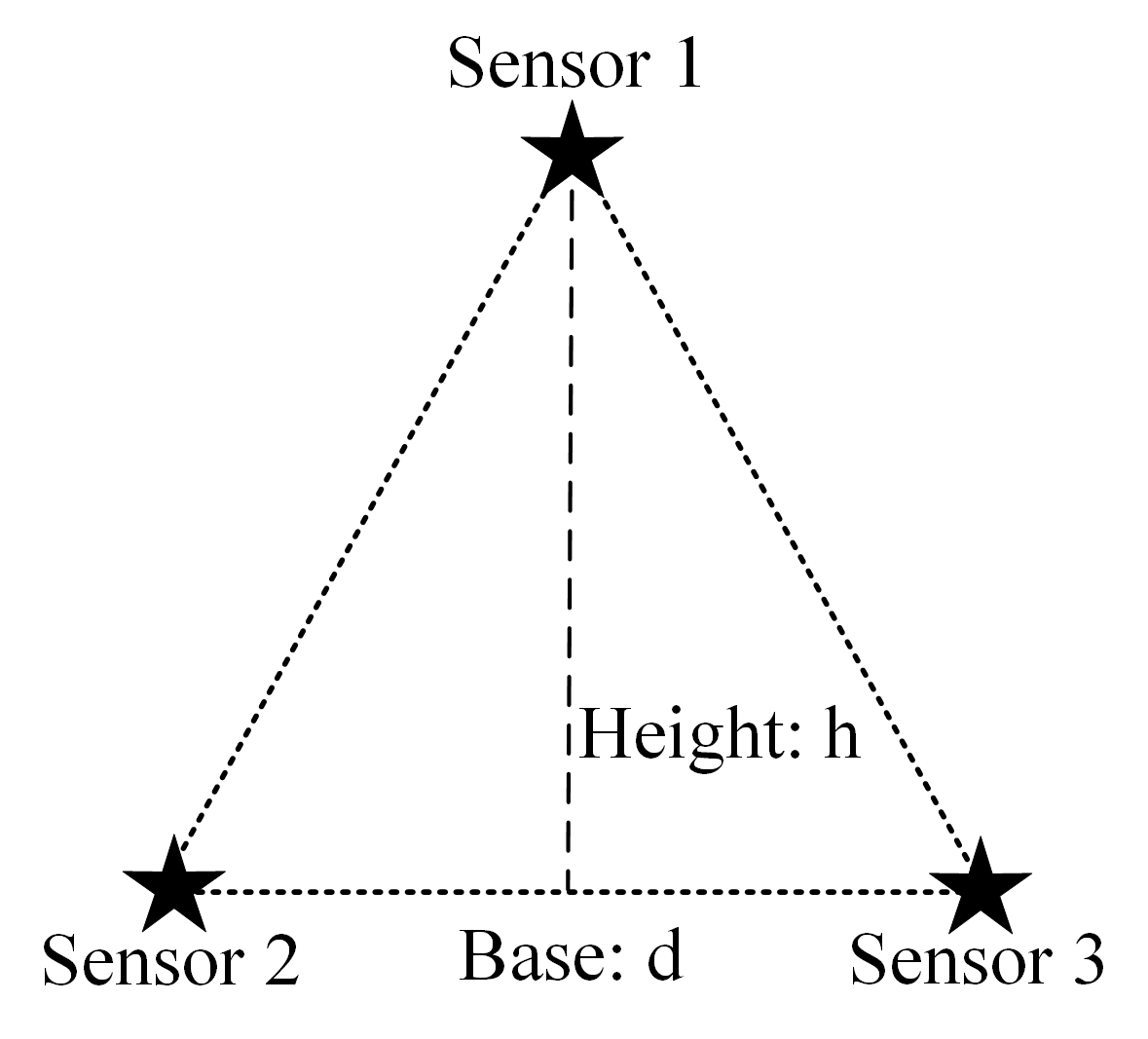}}
	\subfloat[general triangular\label{fig:b}]{
		\includegraphics[width=1.8in]{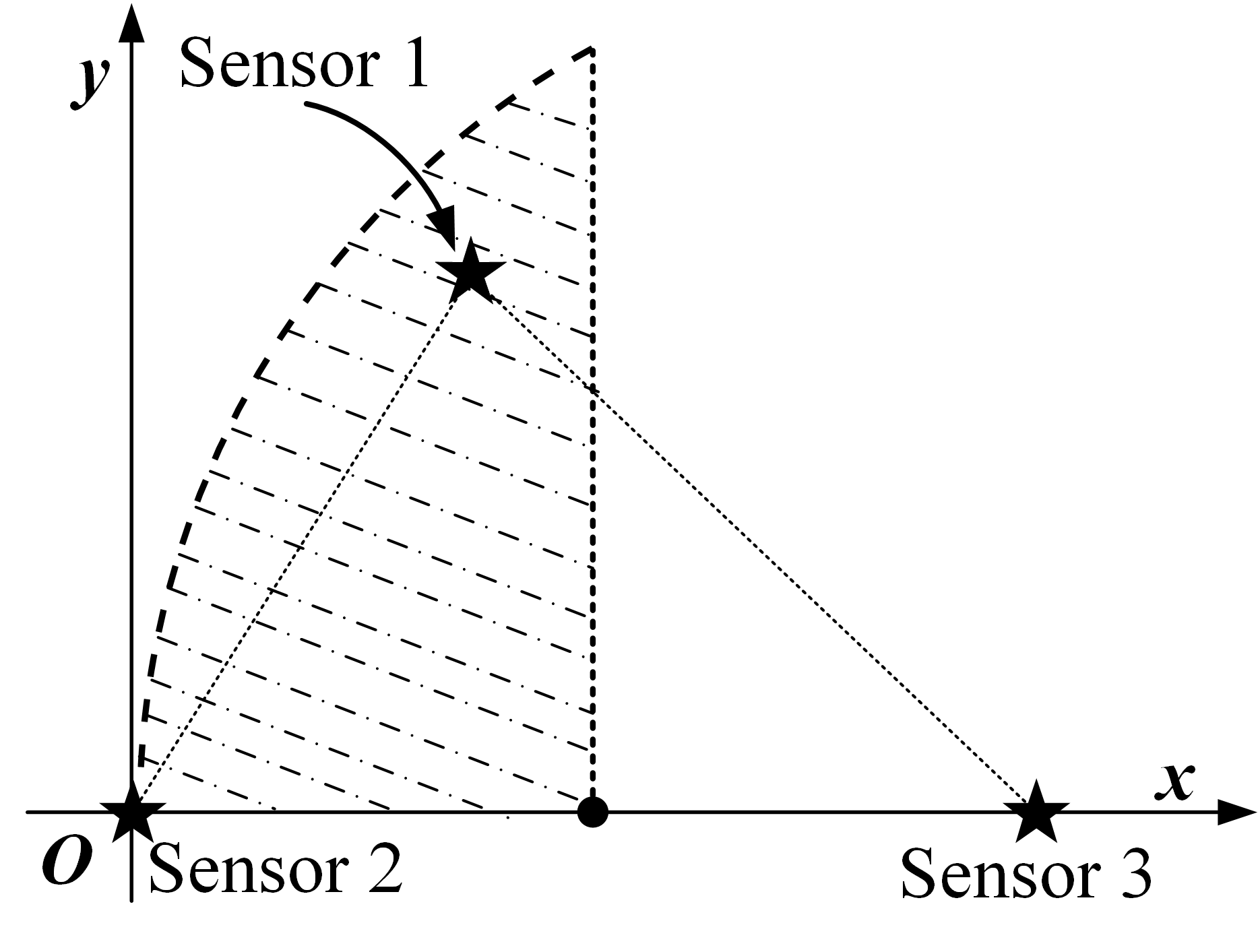}}
	% \caption{Nodes location distribution}
  \caption{The location distribution of three nodes. 
   (a) The node location distribution with three nodes presenting an isosceles triangle layout.
   (b) The node location distribution with three nodes presenting a general triangle layout, where the shaded area is the optional set of node $s_1$.
%   When three nodes  present an isosceles triangle distribution, the position of each node is shown in (a).
% When three nodes present a general triangular distribution, the position of each node is shown in (b), where the shaded area is the optional set of node 1.
}
	\label{Nodes location distribution} 
\end{figure}

When the three nodes present an equilateral triangular layout, it is easy to analyze that the optimal scheduling strategy under the single-step decision mechanism is alternate activation of the three nodes in turn according to the decay of information over time and the equal spatial correlation between individual nodes. 
However, it is essential to note that the above-mentioned alternate activation of nodes presupposes that the distance between nodes is greater than $\frac{{{\lambda }_{t}}}{{{\lambda }_{d}}}\cdot 2\Delta t$. 
The reason is that if the node distance is smaller than this value, the valuable information contained in the sensed data of each node can be covered by the data of other nodes after two moments so that there exists the same increment of information available to two nodes at each decision moment; therefore there is no optimal decision at this time from the perspective of information acquisition.
The following is a preliminary consideration to analyze the possible scheduling scenarios by changing the distance between only two nodes and keeping other conditions unchanged, i.e., keeping the nodes in an isosceles triangular layout. The location distribution of the three nodes is assumed as shown in Fig. \ref{Nodes location distribution}(a).
% \begin{figure}[ht]
%   \centering
%   \includegraphics[width=2.5in]{equilateral}
%   \caption{Node distribution fig}
%   \label{Node distribution}
% \end{figure}
% 三个节点在二维空间的位置分布。当节点呈现等腰三角形分布时，每个节点的位置分布如(a)所示。
% 节点呈现等腰三角分布的节点位置布局
% 当节点呈现一般三角形分布时，每个节点的位置分布如(b)所示，且其中阴影区域为节点1的可选集合。
\subsection{Three-node isosceles triangle layout} %三节点呈等腰三角布局

% \
% \newline
% \indent 
% 本小节主要分析仅改变节点2,3的距离，其他条件保持不变时可能出现的调度情况。
% 由先前分析知三节点呈等边三角形分布时决策调度结果为依次交替激活，即可设节点激活序列为123123123\ldots\ldots，所以从长期来看当每次激活节点3时刻节点1,2,3的最新一次感知数据的AoI取值分别为$2\Delta \tau$，$\Delta \tau$，$3\Delta \tau$，简记为[2,1,3]，此时再激活节点3获取的信息增量记为$info_{[2,1,3]}({{s}_{3}})$。
% 将AoI向量[2,1,3]的取值带入(\ref{remain judgment})式有以下结果：

\begin{figure} [ht]
	\centering
	\subfloat[$info_{[2,\ 1,\ 3]}$\label{fig:a}]{
		\includegraphics[width=1.7in]{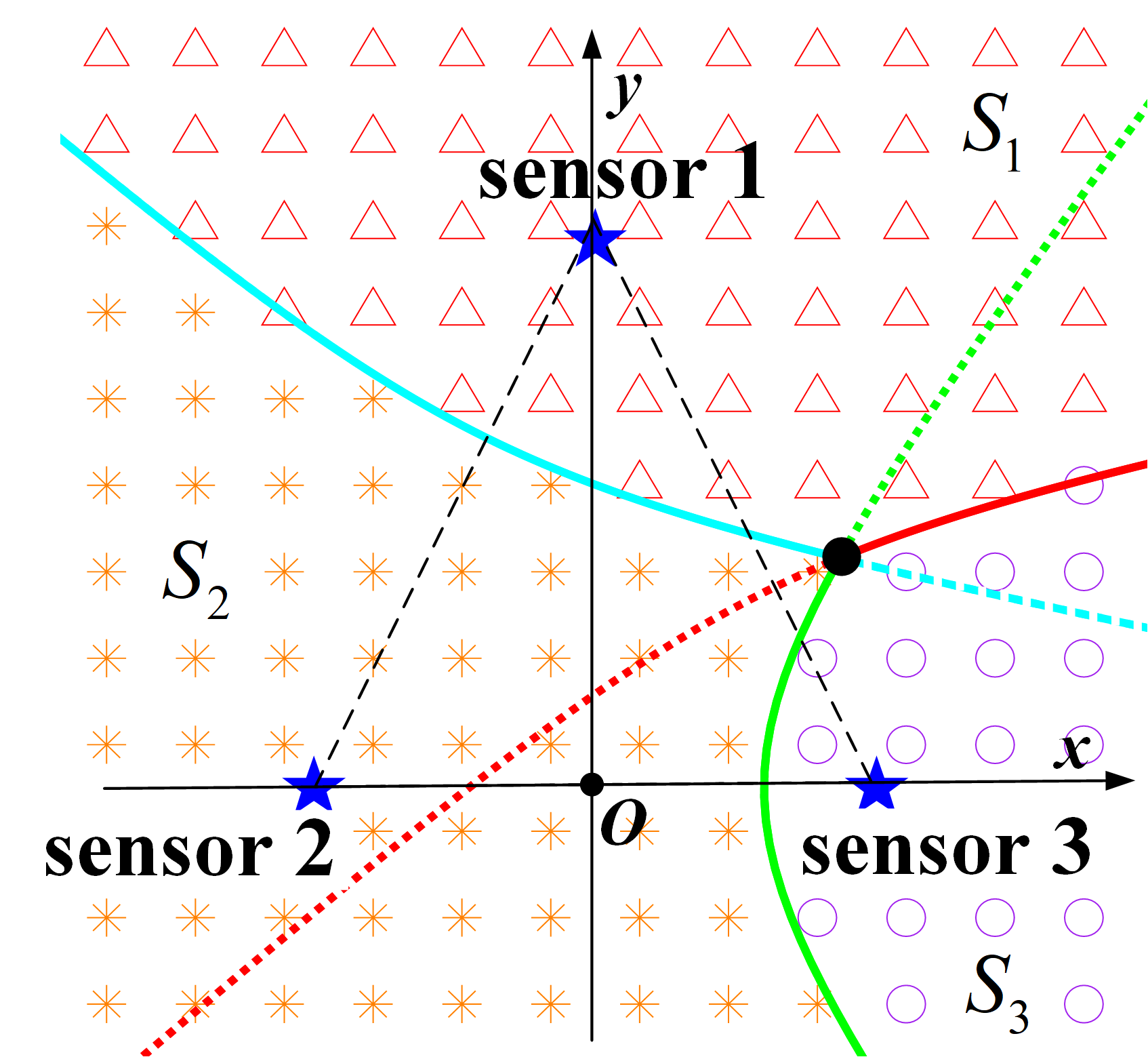}}
	\subfloat[$info_{[2,\ 1,\ 3]}({{s}_{3}})$\label{fig:b}]{
		\includegraphics[width=1.7in]{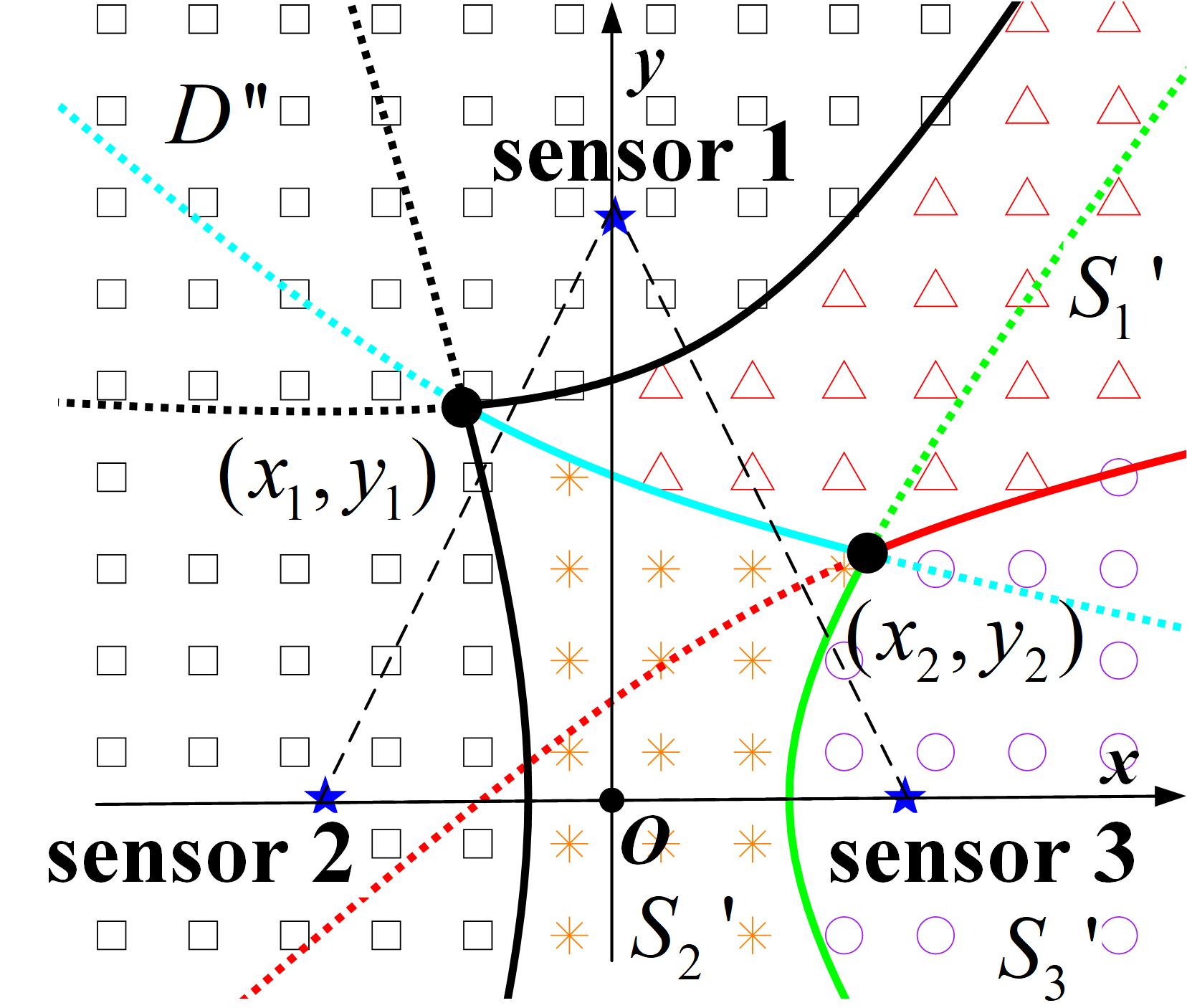}}
  \\
  \subfloat[$info_{[2,\ 1,\ 3]}({{s}_{3}})$\label{fig:c}]{
   \includegraphics[width=1.7in]{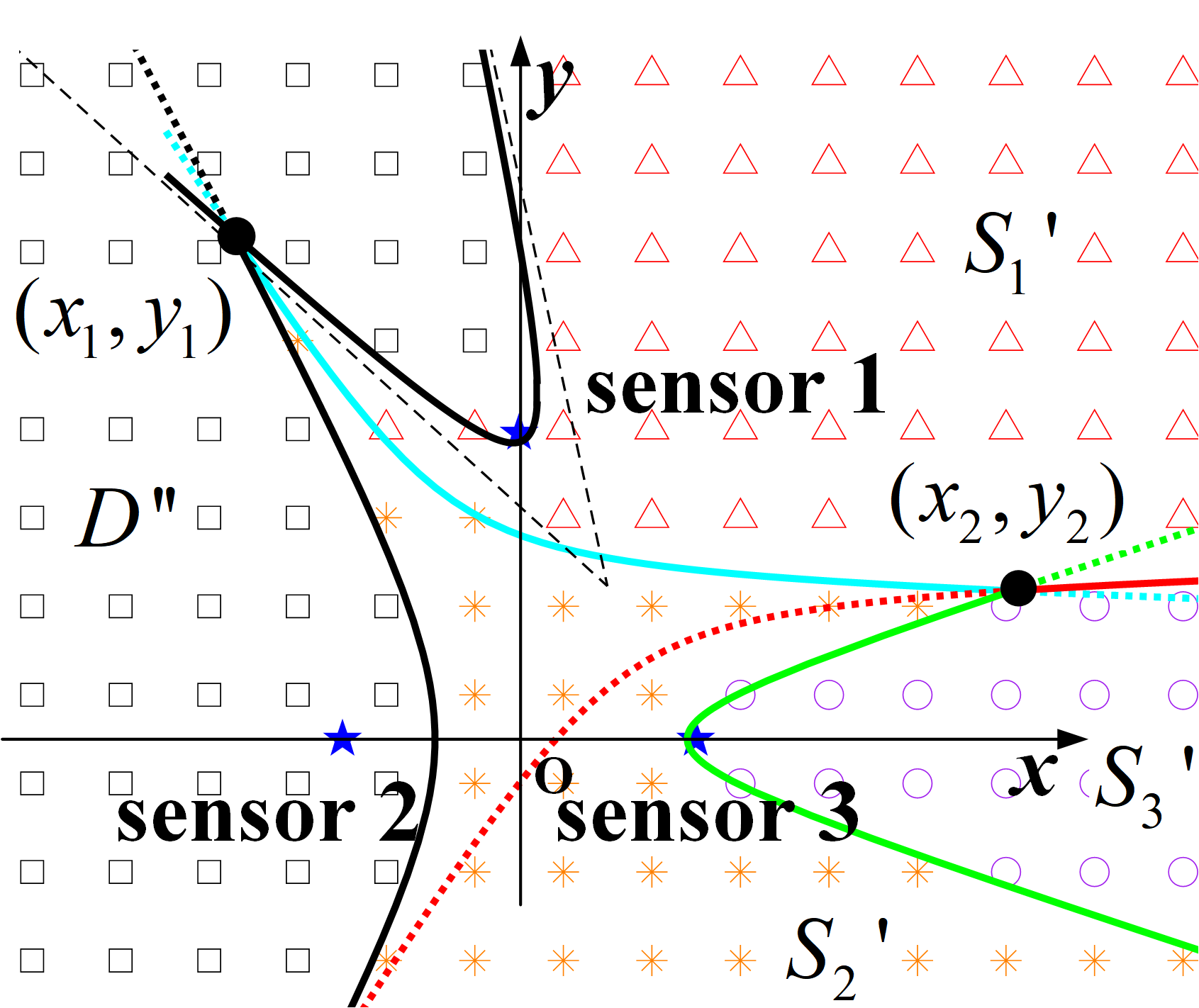}}
  \subfloat[$info_{[2,\ 1,\ 3]}({{s}_{3}})$\label{fig:d}]{
    \includegraphics[width=1.7in]{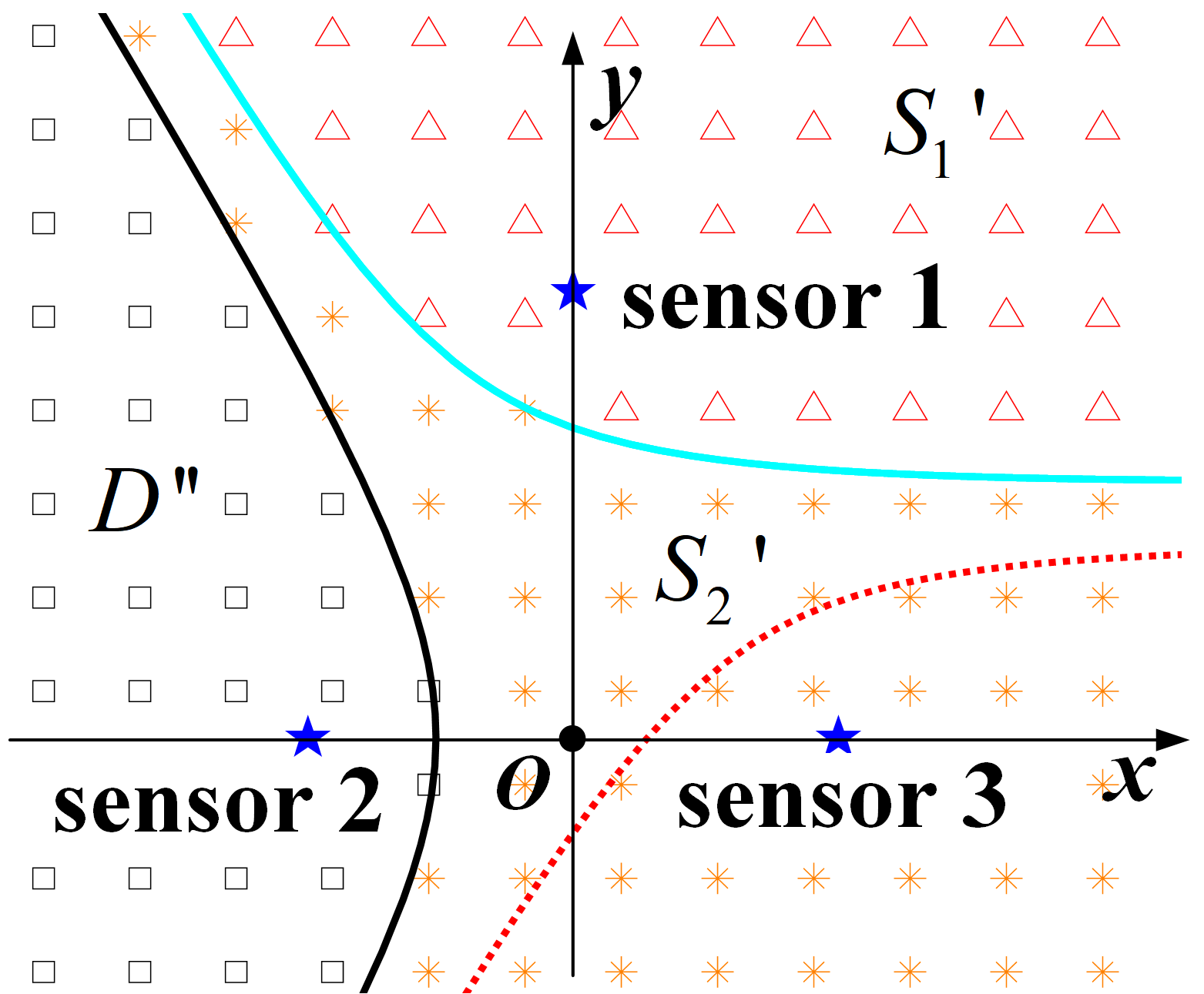}}
	% \caption{Spatio-temporal information map}
  \caption{Spatio-temporal information distribution maps. 
  % The typical information residual distribution map when the AoI vector is [2,1,3] is shown in (a). The typical information distribution map when the AoI vector is [2,1,3] and node $s_3$ is activated at this time is shown in (b). The special case boundary curve $C_{S_1,S_2}$ is not conveniently expressed is shown in (c). \hl{The information} distribution when the node distance is small and the boundary curve existence condition is not met is shown in (d).
  %% 写法2 参考了TWC
  (a) The typical information residual distribution map when the AoI vector is [2,1,3]. 
  (b) The typical information map when the AoI vector is [2,1,3] and node $s_3$ is activated at this time, which can be abbreviated as the map of $info_{[2,1,3]}(s_3)$.
  (c) The special case that boundary curve $C_{S_1,S_2}$ is not conveniently expressed.
  (d) The information map of $info_{[2,1,3]}(s_3)$ when the node distance is small, and the boundary curve existence condition is not met.
  }
	\label{info map} 
\end{figure}

This subsection mainly analyzes the possible scheduling scenarios when only the distance of node $s_2$ and $s_3$ is changed, and other conditions remain unchanged.
From the previous analysis, it is known that when the three nodes are laid out in an equilateral triangle, the scheduling strategy 
is to activate them alternately, and the node activation sequence can be set to $123123\cdots$.
Thus from a long-term perspective, when node $s_3$ is activated, the AoI values of the latest sensing data of nodes $s_1$, $s_2$, $s_3$ are $2\Delta t$, $\Delta t$, $3\Delta t$, respectively, which are abbreviated as $[2,\ 1,\ 3]$. At this time, the amount of incremental information obtained by activating node $s_3$ is recorded as $info_{[2,1,3]}({{s}_{3}})$.
Putting the AoI vector $[2,\ 1,\ 3]$ into Eq. (\ref{remain judgment}), the following results are obtained:
% \begin{equation}
%     \left\{ 
%       \begin{split}
%     & {{S}_{1}}=\{|p-{{p}_{{{s}_{1}}}}|-|p-{{p}_{{{s}_{2}}}}|<\frac{{{\lambda }_{t}}}{{{\lambda }_{d}}}\cdot \Delta t\cap |p-{{p}_{{{s}_{3}}}}|-|p-{{p}_{{{s}_{1}}}}|>\frac{{{\lambda }_{t}}}{{{\lambda }_{d}}}\cdot \Delta t\} \\ 
%   & {{S}_{2}}=\{|p-{{p}_{{{s}_{1}}}}|-|p-{{p}_{{{s}_{2}}}}|>\frac{{{\lambda }_{t}}}{{{\lambda }_{d}}}\cdot \Delta t\cap |p-{{p}_{{{s}_{3}}}}|-|p-{{p}_{{{s}_{2}}}}|>2\cdot \frac{{{\lambda }_{t}}}{{{\lambda }_{d}}}\cdot \Delta t\} \\ 
%   & {{S}_{3}}=\{|p-{{p}_{{{s}_{3}}}}|-|p-{{p}_{{{s}_{1}}}}|<\frac{{{\lambda }_{t}}}{{{\lambda }_{d}}}\cdot \Delta t\cap |p-{{p}_{{{s}_{3}}}}|-|p-{{p}_{{{s}_{2}}}}|<2\cdot \frac{{{\lambda }_{t}}}{{{\lambda }_{d}}}\cdot \Delta t\} \\ 
%   \end{split} 
%   \right.
% \end{equation}

\begin{equation}\label{S1 S2 S3 condition}
  \begin{split}
    & \left\{ \begin{split}
    & {{S}_{1}}:\left\{ \begin{split}
    & |p-{{p}_{{{s}_{2}}}}|-|p-{{p}_{{{s}_{1}}}}|>\frac{{{\lambda }_{t}}}{{{\lambda }_{d}}}\cdot \Delta t \\ 
   & |p-{{p}_{{{s}_{1}}}}|-|p-{{p}_{{{s}_{3}}}}|<\frac{{{\lambda }_{t}}}{{{\lambda }_{d}}}\cdot \Delta t \\ 
  \end{split} \right. \\ 
   & {{S}_{2}}:\left\{ \begin{split}
    & |p-{{p}_{{{s}_{2}}}}|-|p-{{p}_{{{s}_{1}}}}|<\frac{{{\lambda }_{t}}}{{{\lambda }_{d}}}\cdot \Delta t \\ 
   & |p-{{p}_{{{s}_{2}}}}|-|p-{{p}_{{{s}_{3}}}}|<\frac{{{\lambda }_{t}}}{{{\lambda }_{d}}}\cdot 2\Delta t \\ 
  \end{split} \right. \\ 
   & {{S}_{3}}:\left\{ \begin{split}
    & |p-{{p}_{{{s}_{1}}}}|-|p-{{p}_{{{s}_{3}}}}|>\frac{{{\lambda }_{t}}}{{{\lambda }_{d}}}\cdot \Delta t \\ 
   & |p-{{p}_{{{s}_{2}}}}|-|p-{{p}_{{{s}_{3}}}}|>\frac{{{\lambda }_{t}}}{{{\lambda }_{d}}}\cdot 2\Delta t \\ 
  \end{split} .\right. \\ 
  \end{split} \right. \\ 
  \end{split}
\end{equation}
% 显然可以得到信息分布图中不同区域的边界曲线方程分别满足下述条件：
It is evident that the formulas of the boundary curves for different regions in the information distribution map can be obtained, which satisfy the conditions as 
\begin{equation}\label{boundary condition}
  \left\{ \begin{split}
  &C_{S_1,S_2}: |p-{{p}_{{{s}_{2}}}}|-|p-{{p}_{{{s}_{1}}}}|=\frac{{{\lambda }_{t}}}{{{\lambda }_{d}}}\cdot \Delta t \\ 
 &C_{S_1,S_3}: |p-{{p}_{{{s}_{1}}}}|-|p-{{p}_{{{s}_{3}}}}|=\frac{{{\lambda }_{t}}}{{{\lambda }_{d}}}\cdot \Delta t \\ 
 &C_{S_2,S_3}: |p-{{p}_{{{s}_{2}}}}|-|p-{{p}_{{{s}_{3}}}}|=\frac{{{\lambda }_{t}}}{{{\lambda }_{d}}}\cdot 2\Delta t ,\\ 
\end{split} \right.
\end{equation} 
% 且上述边界曲线存在的充要条件分别为：
and the sufficient and necessary conditions for the existence of the above boundary curves are
\begin{equation}\label{Existence condition}
  \left\{ \begin{split}
  & d_{s_1,s_2}>\frac{{{\lambda }_{t}}}{{{\lambda }_{d}}}\cdot \Delta t \\
 & d_{s_1,s_3}>\frac{{{\lambda }_{t}}}{{{\lambda }_{d}}}\cdot \Delta t  \\ 
 & d_{s_2,s_3}>\frac{{{\lambda }_{t}}}{{{\lambda }_{d}}}\cdot 2\Delta t.  
\end{split} \right.
\end{equation} 
% 且若有其中某个条件不满足，则对应的边界曲线则不存在。
% 将等腰三角底边中点作为原点，底边作为x轴方向建立直角坐标系，并设底边长度为d，高为h，则三个节点坐标依次为(0,d),(-d/2,0),(d/2,0)。
% 此时根据\ref{S1 S2 S3 condition},\ref{boundary condition}式即可画出此时系统的信息残余分布图，如图\ref{info map}(a)所示,其中上三角填充区域代表了集合S1,*型填充区域代表了集合S2,圆圈填充区域代表了集合S3。
% 且图\ref{info map}(a)所示的情形是满足\ref{Existence condition}式条件时的情形，当存在其中某个条件不满足时，则对应的边界曲线则不存在，分布地图则相对简化，这里限于篇幅不详细展开。
% 三条边界曲线必然交于一点，可由双曲线几何距离性质证明。
If one of the conditions is not met, the corresponding boundary curve does not exist.
Taking the midpoint of the base of the isosceles triangle as the origin and the base as the $x$-axis direction to establish a rectangular coordinate system, and setting the base length as $d$ and the height of the triangle as $h$, the coordinates of the three nodes are $(0, d)$, $(-\frac{d}{2},0)$, $(\frac{d}{2},0)$, respectively.
Then the information residual distribution map of the system can be drawn according to Eq. (\ref{S1 S2 S3 condition}) and Eq. (\ref{boundary condition}), as shown in Fig. \ref{info map}(a), where the upper triangular, star and circular marker areas represent the sets $S_1'$, $S_2'$, and $S_3'$, respectively, meanwhile, the solid lines are the boundaries of the different set regions, and the dashed lines are the extensions of the boundary curves.
% hl{to be written}
% And Figure 5(a) is the case when the 5 conditions are satisfied, when there is not to meet one of the conditions, then the corresponding boundary curve does not exist, the distribution map is relatively simplified, here limited space not to expand in detail.
The three boundary curves necessarily intersect at a point, which can be proved by the geometric distance property of the hyperbola.
% 当曲线的旋转角theta为0，且曲线中心位于原点时，单条曲线的表达通式为
% Located in the horizontal direction of the x-axis, i.e., the rotation angle $\theta$ is 0, the general formula of the single curve is
When the rotation angle $\theta$ of the curve is 0 and the centre of the curve is at the origin, the general formula of a single curve is
\begin{equation}\label{theta=0,x=gy}
  x=h(y,a,c,sign)=sign\cdot a\cdot \sqrt{1+\frac{{{y}^{2}}}{{{c}^{2}}-{{a}^{2}}}},
\end{equation}
% 其中sign为符号变量，取值为$\pm 1$;$a=\frac{{{\lambda }_{t}}\cdot \Delta t }{2{{\lambda }_{d}}}$;c为双曲线焦距，即两节点之间距离的一半。所以S2与S3分界曲线可以写成：
where $sign$ is a symbolic variable that takes the value $\pm 1$, and $c$ is the hyperbolic focal length, which is half of the distance between the two nodes. Thus the set $S_2$ and $S_3$ boundary curve expression can be written as
\begin{equation}
  C_{S_2,S_3}:x=h(y,2a,\frac{d}{2},1)=2a\cdot \sqrt{1+\frac{{{y}^{2}}}{\frac{{{d}^{2}}}{4}-4{{a}^{2}}}}.
  % x=2a\cdot \sqrt{1+\frac{{{y}^{2}}}{\frac{{{d}^{2}}}{4}-4{{a}^{2}}}}
\end{equation}
% S2,S1分界曲线和S1,S3分界曲线的主要表达通式可由\ref{Rotation curve}式结合\ref{range}可得到\ref{x y express}式
The general expression of $S_1$, $S_2$ boundary curve and $S_1$, $S_3$ boundary curve can be written as Eq. (\ref{x y express}) by combining Eq. (\ref{Rotation curve}) with Eq. (\ref{range}), 
\begin{figure*}%[b]
  \begin{equation}\label{x y express}
    \left\{ \begin{split}
    &y=f(x)={{y}_{o}}-\frac{C}{2B}\cdot (x-{{x}_{o}})\pm \frac{1}{\sqrt{|B|}}\sqrt{1+(\frac{{{C}^{2}}}{4B}-A){{(x-{{x}_{o}})}^{2}}},\ \theta \in [\frac{\pi }{2}-\arctan \frac{b}{a},\frac{\pi }{2}+\arctan \frac{b}{a}]  \\
    &x=g(y)={{x}_{o}}-\frac{C}{2A}\cdot (y-{{y}_{o}})\pm \frac{1}{\sqrt{|A|}}\sqrt{1+(\frac{{{C}^{2}}}{4A}-B){{(y-{{y}_{o}})}^{2}}},\ \theta \in [0,\arctan \frac{b}{a}]\cup [\pi -\arctan \frac{b}{a},\pi ] . \\
  \end{split} \right.
  \end{equation}
\end{figure*}
% 且其中$({{x}_{o}},{{y}_{o}})$为对应曲线中心点即两条渐近线交点坐标，$\theta $为该曲线的旋转角，
where $({{x}_{o}}, {{y}_{o}})$ are the coordinates of the center point of the corresponding curve, i.e., the intersection of the two asymptotes, and $\theta $ is the rotation angle of the curve.
% $a=\frac{{{\lambda }_{t}}\cdot \Delta \tau }{2{{\lambda }_{d}}}$，  
% c为焦距，即节点距离的一半，
In addition, $c$ is the hyperbolic focal length, and%$b=\sqrt{c^2-a^2}$, $A=\frac{{{\cos }^{2}}\theta }{{{a}^{2}}}-\frac{{{\sin }^{2}}\theta }{{{b}^{2}}}$, $B=\frac{{{\sin }^{2}}\theta }{{{a}^{2}}}-\frac{{{\cos }^{2}}\theta }{{{b}^{2}}}$, $C=\sin 2\theta (\frac{1}{{{a}^{2}}}+\frac{1}{{{b}^{2}}})$. 
\begin{equation}
  \left\{
  \begin{split}
    &a=\frac{{{\lambda }_{t}}\cdot \Delta t }{2{{\lambda }_{d}}},\ b=\sqrt{c^2-a^2}\\
    &A=\frac{{{\cos }^{2}}\theta }{{{a}^{2}}}-\frac{{{\sin }^{2}}\theta }{{{b}^{2}}} \\
    &B=\frac{{{\sin }^{2}}\theta }{{{a}^{2}}}-\frac{{{\cos }^{2}}\theta }{{{b}^{2}}}\\
    &C=\sin 2\theta (\frac{1}{{{a}^{2}}}+\frac{1}{{{b}^{2}}}).
  \end{split}
  \right.
\end{equation}
% 为了方便后续表达，可将(\ref{x y express})式的简记为以下形式：
For ease of subsequent expression, the Eq. (\ref{x y express}) is abbreviated to the following form:
\begin{equation}
  \left\{ \begin{split}
    & y=f(x,a,\text{c},{{x}_{o}},{{y}_{o}},\theta ,sign),\theta \in {{\Theta }_{1}} \\ 
   & x=g(y,a,c,{{x}_{o}},{{y}_{o}},\theta ,sign),\theta \in {{\Theta }_{2}}, \\ 
  \end{split} \right.
\end{equation}
% 其中sign取值为正负1，代表\ref{x y express}中两种形式表达中正负号的选取。
% 然后便可得到S2,S1和S1,S3分界曲线,得到\ref{C_S1,S2}和\ref{C_S1,S3}式。
where $sign$ takes the value $\pm 1$, representing the selection of positive or negative signs in the two expressions in Eq. (\ref{x y express}).
Then the $S_1$, $S_2$ and $S_1$, $S_3$ dividing lines can be obtained as  Eq. (\ref{C_S1,S2}) and Eq. (\ref{C_S1,S3}).
% \hl{S1,S2 dividing lines or boundary curve???}
\begin{figure*}
  \begin{equation}\label{C_S1,S2}
    C_{S_1,S_2} : \left\{
      \begin{split}
        &y=f\left(x,a,\frac{\sqrt{{{h}^{2}}+{{d}^{2}}/4}}{2},\ -\frac{d}{4},\frac{h}{2},\arctan (\frac{2h}{d}),1\right), \arctan (\frac{2h}{d}) \in [\frac{\pi }{2}-\arctan \frac{b}{a},\ \frac{\pi }{2}+\arctan \frac{b}{a}] \\
        &x=g\left(y,a,\frac{\sqrt{{{h}^{2}}+{{d}^{2}}/4}}{2},-\frac{d}{4},\frac{h}{2},\arctan (\frac{2h}{d}),1\right),\arctan (\frac{2h}{d}) \in [0,\ \arctan \frac{b}{a}]\cup [\pi -\arctan \frac{b}{a},\ \pi ]  .\\ 
      \end{split}
      \right.
\end{equation}
\end{figure*}
\begin{figure*}
  \begin{equation}\label{C_S1,S3}
    C_{S_1,S_3} : \left\{
      \begin{split}
        &y=f\left(x,a,\frac{\sqrt{{{h}^{2}}+{{d}^{2}}/4}}{2},\frac{d}{4},\frac{h}{2},\pi-\arctan (\frac{2h}{d}),-1\right), \ \pi-\arctan (\frac{2h}{d}) \in [\frac{\pi }{2}-\arctan \frac{b}{a},\frac{\pi }{2}+\arctan \frac{b}{a}] \\
        &x=g\left(y,a,\frac{\sqrt{{{h}^{2}}+{{d}^{2}}/4}}{2},\frac{d}{4},\frac{h}{2},\ \pi-\arctan (\frac{2h}{d}),1\right),\pi-\arctan (\frac{2h}{d}) \in [0,\arctan \frac{b}{a}]\cup [\pi -\arctan \frac{b}{a},\pi ].  \\ 
      \end{split}
      \right.
\end{equation}
\end{figure*}
% 时空信息分布地图。AoI向量为[2,1,3]时的典型信息残余分布如(a)所示。AoI向量为[2,1,3]并此时激活节点3的信息分布图如(b)所示。特殊情况边界曲线不方便表示的示例如(c)所示。节点距离偏小，不满足边界曲线存在条件时的信息分布示意如(d)所示。
% 求解三条曲线的交点则可联立三条曲线方程,即$C_{S1,S2}$,$C_{S1,S3}$,$C_{S2,S3}$求解方程组。
The intersection of the three curves can be solved by combining the equations of the three curves, namely $C_{S_1,S_2}$, $C_{S_1,S_3}$ and $C_{S_2,S_3}$.
% \begin{equation}\label{solve equations}
%   \left\{ \begin{aligned}
%     & x=h(y,2a,\frac{d}{2},1) \\ 
%    & y=f(x,a,\frac{\sqrt{{{h}^{2}}+{{d}^{2}}/4}}{2},\frac{d}{4},\frac{h}{2},\pi -\arctan (\frac{2h}{d}),-1) \\ 
%    & y=f(x,a,\frac{\sqrt{{{h}^{2}}+{{d}^{2}}/4}}{2},-\frac{d}{4},\frac{h}{2},\arctan (\frac{2h}{d}),1) \\ 
%   \end{aligned} \right.
% \end{equation}
% 而联立其中两个曲线方程经过化简合并同类项最终便可得到一个一元四次方程，由于化简过程以及求根公式过于复杂，这里则不详细展开，具体可参考文献(\hl{calculation process write or not})。由于信息增量计算式本身便无法算出精确解析解，所以在方程求解时同样可以利用数值计算求近似数值解，以简化运算步骤。解出的多个解再利用另一曲线公式进行验证，选取满足条件的解集即可。
However, a quartic equation can be obtained by combining two of the curve formulas. Since the simplification process and the solving formula are complicated, they are not detailed here. 
% For details, please refer to the literature(\hl{calculation process write or not}).
In addition, since the incremental information formula cannot calculate the exact analytical result, the numerical calculation can also be used to find the approximate numerical solution when solving intersections. %the equation.% to simplify the operation steps. 
The multiple solutions are then verified using another curve formula, and the solution that satisfies the condition is selected.
% 当此刻激活节点3，由之前的分析方法同样可确定关于获取不到新信息增量的区域即集合D''的条件如下：

When node $s_3$ is activated at this moment, the conditions regarding the region where no new valuable incremental information is obtained, i.e., the set $D''$, can be determined by the previous analysis method as follows:
\begin{equation}\label{}
  \left\{ \begin{split}
    &|p-{{p}_{{{s}_{3}}}}|-|p-{{p}_{{{s}_{1}}}}|\ge\frac{{{\lambda }_{t}}}{{{\lambda }_{d}}}\cdot AoI_{s_1}=\frac{{{\lambda }_{t}}}{{{\lambda }_{d}}}\cdot2\Delta t \\ 
    &|p-{{p}_{{{s}_{3}}}}|-|p-{{p}_{{{s}_{2}}}}|\ge\frac{{{\lambda }_{t}}}{{{\lambda }_{d}}}\cdot AoI_{s_2}=\frac{{{\lambda }_{t}}}{{{\lambda }_{d}}}\cdot\Delta t  .
\end{split} \right.
\end{equation} 
% 所以当满足条件${{d}_{{{S}_{3}},{{S}_{1}}}}\ge \frac{{{\lambda }_{t}}}{{{\lambda }_{d}}}\cdot \Delta t$和${{d}_{{{S}_{3}},{{S}_{2}}}}\ge \frac{{{\lambda }_{t}}}{{{\lambda }_{d}}}\cdot 2\Delta t$时，可得到D''的两条边界曲线，如图\ref{info map}(b)所示，
% % 上三角区域代表了集合S1',*型区域代表了集合S2',圆圈区域代表了集合S3'，
% 矩形标记区域为采集不到新信息的区域D''。当上述两个条件若有一个不满足时，对应的边界曲线则不存在，对应的信息残余地图则相对更简化，限于篇幅这里不详细展开，感兴趣的读者可自行分析。
Then combining the constraints of Eq. (\ref{Existence condition}), when the conditions ${{d}_{{{s}_{1}},{{s}_{3}}}}\ge \frac{{{\lambda }_{t}}}{{{\lambda }_{d}}}\cdot 2\Delta t$ and ${{d}_{{{s}_{2}},{{s}_{3}}}}\ge \frac{{{\lambda }_{t}}}{{{\lambda }_{d}}}\cdot 2\Delta t$ are met, two boundary curves of $D''$ can be obtained as shown in Fig. \ref{info map}(b), where the rectangular, upper triangular, star, and circular marker regions represent the sets $D''$, $S_1'$, $S_2'$, and $S_3'$, respectively, and the set $S_j'$ is defined as shown in Eq. (\ref{set S'}).
% where the rectangular marked area is the set $D''$.
% where the upper triangle marked area represents the set $S_1'$, the star marked area represents the set $S_2'$, and the circle marked area represents the set $S_3'$. 
% \hl{write or not}
% When one of the above conditions is not met, the corresponding boundary curve does not exist, and the corresponding information residual map is relatively more simplified, which is not expanded in detail here for the interested readers to analyze by themselves.
% \begin{figure}[ht]
%   \centering
%   \includegraphics[width=2.8in]{enable node3v2.png}
%   \caption{获取信息}
%    \label{enable node 3}
% \end{figure}
% 节点2附近矩形标记区域的边界曲线公式$C_{S_2',D''}$可由(\ref{theta=0,x=gy})得到：
Moreover, the boundary curve formula $C_{S_2',\ D''}$ of the rectangular marked region near node $s_2$ can be obtained from Eq. (\ref{theta=0,x=gy}) as
\begin{equation}
  C_{S_2',D''}:x=h(y,a,\frac{d}{2},-1)\text{=}-a\cdot \sqrt{1+\frac{{{y}^{2}}}{\frac{{{d}^{2}}}{4}-{{a}^{2}}}},
  % x=-a\cdot \sqrt{1+\frac{{{y}^{2}}}{\frac{{{d}^{2}}}{4}-{{a}^{2}}}}
\end{equation}
% 而位于节点1附近矩形标记区域的边界方程$C_{S_1',D''}$可由(\ref{x y express})式得到，即
and the boundary curve $C_{S_1',D''}$ of the rectangular marked area near node $s_1$ can be obtained from Eq. (\ref{x y express}).
% \begin{equation}
%     y=f(x,2a,\frac{\sqrt{{{h}^{2}}+{{d}^{2}}/4}}{2},\frac{d}{4},\frac{h}{2},\pi -\arctan (\frac{2h}{d}),1)
% \end{equation}
% 且需满足条件$\pi -\arctan (\frac{2h}{d})\in {{\Theta }_{1}}$,当不满足这个条件时出现的情况将在后续说明。另外上式需要注意的是a的取值变为2a时，涉及a的所有取值，包括b,A,B,C等都要依次变化。
% 联立曲线的公式可求得对应的曲线交点坐标，如该图\ref{info map} (b)所示，分别为设为(x1,y1),(x2,y2)。设在位置(x,y)处能够获取的信息增量为
The coordinates of the intersection points of curves can be obtained by combining the formulas of the curves, as shown in the Fig. \ref{info map}(b), which are set as $(x_1,y_1)$, $(x_2,y_2)$, respectively. Let the valuable incremental information that can be obtained at the position $(x,y)$ be
\begin{equation}
  \begin{split}
   &info_{{{s}_{e}},{{s}_{p}}}(x,y)\\
   =&-\frac{1}{2}\log \frac{1-{{e}^{-{{2\lambda }_{d}}\cdot \sqrt{{{(x-{{x}_{{{s}_{e}}}})}^{2}}+{{(y-{{y}_{{{s}_{e}}}})}^{2}}}}}}{1-{{e}^{-{{2\lambda }_{d}}\cdot \sqrt{{{(x-{{x}_{{{s}_{p}}}})}^{2}}+{{(y-{{y}_{{{s}_{p}}}})}^{2}}}-{{2\lambda }_{t}}\cdot AoI_{{s}_{p}}}}} \\
  \end{split}
\end{equation}
% \begin{equation}
%   \begin{split}
%     &  inf{{o}_{gain}}(p,s_i)=h({{Y}_{p}}|{{X}_{past}^*})-h({{Y}_{p}}|{{X}_{{{s}_{i}}}}) \\ 
%    & =\left[ h({{Y}_{p}})-h({{Y}_{p}}|{{X}_{{{s}_{i}}}}) \right]-\left[ h({{Y}_{p}})-h({{Y}_{p}}|{{X}_{past}^*}) \right] \\ 
%   %  & =-\frac{1}{2}\log \frac{1-{{\rho }_{{{s}_{i}}}}^{2}}{1-{{\rho }^{*}}^{2}} \\ 
%    & =-\frac{1}{2}\log \frac{1-{{e}^{-2{{\lambda }_{d}}\cdot |p-{{p}_{{{s}_{i}}}}\text{ }\!\!|\!\!\text{ }}}}{1-{{e}^{-2{{\lambda }_{d}}\cdot |p-{{p}^{*}}(p)|-2{{\lambda }_{t}}\cdot  AoI^*(p)}}}  
%   \end{split}
% \end{equation}
% 其中$S_{e}$当前激活的节点，$S_{p}$为在该位置具有信息残余的节点。则在S1'区域内能够获取的信息增量为：
where $s_{e}$ is the currently active node, and $s_{p}$ is the node with information residuals at that location. %Then the valuable incremental information that can be obtained in the region $S_1'$, $S_2'$ and $S_3'$ is
Then the valuable information increments that can be acquired in regions $S_1'$,  $S_2'$ and $S_3'$ are shown in Eq. (\ref{info S1'}), Eq. (\ref{info S2'}) and Eq. (\ref{info S3'}), respectively.
% Then the valuable incremental information that can be obtained in the $S_1'$, $S_2'$ and $S_3'$ regions, respectively, is shown in Eq. (\ref{info S1'}), Eq. (\ref{info S2'}) and Eq. (\ref{info S3'}).
% \begin{figure*}
  \begin{equation}\label{info S1'}
    \begin{split}
       \iint\limits_{{{S}_{1}}'}{{info_{{{s}_{3}},{{s}_{1}}}(x,y)}}dxdy 
      =&\int_{{{x}_{1}}}^{{{x}_{2}}}{dx\int_{C_{S_1',S_2'}}^{C_{S_1',D''}}{info_{{{s}_{3}},{{s}_{1}}}(x,y)}dy} \\ 
      +&\int_{{{x}_{2}}}^{\inf }{dx\int_{C_{S_1',S_3'}}^{C_{S_1',D''}}{info_{{{s}_{3},{{s}_{1}}}}(x,y)}dy}, \\ 
      % =&\int_{{{x}_{1}}}^{{{x}_{2}}}{dx\int_{f(x,a,\frac{\sqrt{{{h}^{2}}+{{d}^{2}}/4}}{2},- \frac{d}{4},\frac{h}{2},\arctan (\frac{2h}{d}),1)}^{f(x,2a,\frac{\sqrt{{{h}^{2}}+{{d}^{2}}/4}}{2},\frac{d}{4},\frac{h}{2},\pi -\arctan (\frac{2h}{d}),1)}{info{_{{{s}_{3}},{{s}_{1}}}}(x,y)}dy}  \\
      % +&\int_{{{x}_{2}}}^{\inf }{dx\int_{f(x,a,\frac{\sqrt{{{h}^{2}}+{{d}^{2}}/4}}{2},\frac{d}{4},\frac{h}{2},\pi -\arctan (\frac{2h}{d}),-1)}^{f(x,2a,\frac{\sqrt{{{h}^{2}}+{{d}^{2}}/4}}{2},\frac{d}{4},\frac{h}{2},\pi -\arctan (\frac{2h}{d}),1)}{info{_{{{s}_{3}},{{s}_{1}}}}(x,y)}dy} .\\
    \end{split}
  \end{equation}
% \end{figure*}
% \begin{split}
%   & \iint\limits_{{{S}_{1}}'}{\left( {{e}^{-{{\lambda }_{d}}\cdot |p-{{p}_{{{s}_{_{3}}}}}|}}-{{e}^{-{{\lambda }_{d}}\cdot |p-{{p}_{{{s}_{1}}}}|-{{\lambda }_{t}}\cdot |t-{{t}_{{{S}_{1}}}}|}} \right)}d\sigma  \\ 
%   =&\int_{{{x}_{1}}}^{{{x}_{2}}}{dx\int_{C_{S_1,\ S_2}}^{C_{S_1',\ D''}}{info_{{{S}_{3}},{{S}_{1}}}(x,y)}dy} \\ 
%   +&\int_{{{x}_{2}}}^{\inf }{dx\int_{C_{S_1,\ S_3}}^{C_{S_1',\ D''}}{info_{{{S}_{3},{{S}_{1}}}}(x,y)}dy} \\ 
%   =&\int_{{{x}_{1}}}^{{{x}_{2}}}{dx\int_{f(x,a,\frac{\sqrt{{{h}^{2}}+{{d}^{2}}/4}}{2},- \frac{d}{4},\frac{h}{2},\arctan (\frac{2h}{d}),1)}^{f(x,2a,\frac{\sqrt{{{h}^{2}}+{{d}^{2}}/4}}{2},\frac{d}{4},\frac{h}{2},\pi -\arctan (\frac{2h}{d}),1)}{info{_{{{S}_{3}},{{S}_{1}}}}(x,y)}dy} \\ 
%   +&\int_{{{x}_{2}}}^{\inf }{dx\int_{f(x,a,\frac{\sqrt{{{h}^{2}}+{{d}^{2}}/4}}{2},\frac{d}{4},\frac{h}{2},\pi -\arctan (\frac{2h}{d}),-1)}^{f(x,2a,\frac{\sq(rt{{{h}^{2}}+{{d}^{2}}/4}}{2},\frac{d}{4},\frac{h}{2},\pi -\arctan (\frac{2h}{d}),1)}{info{_{{{S}_{3}},{{S}_{1}}}}(x,y)}dy} .\\
% \end{split}
% \end{equation}
% 在S2'区域能够获取的信息增量为：
% The valuable incremental information that can be obtained in the $S_2'$ region is
% \begin{figure*}
  \begin{equation}\label{info S2'}
    \begin{split}
     \iint\limits_{{{S}_{2}}'}{{info{_{{{s}_{3}},{{s}_{2}}}}(x,y)}}dxdy  
    =&\int_{-\infty }^{{{y}_{2}}}{dy\int_{C_{S_2',D''}}^{C_{S_2',S_3'}}{info{_{{{s}_{3}},{{s}_{2}}}}(x,y)}dx}  \\
    +&\int_{{{x}_{1}}}^{{{x}_{2}}}{dx\int_{y_2}^{C_{S_1',S_2'}}{info{_{{{s}_{3}},{{s}_{2}}}}(x,y)}dy} \\
    -&\int_{{{y}_{2}}}^{{{y}_{1}}}{dy\int_{{{x}_{1}}}^{C_{S_2',D''}}{info{_{{{s}_{3}},{{s}_{2}}}}(x,y)}dx}, \\
    \end{split}
  \end{equation}
% \end{figure*}
% \begin{equation}\label{info S2'}
%   \begin{split}
%   & \iint\limits_{{{S}_{2}}'}{\left( {{e}^{-{{\lambda }_{d}}\cdot |p-{{p}_{{{s}_{3}}}}|}}-{{e}^{-{{\lambda }_{d}}\cdot |p-{{p}_{{{s}_{2}}}}|-{{\lambda }_{t}}\cdot |t-{{t}_{{{s}_{2}}}}|}} \right)}d\sigma  \\ 
%   =&\int_{-\infty }^{{{y}_{2}}}{dy\int_{h(y,a,\frac{d}{2},-1)}^{h(y,2a,\frac{d}{2},1)}{info{_{{{S}_{3}},{{S}_{2}}}}(x,y)}dx} \\ 
%   +&\int_{{{x}_{1}}}^{{{x}_{2}}}{dx\int_{y_2}^{f(x,a,\frac{\sqrt{{{h}^{2}}+{{d}^{2}}/4}}{2},-\frac{d}{4},\frac{h}{2},\arctan (\frac{2h}{d}),1)}{info{_{{{S}_{3}},{{S}_{2}}}}(x,y)}dy} \\ 
%   -&\int_{{{y}_{2}}}^{{{y}_{1}}}{dy\int_{{{x}_{1}}}^{h(y,a,\frac{d}{2},-1)}{info{_{{{S}_{3}},{{S}_{2}}}}(x,y)}dx} .\\ 
%   \end{split}
% \end{equation}
% 在S3'区域内能够获取的信息增量为：
% The incremental information that can be obtained in the $S_3'$ region is
% \begin{figure*}
  \begin{equation}\label{info S3'}
    \begin{split}
     \iint\limits_{{{S}_{3}}'}{{info{_{{{s}_{3}},{{s}_{3}}}}(x,y)}}dxdy  
      =&\int_{-\infty }^{{{y}_{2}}}{dy\int_{C_{S_2',S_3'}}^{+\infty }{info{_{{{s}_{3}},{{s}_{3}}}}(x,y)}dx} \\
      +&\int_{{{x}_{2}}}^{\inf }{dx\int_{{{y}_{2}}}^{C_{S_1',S_3'}}{info{_{{{s}_{3}},{{s}_{3}}}}(x,y)}dy} ,\\ 
    \end{split}
  \end{equation}
  where 
  \begin{equation}\label{}
    \left\{ \begin{split}
    & C_{S_1',S_2'}:f(x,a,\frac{\sqrt{{{h}^{2}}+{{d}^{2}}/4}}{2},- \frac{d}{4},\frac{h}{2},\arctan (\frac{2h}{d}),1)\\ 
    & C_{S_1',D''} :f(x,2a,\frac{\sqrt{{{h}^{2}}+{{d}^{2}}/4}}{2},\frac{d}{4},\frac{h}{2},\pi -\arctan (\frac{2h}{d}),1) \\ 
    & C_{S_1',S_3'}:f(x,a,\frac{\sqrt{{{h}^{2}}+{{d}^{2}}/4}}{2},\frac{d}{4},\frac{h}{2},\pi -\arctan (\frac{2h}{d}),-1) \\
    & C_{S_2',S_3'}:h(y,2a,\frac{d}{2},1) \\
    & C_{S_2',D''}:h(y,a,\frac{d}{2},-1) .\\
    % & C_{S_3',D''}:
    \end{split} \right.
  \end{equation}
% 再利用(\ref{Node information increment})式
% 再将上述结果累加求和即可得到此刻激活节点3获取的信息总增量表达式。
% 其中需要注意的是，上面的表达式中曲线方程，$C_{S_1,S_2}$,$C_{S_1,S_3}$,$C_{S_1',D''}$皆采取了第一种表达形式，即y=f(x)的形式，这适用于大部分的节点布局情形，但不适用于所有的。
The above results are then summed to obtain the expression for the total incremental information of activating node $s_3$ at this moment. However, it should be noted that the curve formulas in the above expressions, $C_{S_1',S_2'}$, $C_{S_1',S_3'}$ and $C_{S_1',D''}$ all take the first expression form $y=f(x)$, which is applicable to most of the node layout cases, but not to all. 
% For example, when the curve cannot be expressed in the form of $y=f(x)$, the expression can be rewritten by changing the order of integration, which will not go into detail here due to limited space. 
For example, when the curve $C_{S_1',\ D''}$ is only conveniently expressed in the form $x = g(y)$, the information integral expression for the region $S_1'$ can be rewritten according to the same method by changing the order of integration. 
In addition, there are a few special layouts where a curve cannot be uniquely expressed as either of the two forms $y=f(x)$ and $x=g(y)$. 
% 例如当$\frac{\sqrt{{{h}^{2}}+{{d}^{2}}/4}}{2}<\sqrt{2}\cdot 2a$时，位于节点1附近采集不到信息的区域边界曲线无法写出唯一的函数表达式，具体可由图\ref{Cannot represent}示意所示，显然由于曲线的弧度较大，该曲线的两条渐进线与横向纵向的直线最多都有两个交点存在，所以其在当前坐标系下无法唯一的表示为两种函数表达形式中的任一种，则此时在这片区域内获取的信息增量计算式则需要重新改写。
% 此时一种解决方案是尝试针对特定积分区域调整坐标系的角度，方便积分表达，限于篇幅的原因，这里不再详细展开。
% 另外若节点之间距离较小，不满足边界曲线的存在条件，则信息地图相对简化，例如图5（d）中示例的那样，这里也不再赘述。
For example, when $\frac{\sqrt{{{h}^{2}}+{{d}^{2}}/4}}{2}<\sqrt{2}\cdot 2a$, the regional boundary curve $C_{S_1',D''}$ 
% located near node $s_1$ where no information increment is obtained 
cannot be written as a unique functional expression, as schematically shown in Fig. \ref{info map}(c).  Since the curve has a large radian and the two asymptotic lines of the curve have at most two intersections with the horizontal or vertical lines, it cannot be uniquely expressed as either of the two functional expressions in the current coordinate system. 
% Then the expressions of incremental information obtained in this region needs to be rewritten. 
It is one solution to adjust the angle of the coordinate system to facilitate the integration expression. For example, for Fig. \ref{info map}(c), set the direction of the line where nodes s1 and s3 are located as the horizontal or vertical axis of the right angle coordinate system, and then choose the appropriate order of integration.
% which is not expanded in detail here for reasons of limited space. 
Moreover, if the distance between nodes is small and does not meet the conditions for the existence of boundary curves, the information map is relatively simplified as exemplified in Fig. \ref{info map}(d) when ${{d}_{{{s}_{1}},{{s}_{3}}}}< \frac{{{\lambda }_{t}}}{{{\lambda }_{d}}}\cdot 2\Delta t$ and ${{d}_{{{s}_{2}},{{s}_{3}}}}< \frac{{{\lambda }_{t}}}{{{\lambda }_{d}}}\cdot 2\Delta t$.
% \begin{figure}[ht]
%   \centering
%   \includegraphics[width=2.8in]{无法表示.png}
%   \caption{无法表示}
%   \label{Cannot represent}
% \end{figure}
% 例如,针对图5c，将节点s1和s3所在直线方向设为直角坐标系的横轴或纵轴，然后选择合适的积分顺序。
% 理论上如果逐渐减小节点2,3之间的距离，其他距离等参数保持不变，则由于节点2,3空间相关性的增加，导致在激活2节点后激活3节点能够获取的信息增量应该逐渐减少，并且由于节点2,3联合覆盖的有效区域随着距离的缩短逐步缩小，所以此时激活节点1能够获取的信息增量收益可能逐步上升。
% 所以在理论上应该会出现此刻激活节点1和节点3获取的信息增量相同即满足条件$inf{{o}_{[2,1,3]}}({{s}_{3}})=inf{{o}_{[2,1,3]}}({{s}_{1}})$的临界情况，并如果在该临界情况下继续缩小节点2,3的距离，由之前的分析可知此时将导致在单步决策机制下此刻的优化调度为1213……循环往复。
% \hl{**********Grammerly right here************}

Theoretically, if the distance between nodes $s_2$ and $s_3$ is gradually reduced and other parameters remain unchanged, the incremental information that can be obtained by activating node $s_3$ after activating node $s_2$ should gradually decrease due to the increase of spatial correlation of nodes $s_2$ and $s_3$. Meanwhile, the incremental information that can be obtained by activating node $s_1$ gradually increases because the effective area jointly covered by nodes $s_2$ and $s_3$ is gradually reduced with the shortening of their distance. 
Thus theoretically, there should be a critical case where the information increments obtained by activating node $s_1$ and node $s_3$ are the same, i.e., the condition $info_{[2,1,3]}({{s}_{3}})=info_{[2,1,3]}({{s}_{1}})$ is met. If the distance between nodes $s_2$ and $s_3$ continues to be reduced in this critical case, it is clear from the previous analysis that this will lead to optimal scheduling of 1213… cycles at this moment under the single-step decision mechanism.

% 根据相同的分析过程可以写出此时激活节点s1能够获取的信息增量公式。为了曲线表达的便利性，此时可将节点1，节点2连成的直线设为x轴，中点为原点，建立直角坐标系。
% 激活节点1能够获取信息增量的不同区域分布则如图\ref{info map node1} 所示。

\begin{figure} [ht]
	\centering
	\subfloat[$info_{[2,\ 1,\ 3]}({{s}_{1}})$\label{fig:a}]{
		\includegraphics[width=1.7in]{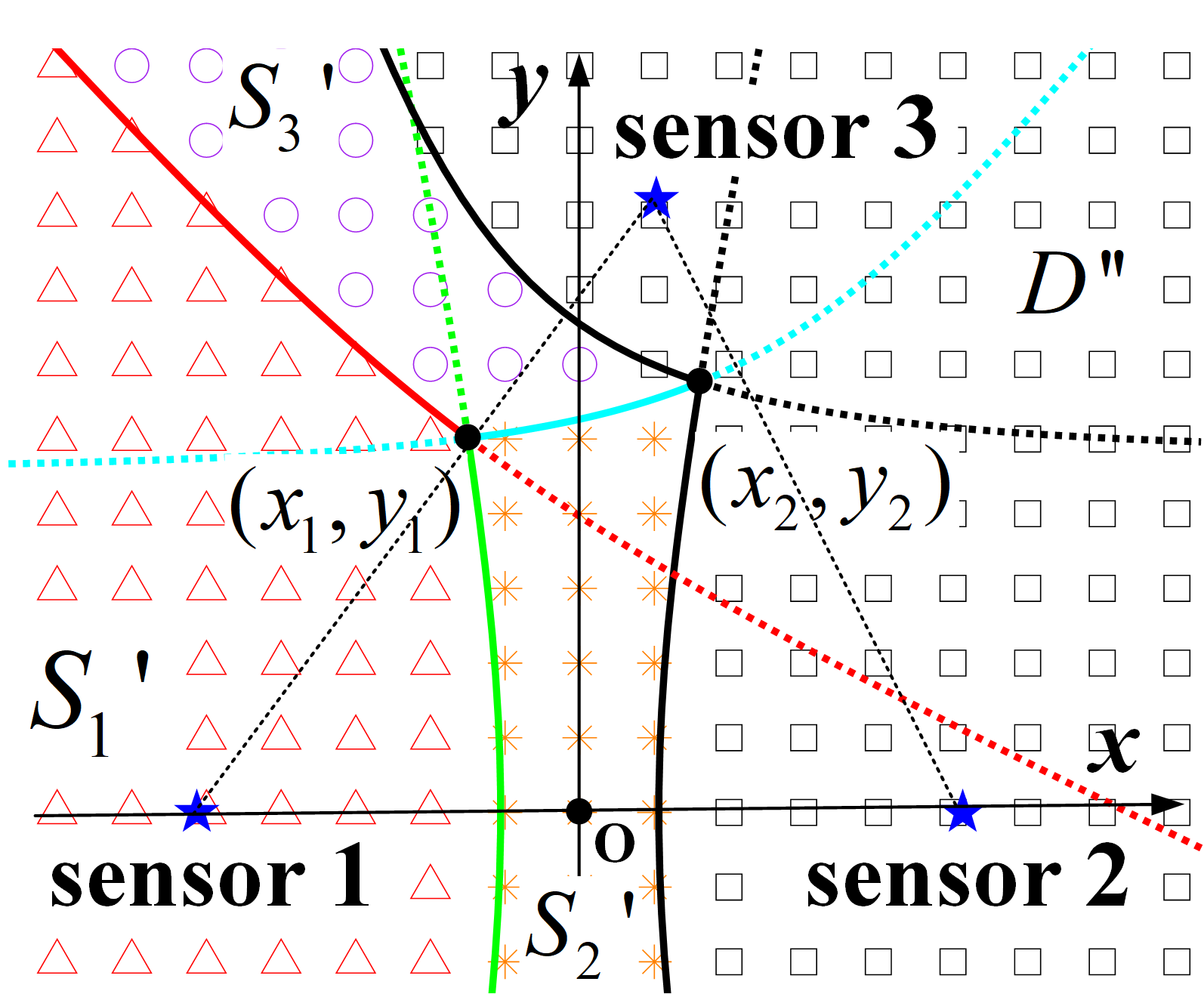}}
	\subfloat[$info_{[inf,\ 2,\ 1]}({{s}_{2}})$ \label{fig:b}]{
		\includegraphics[width=1.7in]{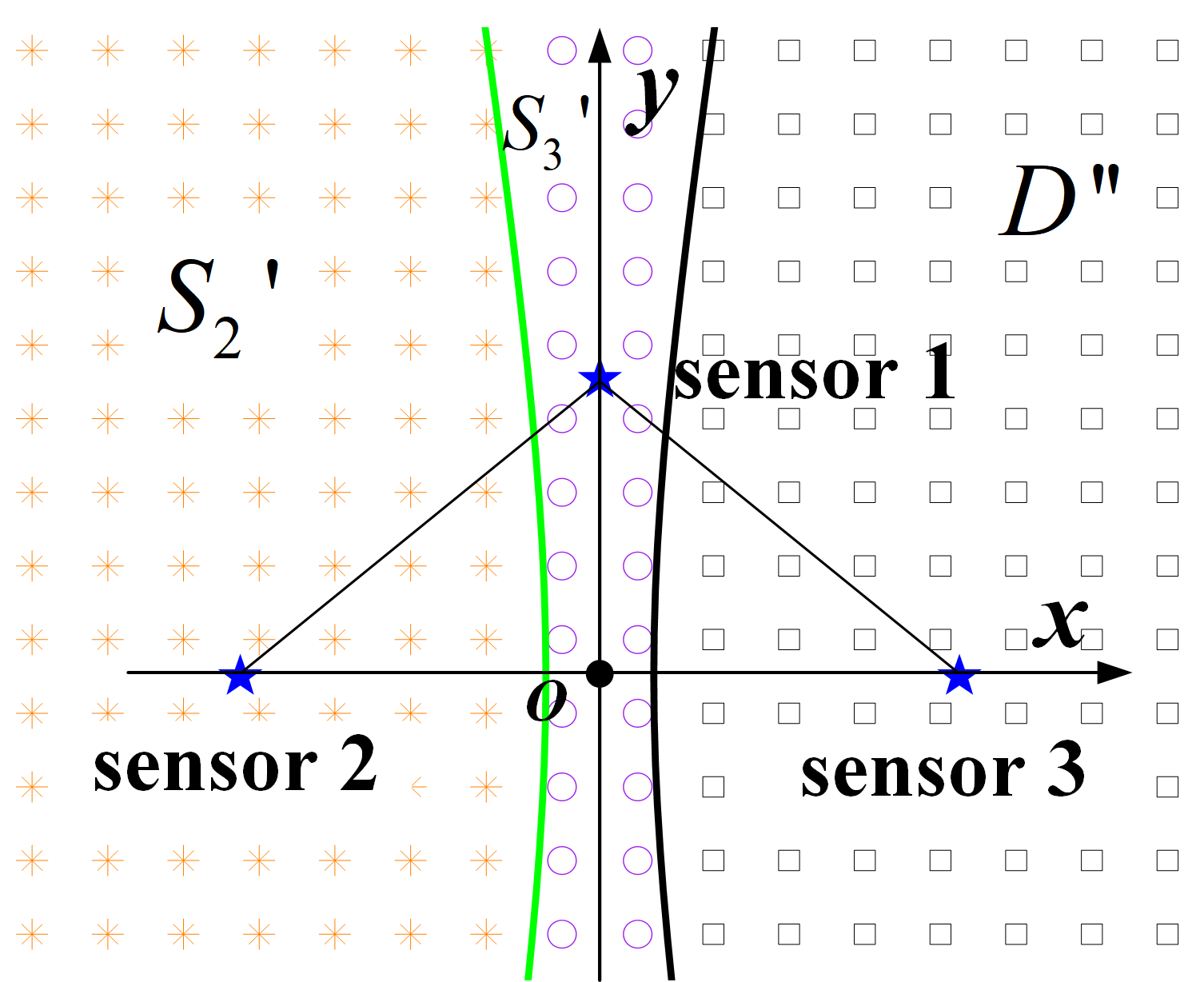}}%{dis_3.png}}
  \\
  \subfloat[$info_{[inf,\ 2,\ 1]}({{s}_{1}})$\label{fig:c}]{
    \includegraphics[width=1.7in]{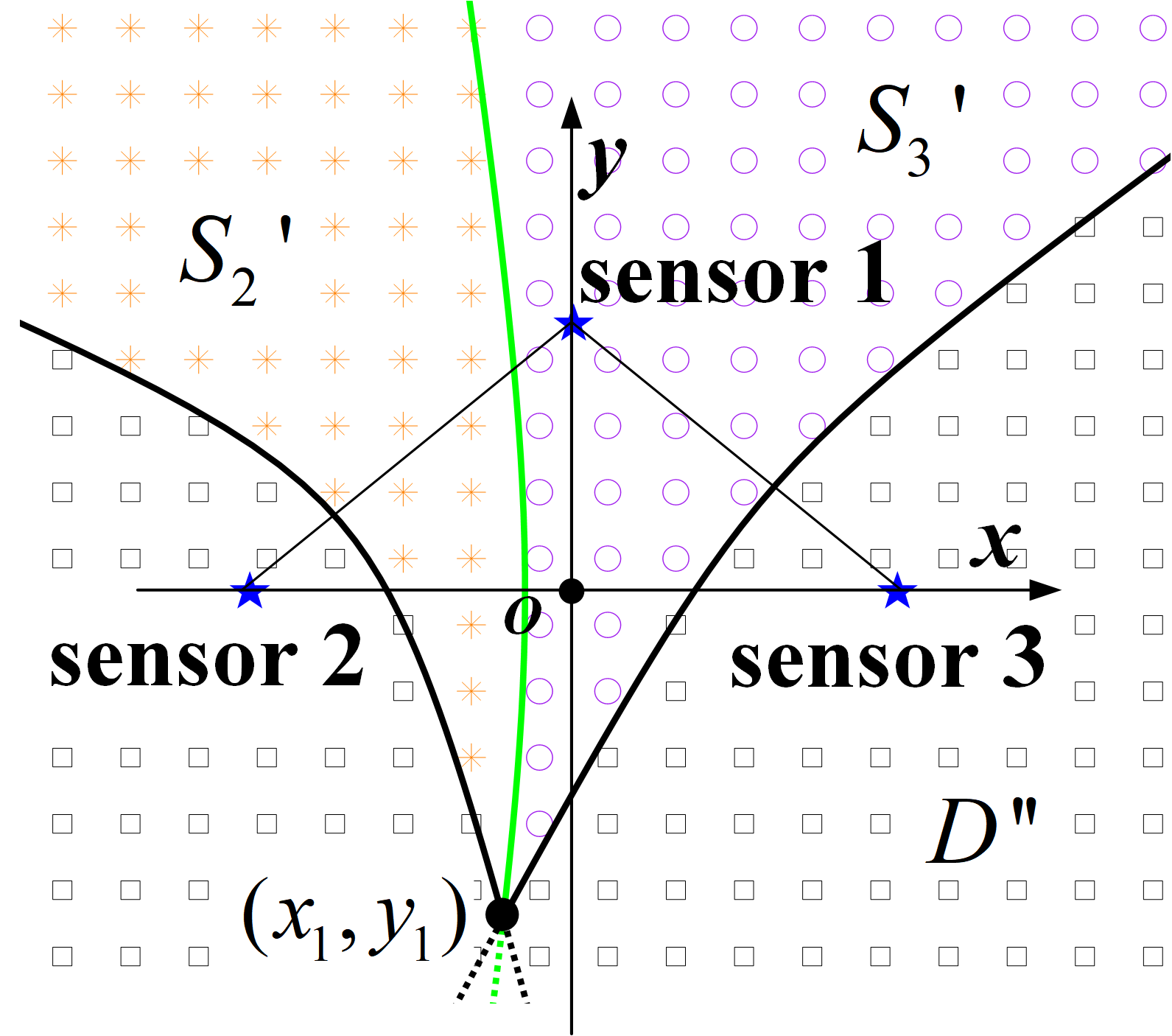}}
  \subfloat[$info_{[inf,\ 2,\ 1]}({{s}_{1}})$\label{fig:d}]{
    \includegraphics[width=1.7in]{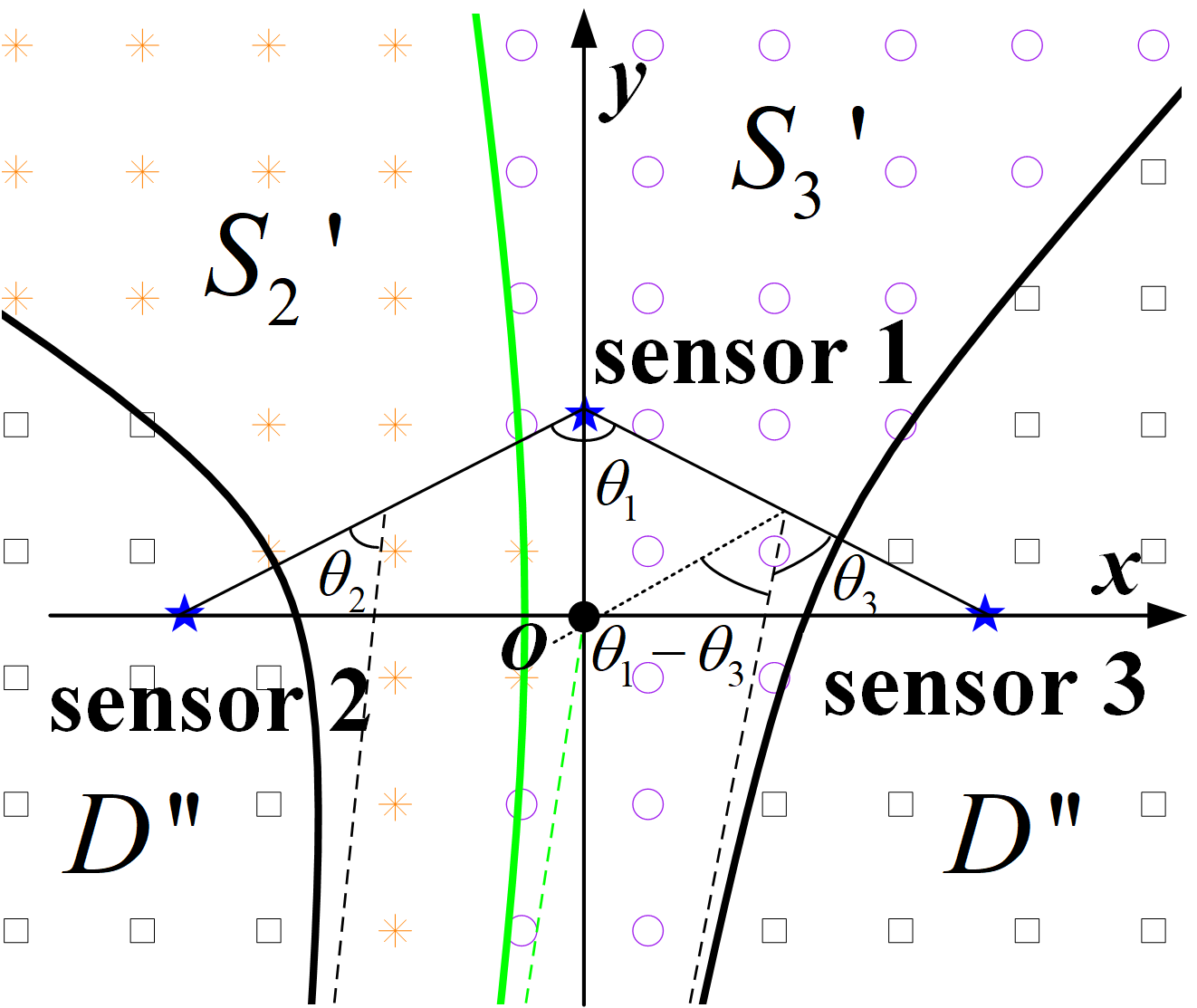}}
	% \caption{Spatio-temporal information map}
  \caption{Spatio-temporal information distribution maps. 
  % %% 写法1，每个都是一个句子
  % The typical information distribution map when the AoI vector is [2, 1, 3] and node $s_1$ is activated at this time is shown in (a). 
  % The typical information distribution map when the AoI vector is [inf,2,1] and node $s_2$ is activated at this time is shown in (b).
  % % [inf,2,1] (𝑠1)
  % The typical information distribution when the AoI vector is [inf,2,1] and node $s_1$ is activated at this time and the three boundary curves have intersection points is shown in (c). 
  % The critical case when the AoI vector is [inf,2,1] and the node s1 is activated at this time and the three boundary curves are asymptotically parallel is shown in (d).
  %%%  写法2，参考率一篇TWC的文章
  (a) The typical information map of $info_{[2,\ 1,\ 3]}({{s}_{1}})$. %when the AoI vector is [2, 1, 3] and node $s_1$ is activated at this time.
  (b) The typical information map of $info_{[inf,\ 2,\ 1]}({{s}_{2}})$. %when the AoI vector is [inf,2,1] and node $s_2$ is activated at this time.
  (c) The information map of $info_{[inf,\ 2,\ 1]}({{s}_{1}})$ with three curves having an intersection point. %when the AoI vector is [inf,2,1] and node $s_1$ is activated at this time and 
  (d) The information map of $info_{[inf,\ 2,\ 1]}({{s}_{1}})$ with the asymptotes of three curves being parallel to each other. 
}
  % 可以简记为$info_{[2,\ 1,\ 3]}({{s}_{1}})$的地图
  % 当AoI向量为[inf,2,1]并此时激活节点s1且三条边界曲线有交点时的典型信息分布图如(c)所示。当AoI向量为[inf,2,1]并此时激活节点s1时且三条边界曲线渐近线平行的临界情形示例如(d)所示。
	% \label{info map} 
	\label{info map node1} 
\end{figure}

For the convenience of curve expression, the straight line formed by node $s_1$ and $s_2$ can be set as the $x$ axis, and the midpoint of $s_1$ and $s_2$ is the origin to establish a rectangular coordinate system. According to the same analysis process, the information increment formula that can be obtained by activating node $s_1$ at this time can be written. 
The distribution of the different areas where activation node $s_1$ can obtain information increments is shown in Fig. \ref{info map node1}(a).
% \begin{figure}[ht]
%   \centering
%   \includegraphics[width=2.8in]{sense enable node 1.png}
%   \caption{info121}
%   %\label{fig_1}
% \end{figure}
% 显然其与图\ref{info map} (b)非常相似，由之前的思路可以得到\ref{info 213 1}式的结果。
Obviously, it is very similar to Fig. \ref{info map}(b), and the Eq. (\ref{info 213 1}) and Eq. (\ref{curve 213_1}) can be obtained from the previous analysis. 
\begin{figure*}
  \begin{equation}\label{info 213 1}
    \begin{split}
      inf{{o}_{[2,1,3]}}({{s}_{1}})  
    =& \int_{-\infty }^{{{y}_{1}}}{dy\int_{-\inf }^{C_{S_1',S_2'}}{info{_{{{s}_{1}},{{s}_{1}}}}(x,y)}dx}  
    +  \int_{-\inf }^{{{x}_{1}}}{dx\int_{{{y}_{1}}}^{C_{S_1',S_3'}}{info{_{{{s}_{1}},{{s}_{1}}}}(x,y)}dy} \\
    + & \int_{-\infty }^{{{y}_{1}}}{dy\int_{C_{S_1',S_2'}}^{C_{S_2',D''}}{info{_{{{s}_{1}},{{s}_{2}}}}(x,y)}dx}   
    + \int_{{{x}_{1}}}^{{{x}_{2}}}{dx\int_{{{y}_{1}}}^{C_{S_2',S_3'}}{info{_{{{s}_{1}},{{s}_{2}}}}(x,y)}dy} 
    -\int_{{{y}_{1}}}^{{{y}_{2}}}{dy\int_{C_{S_2',D''}}^{{{x}_{2}}}{info{_{{{s}_{1}},{{s}_{2}}}}(x,y)}dx}  \\
    +&\int_{-\inf }^{{{x}_{1}}}{dx\int_{C_{S_1',S_3'}}^{C_{S_3',D''}}{info{_{{{s}_{1}},{{s}_{3}}}}(x,y)}dy}  
    +\int_{{{x}_{1}}}^{{{x}_{2}}}{dx\int_{C_{S_2',S_3'}}^{C_{S_3',D''}}{info{_{{{s}_{1}},{{s}_{3}}}}(x,y)}dy} \\ 
    \end{split}
  \end{equation}
\end{figure*}
% where 
\begin{figure*}
  \begin{equation}\label{curve 213_1}
    \left\{ \begin{split}
    & C_{S_1',S_2'}:h(y,a,\frac{\sqrt{{{h}^{2}}+{{d}^{2}}/4}}{2},-1) \\ 
    & C_{S_1',S_3'}:f(x,a,\frac{\sqrt{{{h}^{2}}+{{d}^{2}}/4}}{2},-\frac{{{d}^{2}}}{2\sqrt{4{{h}^{2}}+{{d}^{2}}}},\frac{hd}{\sqrt{4{{h}^{2}}+{{d}^{2}}}},\arctan (\frac{4hd}{4{{h}^{2}}-{{d}^{2}}}),1) \\
    & C_{S_2',S_3'}:f(x,2a,\frac{d}{2},\frac{4{{h}^{2}}-{{d}^{2}}}{4\sqrt{4{{h}^{2}}+{{d}^{2}}}},\frac{hd}{\sqrt{4{{h}^{2}}+{{d}^{2}}}},\pi -\arctan (\frac{2h}{d}),1) \\
    & C_{S_2',D''}:h(y,a,\frac{\sqrt{{{h}^{2}}+{{d}^{2}}/4}}{2},1)\\
    & C_{S_3',D''}:f(x,3a,\frac{\sqrt{{{h}^{2}}+{{d}^{2}}/4}}{2},-\frac{{{d}^{2}}}{2\sqrt{4{{h}^{2}}+{{d}^{2}}}},\frac{hd}{\sqrt{4{{h}^{2}}+{{d}^{2}}}},\arctan (\frac{4hd}{4{{h}^{2}}-{{d}^{2}}}),1). \\ 
    \end{split} \right.
  \end{equation}
\end{figure*}
It is also important to note that the above integral formula may not be applicable to all layout conditions,  which is subject to the following restrictions as Eq. (\ref{restrictions}). 
\begin{figure*}
  \begin{equation}\label{restrictions}
    \left\{ \begin{split}
      & \arctan \frac{4hd}{4{{h}^{2}}-{{d}^{2}}}\in [\frac{\pi }{2}-\arctan \frac{\sqrt{\frac{4{{h}^{2}}+{{d}^{2}}}{16}-9{{a}^{2}}}}{3a},\frac{\pi }{2}+\arctan \frac{\sqrt{\frac{4{{h}^{2}}+{{d}^{2}}}{16}-9{{a}^{2}}}}{3a} ]\\ 
    & \pi -\arctan \frac{2h}{d}\in [\frac{\pi }{2}-\arctan \frac{\sqrt{\frac{4{{h}^{2}}+{{d}^{2}}}{16}-4{{a}^{2}}}}{2a},\frac{\pi }{2}+\arctan \frac{\sqrt{\frac{4{{h}^{2}}+{{d}^{2}}}{16}-4{{a}^{2}}}}{2a} ] .\\ 
    \end{split} \right.
  \end{equation}
\end{figure*}
% 同样也会出现先前所述曲线方程无法唯一表示的情形，应对的方法同样是调整坐标系的角度，重写积分式，这里则不详细展开。
There are also cases where the curve formulas cannot be uniquely represented, and the method to deal with this is to adjust the coordinate system angle and rewrite the integral formula.
% \hl{to be modified}
% \begin{figure}[ht]
%   \centering
%   \includegraphics[width=2.0in]{节点1曲线无法表示.png}
%   \caption{节点1曲线无法表示}
%   % \label{fig_2}
% \end{figure}
% 此外当节点1,3之间距离小于$\frac{{{\lambda }_{t}}}{{{\lambda }_{d}}}\cdot 3\Delta t$时，此时在节点3附近区域不存在采集不到信息的区域，信息获取分布图变为下图\ref{d13<3}所示的情形，且可由几何性质可以证明节点1,3附近的两条曲线必然不存在交点。这时由于节点3附近不存在采集不到信息的区域，上述积分式关于部分积分边界和积分限作相应调整，具体这里不再详细赘述。\hl{Write or not}
% \begin{figure}[ht]
%   \centering
%   \includegraphics[width=2.5in]{dis_3.png}
%   \caption{<3}
%   \label{d13<3}
% \end{figure}
% 联立方程$inf{{o}_{[2,1,3]}}({{s}_{3}})-inf{{o}_{[2,1,3]}}({{s}_{1}})==0$即为上述两种调度情况的边界情形，显然由于积分式的繁琐，很难计算准确解析解，只能用数值计算的方式求近似数值解。
% 具体求解算法可采用二分法，即先确定解的大概范围，然后取区间中点，计算该点处方程的取值，再选取区间两侧方程取值异号的区间。重复该操作直到误差满足一定限制，则可将区间中点定为方程近似零点。具体计算结果将在下一章节给出。
% 而求解出的离散的d和h的取值对应了周期调度序列123和1213的边界分布。

% The Eq. (\ref{Eq d h}) is 
The boundary situation of the above two scheduling situations meets the condition as
\begin{equation}\label{Eq d h}
  info_{[2,1,3]}({{s}_{3}})-info_{[2,1,3]}({{s}_{1}})=0.
\end{equation}
Obviously, due to the cumbersome integral formula, it is challenging to seek the exact analytical solution, and only the approximate numerical solution can be obtained by numerical calculation.
The specific solution algorithm can use the dichotomy method, that is, first determine the approximate range of the solution, second take the midpoint of the range and calculate the value of the equation at this point, then select the range where the value of the equation on two sides of the range with different signs, which means that one is greater than zero and the other is less than zero. 
Repeat this operation until the error meets certain limits, then the midpoint of the interval can be set as the approximate solution. The specific calculation results will be given in Section V.
The obtained discrete values of $d$ and $h$ correspond to the boundary distributions of the periodic scheduling sequences $123$ and $1213$.

% 相反，若逐渐不断增大节点2,3之间的距离，由于空间相关性的逐渐减弱，他们之间交替激活能够获取的信息增量逐渐增多。而节点1和节点2,3的距离暂时没变，且由于节点2,3的有效覆盖区域随着距离的增大而逐步扩大，所以可能节点2,3被激活的占比逐渐增大，节点1被激活的占比逐渐减少。
On the contrary, if the distance between node $s_2$ and $s_3$ gradually increases, the incremental information obtained by alternate activation between them gradually increases due to the gradual weakening of spatial correlation. 
While the distance between node $s_1$ and node $s_2$, $s_3$ remains unchanged and the effective coverage area of node $s_2$ and $s_3$ gradually expands with the increase of distance, it is possible that the proportion of node $s_2$ and $s_3$ being activated gradually increases and the proportion  of node $s_1$ being activated gradually decreases.
% 首先考虑一种较为极端的情况，即节点2,3之间距离非常大，即等效于等腰三角形底边远远大于高的情况。
% 由于此时节点2,3距离较远，则交替激活时节点2,3分别能够获取的信息增量小于等于$inf{{o}_{[\inf ,1,2]}}({{s}_{3}}),inf{{o}_{[\inf ,2,1]}}({{s}_{2}})$,且两者取值相等，当节点1的数据信息完全被节点2,3的残余信息所覆盖时取等号。
% 经过一定时刻激活节点1，此时获取的信息增量小于等于$inf{{o}_{[\inf ,1,2]}}({{s}_{1}})$,$inf{{o}_{[\inf ,2,1]}}({{s}_{1}})$，且当节点1的残余信息完全被节点2或3的新信息覆盖时，取等号。如果此时满足条件：
First, a quite extreme case is considered where the distance between node $s_2$ and $s_3$ is very large, which is equivalent to the case where the base of an isosceles triangle is much larger than the height.
Since node $s_2$ and $s_3$ are far away from each other, the incremental information that node $s_2$ and $s_3$ can obtain when they are alternately activated is less than or equal to $inf{{o}_{[\inf ,1,2]}}({{s}_{3}})$ or $inf{{o}_{[\inf ,2,1]}}({{s}_{2}})$. When the residual information of node $s_1$ is completely covered by the new data of node $s_2$ or $s_3$, the equal sign is taken. 

% \begin{figure*}
  \begin{equation}\label{if node1 activate}
    \begin{split}
      inf{{o}_{[\inf ,2,1]}}({{s}_{1}})
      <inf{{o}_{[\inf ,2,1]}}({{s}_{2}}).
      % \begin{split}
      %   &inf{{o}_{[\inf ,1,2]}}({{s}_{1}})=inf{{o}_{[\inf ,2,1]}}({{s}_{1}})\\
      %   &<inf{{o}_{[\inf ,1,2]}}({{s}_{3}})=inf{{o}_{[\inf ,2,1]}}({{s}_{2}}).
    \end{split}
  \end{equation}
% \end{figure*}

% 即无论何时激活节点1，其能够获取的信息增量都不超过交替激活2,3的信息量，所以在单步信息优化决策机制中，此时节点1在系统中则永远不会被激活。而这种情形的临界条件即令上述不等式取等号，求出满足该等式条件的布局约束。具体的，不等式右边可按照先前的分析思路写成：
If the constraint of Eq. (\ref{if node1 activate}) is met, whenever node $s_1$ is activated, the amount of incremental information node it can obtain does not exceed that of alternate activation of $s_2$ and $s_3$. Thus, node $s_1$ will never be activated in the system with the single-step decision mechanism. 
The critical condition for this situation can be obtained by making the above inequality take an equal sign and finding the layout constraints that meet the condition of the equation. 
Specifically, according to the previous analysis ideas as shown in Fig. \ref{info map node1}(b), the right-hand side of the inequality can be written as follows:
\begin{equation}\label{232}
  \begin{split}
     %inf{{o}_{[\inf ,1,2]}}({{s}_{3}})=
    inf{{o}_{[\inf ,2,1]}}({{s}_{2}}) 
   = &  \int_{-\infty }^{+\infty }{dy}\int_{-\infty}^{C_{S_2',S_3'}}{info{_{{{s}_{2}},{{s}_{2}}}}(x,y)dx},\\ 
    + &\int_{-\infty }^{+\infty }{dy\int_{C_{S_2',S_3'}}^{C_{S_3',D''}}{info{_{{{s}_{2}},{{s}_{3}}}}(x,y)dx}} \\
    % = & \int_{-\infty }^{+\infty }{dy\int_{h(y,a,\frac{d}{2},-1)}^{h(y,a,\frac{d}{2},1)}{info{_{{{s}_{2}},{{s}_{2}}}}(x,y)dx}} \\ 
    % + &\int_{-\infty }^{+\infty }{dy}\int_{h(y,a,\frac{d}{2},1)}^{+\infty }{info{_{{{s}_{2}},{{s}_{3}}}}(x,y)dx} ,\\
  \end{split}
\end{equation}
where 
\begin{equation}\label{}
  \left\{ \begin{split}
  & C_{S_2',S_3'}:h(y,a,\frac{d}{2},-1) \\
  & C_{S_3',D''}:h(y,a,\frac{d}{2},1).\\
  \end{split} \right.
\end{equation}
% 等式左边依据先前的分析方法可以写成
Meanwhile, the left-hand side of the inequality can be written as
\begin{equation}\label{231}
  \begin{split}
     %info_{[\inf ,1,2]}({{s}_{1}})=
    inf{{o}_{[\inf ,2,1]}}({{s}_{1}})  
   =& \int_{{{y}_{1}}}^{+\infty }{dy\int_{C_{S_2',D''}}^{C_{S_2',S_3'}}{info{_{{{s}_{1}},{{s}_{2}}}}(x,y)dx}} \\ 
   +& \int_{{{y}_{1}}}^{+\infty }{dy}\int_{C_{S_2',S_3'}}^{C_{S_3',D''}}{info{_{{{s}_{1}},{{s}_{3}}}}(x,y)dx}\\ 
  %  =& \int_{{{y}_{1}}}^{+\infty }{dy\int_{g(y,2a,\frac{\sqrt{{{h}^{2}}+{{d}^{2}}/4}}{2},-\frac{d}{4},\frac{h}{2},\arctan (\frac{2h}{d}),-1)}^{h(y,a,\frac{d}{2},-1)}{info{_{{{s}_{1}},{{s}_{2}}}}(x,y)dx}} \\ 
  %  +& \int_{{{y}_{1}}}^{+\infty }{dy}\int_{h(y,a,\frac{d}{2},-1)}^{g(y,a,\frac{\sqrt{{{h}^{2}}+{{d}^{2}}/4}}{2},\frac{d}{4},\frac{h}{2},\pi -\arctan (\frac{2h}{d}),1)}{info{_{{{s}_{1}},{{s}_{3}}}}(x,y)dx} ,\\
  \end{split}
\end{equation}
where 
\begin{equation}\label{}
  \left\{ \begin{split}
  & C_{S_2',S_3'}:h(y,a,\frac{d}{2},-1) \\
  & C_{S_2',D''}:g(y,2a,\frac{\sqrt{{{h}^{2}}+{{d}^{2}}/4}}{2},-\frac{d}{4},\frac{h}{2},\arctan (\frac{2h}{d}),-1)\\
  & C_{S_3',D''}:g(y,a,\frac{\sqrt{{{h}^{2}}+{{d}^{2}}/4}}{2},\frac{d}{4},\frac{h}{2},\pi -\arctan (\frac{2h}{d}),1),\\
  \end{split} \right.
\end{equation}
% 其中y1为三条曲线交点纵坐标，但交点也不是所有情形都存在. 当三条曲线不存在交点时，则将积分限中的y1取负无穷便可。由三条曲线渐进线平行时可利用相关角度关系推导出有无交点的临界条件，如图\ref{curve parallel}所示，即有
and $y_1$ is the vertical coordinate of the intersection point of the three curves as illustrated in Fig. \ref{info map node1}(c). However, the intersection point does not exist in all cases. When there is no intersection of the three curves, the $y_1$ in the integral limit is taken as negative infinity.
When the asymptotes of the three curves are parallel, the relevant angle relationship can be used to deduce the critical condition of whether there is an intersection, as shown in Fig. \ref{info map node1}(d) and Eq. (\ref{parallel angle}). %that is, there is
% \begin{equation}
%   \begin{split}
%      {{\theta }_{2}}=&{{\theta }_{1}}-{{\theta }_{3}} \\ 
%      \Leftrightarrow \arctan \left( \frac{\sqrt{\frac{{{h}^{2}}+{{d}^{2}}/4}{4}-4{{a}^{2}}}}{2a} \right)=&2\arctan \left( \frac{d}{2h} \right)\\
%      -&\arctan \left( \frac{\sqrt{\frac{{{h}^{2}}+{{d}^{2}}/4}{4}-{{a}^{2}}}}{a} \right) \\ 
%   \end{split}
% \end{equation}
%单排公式
\begin{figure*}
  \begin{equation}\label{parallel angle}
     \arctan \left( \frac{\sqrt{\frac{{{h}^{2}}+{{d}^{2}}/4}{4}-4{{a}^{2}}}}{2a} \right)={{\theta }_{2}}={{\theta }_{1}}-{{\theta }_{3}}=2\arctan \left( \frac{d}{2h} \right)
     -\arctan \left( \frac{\sqrt{\frac{{{h}^{2}}+{{d}^{2}}/4}{4}-{{a}^{2}}}}{a} \right). 
  \end{equation}
\end{figure*}
% 且上式取值左边大于右边则存在交点，反之则不存在。
If the left-hand side of the Eq. (\ref{parallel angle}) is greater than the right-hand side, there is an intersection, and vice versa.
% ，\hl{To be written 0511}而三条曲线是否存在交点的临界条件如下：*********当满足*************时，即不存在交点，反之则存在交点。
% \begin{figure}[ht]
%   \centering
%   \includegraphics[width=3.0in]{curve parallel.png}
%   \caption{curve parallel}
%   \label{curve parallel}
% \end{figure}

% 联立\ref{232}和\ref{231}两式则可确定临界情形的布局约束。
% 同样上述积分表达式适用于大部分布局情形，即由于部分布局情形的特殊可能导致上述积分表示不完全适用。例如当节点距离较小时，同样会出现采集不到信息区域的缩小，导致该区域的边界曲线也在变化，这里限于篇幅则不再详细展开。
% 下一章节节将详细展开利用数值计算得到的数值结果。
The layout constraints for the critical case can be determined by combining Eq. (\ref{232}) and Eq. (\ref{231}). 
Again the above integral expressions apply to most layout cases, which means that the above integral expression may not be fully applicable due to the few exceptional layout cases. 
% \hl{To be supplemented}
% For example, when the node distance is small, there will also be a reduction of the region where no information is collected, resulting in the boundary curve of the region also changing, which will not be expanded in detail here due to the limitation of space.
% The next section expand the numerical results obtained by numerical calculation in detail.

% 此外，若当不等式\ref{if node1 activate}不成立时，节点s_1经过一定时刻必然再次激活，且随着节点2,3距离的缩小，节点1被激活的频率逐渐升高。利用指数相关函数的性质可分析出每次节点1数据的信息经过多少个时刻会被节点2或3的新数据覆盖，即可得到节点1激活频率的最长时间间隔不超过$\left\lceil \frac{{{d}_{{{S}_{1}},{{S}_{other}}}}}{\frac{{{\lambda }_{t}}}{{{\lambda }_{d}}}\cdot \Delta t} \right\rceil $。
On the other hand, if the inequality in Eq. (\ref{if node1 activate}) does not hold, node $s_1$ is bound to be activated again after certain moments, and the frequency of node $s_1$ being activated gradually increases as the distance of nodes $s_2$, $s_3$ shrinks. 
Using the properties of the exponential correlation function, it can be analyzed how many times the information of the data of node $s_1$ will be covered by the new data of node $s_2$ or $s_3$, and it can be obtained that the maximum time interval of the activation frequency of node $s_1$ does not exceed 
$\left\lceil \frac{{{\lambda }_{d}\cdot{d}_{{{s}_{1}},{{s}_{2}}}}}{{{{{\lambda }_{t}}}}{{}}\cdot \Delta t} \right\rceil $.

% 而具体节点s_1激活频率不同的各种情况之间的临界情况，主要求解思路与前面大部分相同，令两积分式相等寻找节点布局距离之间的关系。
% 综上所述，当节点呈现等腰三角形分布时所出现的不同调度结果和对应的条件如表1所示，其中展示了各种调度结果对应的节点典型位置布局，以及相关边界分布的条件。
Regarding the critical situation between various cases with different activation frequencies of nodes $s_1$, the main solution idea is the same as the previous ones, which let the two incremental information integral formulas be equal to find the relationship between node layout distances. 

From the above analysis, it can be seen that if the information increment is calculated by dividing the integral equation into different regions, although an analytical integral equation is obtained, the process is tedious and the exact analytical results cannot be obtained. 
In addition, when traversing the calculation for different node layouts, due to the coordinate system and curve angle, it leads to that the integral equation need to be rewritten according to different situations. 
Therefore, when solving the equation, the information incremental calculation equation can be written as Eq. (\ref{numerical calculation}), and the approximate numerical solution can be solved with the help of computer for numerical calculation.

% 从上述分析可以看出，若将信息增量分不同区域积分式进行计算，虽然得到了完全解析的积分式，但过程较为繁琐，且无法得到精确解析结果，另外针对不同节点布局遍历计算时，由于坐标系和曲线角度的原因，导致写出的积分式需要根据不同情况改写。所以求解方程时，可将信息增量计算式写成以下形式，并借助计算机进行数值计算，求解近似数值解。

\begin{figure*}
  \begin{equation}\label{numerical calculation}
  \begin{split}
   {{I}_{gain}}
%   =\iint\limits_{D}{\max \{inf{{o}_{gain}}(p),0\}}d\sigma  \\ 
%  & =\iint\limits_{D}{\max \left\{ -\frac{1}{2}\log \frac{1-{{e}^{-2{{\lambda }_{d}}\cdot |p-{{p}_{{{s}_{e}}}}\text{ }\!\!|\!\!\text{ }}}}{1-{{e}^{-2{{\lambda }_{d}}\cdot |p-{{p}^{*}}(p)|-2{{\lambda }_{t}}\cdot |t-{{t}^{*}}(p)|}}},0 \right\}}d\sigma  \\ 
  =\int_{-\infty }^{+\infty }{\int_{-\infty }^{+\infty }{\max \left\{ -\frac{1}{2}\log \frac{1-{{e}^{-2{{\lambda }_{d}}\cdot \sqrt{{{(x-{{x}_{{{s}_{e}}}})}^{2}}+{{(y-{{y}_{{{s}_{e}}}})}^{2}}}}}}{\min \left\{ 1-{{e}^{-2{{\lambda }_{d}}\cdot \sqrt{{{(x-{{x}_{{{s}_{i}}}})}^{2}}+{{(y-{{y}_{{{s}_{i}}}})}^{2}}}-2{{\lambda }_{t}}\cdot Ao{{I}_{{{s}_{i}}}}}}\Big|i=1,2,3 \right\}},0 \right\}dxdy}} \\ 
  \end{split}
\end{equation}
\end{figure*}

In summary, the different scheduling results and corresponding conditions with three nodes presenting an isosceles triangle layout are shown in Table \ref{table1}, which shows the typical location layout of the nodes corresponding to various scheduling results, and the conditions of the associated boundary distribution.
% 例如三个节点交替激活和与节点1激活占比小于1/3的边界分布方程：
% \begin{equation}
%   inf{{o}_{[3,2,1]}}({{s}_{1}})=inf{{o}_{[3,2,1]}}({{s}_{2}})
% \end{equation}

% 节点1激活间隔3次与4次的边界分布：
% \begin{equation}
%   inf{{o}_{[4,1,2]}}({{s}_{1}})=inf{{o}_{[4,1,2]}}({{s}_{3}})
% \end{equation}
% 其他边界分布同理。
% 为了数值计算的便捷性，信息增量的计算方法也可以采用下一章的simpson数值近似积分方法和二分逼近求零点法求解不同调度情况之间的理论数值边界\hl{To be written(or not)}。

\begin{table*}[!ht]
  \caption{Different scheduling situations with the single-step mechanism when three nodes present an isosceles triangle layout. \label{table1}}
  \center
  \begin{tabular}{|c|c|c|}\hline
  Typical layout& Scheduling results &  Conditions \\ \hline
  \multirow{4}{*}{ }&$23\ 23\cdots$& $inf{{o}_{[\inf,\ 2,\ 1]}}({{s}_{2}})>inf{{o}_{[\inf ,2,1]}}({{s}_{1}})$ \\\cline{2-3}

  \multirow{4}{*}{\includegraphics[width=2in]{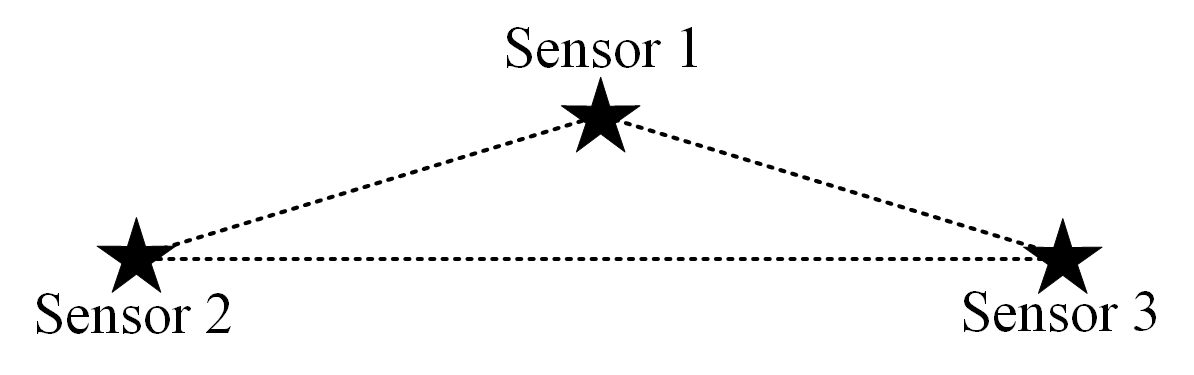}}& boundary& $inf{{o}_{[\inf,2,1]}}({{s}_{2}})=inf{{o}_{[\inf ,2,1]}}({{s}_{1}})$ \\ \cline{2-3}

  \multirow{4}{*}{} & $\vdots$& $\vdots$\\\cline{2-3}

  \multirow{4}{*}{}& $1\ 232\ 1\ 323\cdots$& \tabincell{c}{$inf{{o}_{[3,2,1]}}({{s}_{1}})<inf{{o}_{[3,2,1]}}({{s}_{2}})$ \\ 
  $inf{{o}_{[2,1,3]}}({{s}_{3}})>inf{{o}_{[2, 1, 3]}}({{s}_{1}})$  \\
  $inf{{o}_{[1,3,2]}}({{s}_{2}})>inf{{o}_{[1, 3, 2]}}({{s}_{3}})$\\
  % $inf{{o}_{[4,\ 1,\ 2]}}({{s}_{1}})>inf{{o}_{[4,\ 1,\ 2]}}({{s}_{3}})$\\
  $inf{{o}_{[4,2,1]}}({{s}_{1}})>inf{{o}_{[4,2,1]}}({{s}_{2}})$ } \\ \hline

  & boundary& $inf{{o}_{[3,2,1]}}({{s}_{1}})=inf{{o}_{[3,2,1]}}({{s}_{2}})$  \\ \hline

  \multirow{5}{*}{\includegraphics[width=1.0in]{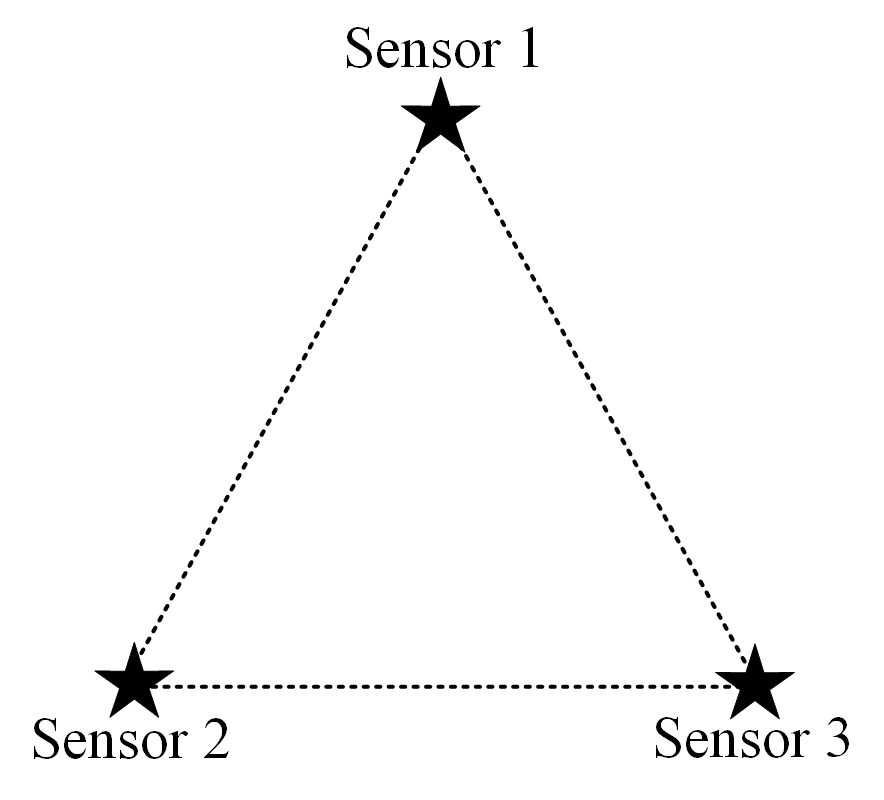}} & & \\
  \multirow{5}{*}{}& & \\
  \multirow{5}{*}{}& $123\ 123\cdots$& \tabincell{c}{$inf{{o}_{[3,2,1]}}({{s}_{1}})>inf{{o}_{[3,2,1]}}({{s}_{2}})$ \\ $inf{{o}_{[2,1,3]}}({{s}_{3}})>inf{{o}_{[2,1,3]}}({{s}_{1}})$  \\$inf{{o}_{[1,3,2]}}({{s}_{2}})>inf{{o}_{[1,3,2]}}({{s}_{3}})$} \\ 
  \multirow{5}{*}{}& & \\
  \multirow{5}{*}{}& & \\\hline

  & boundary& $info_{[2,1,3]}({{s}_{3}})=info_{[2,1,3]}({{s}_{1}})$ \\ \hline

   \multirow{11}{*}{\includegraphics[width=1.0in]{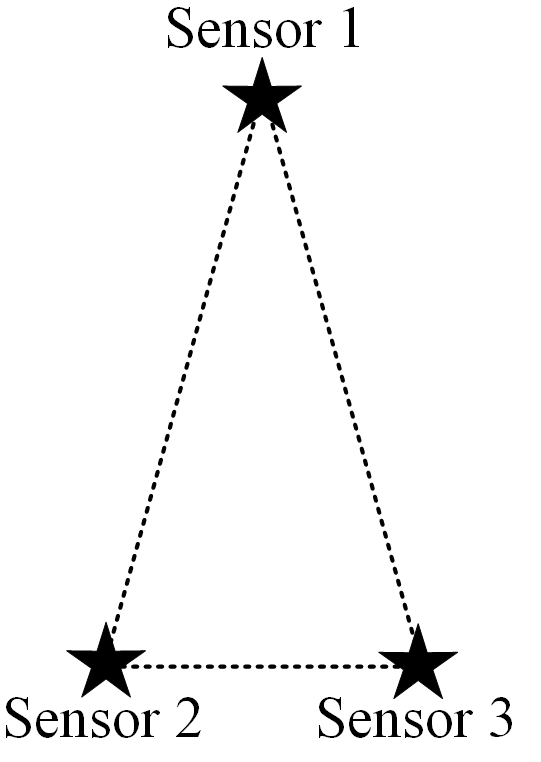}} & & \\  
  \multirow{11}{*}{} & & \\
  \multirow{11}{*}{}& & \\
  % \multirow{11}{*}{}& & \\
  \multirow{11}{*}{}& & \\
  \multirow{11}{*}{}& $1213\ 1213\cdots$ & \tabincell{c}{$info_{[2,1,3]}({{s}_{3}})<info{_{[2,1,3]}}({{s}_{1}})$\\ 
  $info_{[1,2,4]}({{s}_{3}})>info{_{[1,2,4]}}({{s}_{2}})$\\  %条件多余
  $info_{[1,4,2]}({{s}_{2}})>info{_{[1,4,2]}}({{s}_{3}})$\\
  
  $info_{[2,3,1]}({{s}_{2}})<info{_{[2,3,1]}}({{s}_{1}})$\\
  
  } \\  
  \multirow{11}{*}{}& & \\
  \multirow{11}{*}{}& & \\
  % \multirow{11}{*}{}& & \\
  % \multirow{11}{*}{}& & \\
  \multirow{11}{*}{}& & \\\hline
\end{tabular}
\end{table*}

% \begin{tabular}{|c|c|}
%   \hline
%   1 & the first line \\
%   \hline
%   2 & \tabincell{c}{haha\\ heihei\\zeze} \\
%   \hline
% \end{tabular}

% For the convenience of numerical computation, the increment information can also be computed using the \hl{Simpson} numerical approximate integration method in the next chapter to solve the theoretical numerical boundary between different scheduling cases. 
% \hl{To be written (or not)} 
\subsection{Three-node general triangular layout}%三节点一般三角布局
% \
% \newline
% \indent 
% 前一小节初步分析了三个节点呈现等腰三角形分布情形时的单步信息最优机制下可能出现的优化调度结果，本节则主要将上述结果拓展到节点呈现一般三角形分布的情形。
% 当三个节点呈现一般三角形分布情形时，由于描述一个一般三角形至少需要三个变量。由于三维图的不直观性和理论分析的不易性，本部分的调度分析主要针对固定其中一个变量的取值，改变另外两个量取值遍历求其优化调度结果。
% %另外可改变固定量的不同取值，从而逐步遍历大部分节点一般分布布局下的优化调度结果。
% 本节具体的分析思路主要是固定节点分布中的最大节点距离，且距离最大的两个节点位置不变，遍历另一节点的不同位置，分析其可能出现的调度结果。例如假设节点2,3之间距离d最大，且分别位于二维直角坐标系中的原点和（d,0）两点处，且根据三角形的水平对称性可减少一半的遍历区域。则节点1需要遍历区域如图\ref{遍历区域}所示，且图中阴影区域为节点1的可选位置。
The previous subsection preliminarily analyzes the possible optimal scheduling results under the single-step information-optimal mechanism when the three nodes exhibit an isosceles triangle distribution. This subsection extends the above results to cases where the nodes exhibit a general triangle distribution. 
When three nodes present a general triangular distribution, at least three variables are required to describe a general triangle. Due to the unintuitiveness of the three-dimensional diagram and the difficulty of theoretical analysis, the scheduling analysis in this subsection focuses on keeping the value of one of the variables unchanged and changing the values of the other two variables to traverse and obtain the optimal scheduling results.
% In addition, different values ​​of the fixed amount can be changed, so as to gradually traverse the optimal scheduling results under the general distribution layout of most nodes. 
The specific analysis idea in this subsection is mainly to fix the maximum node distance and keep the positions of the two nodes with the largest distance unchanged, traversing the different positions of the other node and analyzing the possible scheduling results.
For example, it is assumed that the distance $d$ between nodes $s_2$ and $s_3$ is the largest, and they are located at $(0,0)$ and $(d, 0)$ respectively in the 2D rectangular coordinate system. The traversed area can be reduced by half according to the horizontal symmetry of the triangle. Then node $s_1$ needs to traverse the area as illustrated in the Fig. \ref{Nodes location distribution}(b), where the shaded area in the figure is the optional position of node $s_1$.
% \begin{figure}[ht]
%   \centering
%   \includegraphics[width=2.8in]{一般布局遍历区域.png}
%   \caption{节点1遍历区域}
%   \label{遍历区域}
% \end{figure}
% 当节点s1的位置在遍历区域的边界上选择时，三个节点呈现等腰三角形。根据先前的分析可粗略获得可能出现的调度情形，例如在节点2,3的中垂线上，随着节点1纵向高度的增加，其最终节点1激活频率占比可能由0逐渐趋于1/3。
% 另外在遍历区域的圆弧边界上，假如从顶端向左下方遍历，由之前的分析思路可知，调度情况可能由开始的交替激活，然后在满足节点距离最小限制下，调度情形变为节点3激活占比为50\%,而另外两个节点激活占比分别为四分之一的情况。

% 所以根据上述思路，节点位置可以由横轴正向方向遍历，且在每个横轴取值固定情况下依顺序遍历纵向取值，并分别寻找下列方程的数值解，为判断不同区域的优化调度提供依据
When the position of node $s_1$ is selected on the boundary of the traversal region, the three nodes present an isosceles triangle. According to the previous analysis, the possible scheduling situation can be roughly obtained. %For example, on the perpendicular bisector of nodes $s_2$ and $s_3$, as the longitudinal height of node $s_1$ increases, the activation frequency ratio of node $s_1$ may gradually tend to $\frac{1}{3}$ from $0$. 
For example, as the vertical height of node $s_1$ increases on the perpendicular bisector of nodes $s_2$ and $s_3$, the final activation proportion of node $s_1$ may gradually increase from $0$ to $\frac{1}{3}$. 
In addition,  if traversing from the top to the lower left on the arc boundary, the scheduling situation may change from alternating activation at the beginning to a situation where node $s_3$ activation accounts for a half and the proportion of the other two nodes is one fourth under the minimum node distance constraint, as shown in the previous analysis. 

% 根据上述分析可得到一般三角布局下的可能调度结果，如表2所示。
Based on the above analysis, the possible scheduling results under the general triangular layout can be obtained, as shown in Table \ref{table3}.
\begin{table*}[!ht]
  \caption{Different scheduling situations with the single-step mechanism when three nodes present an general triangle layout. \label{table3}}
  \center
  \begin{tabular}{|c|c|c|}\hline
    \tabincell{c}{Typical\\ layout}& \tabincell{c}{Scheduling \\results} &  Conditions \\ \hline
  \multirow{4}{*}{\includegraphics[width=1.5in]{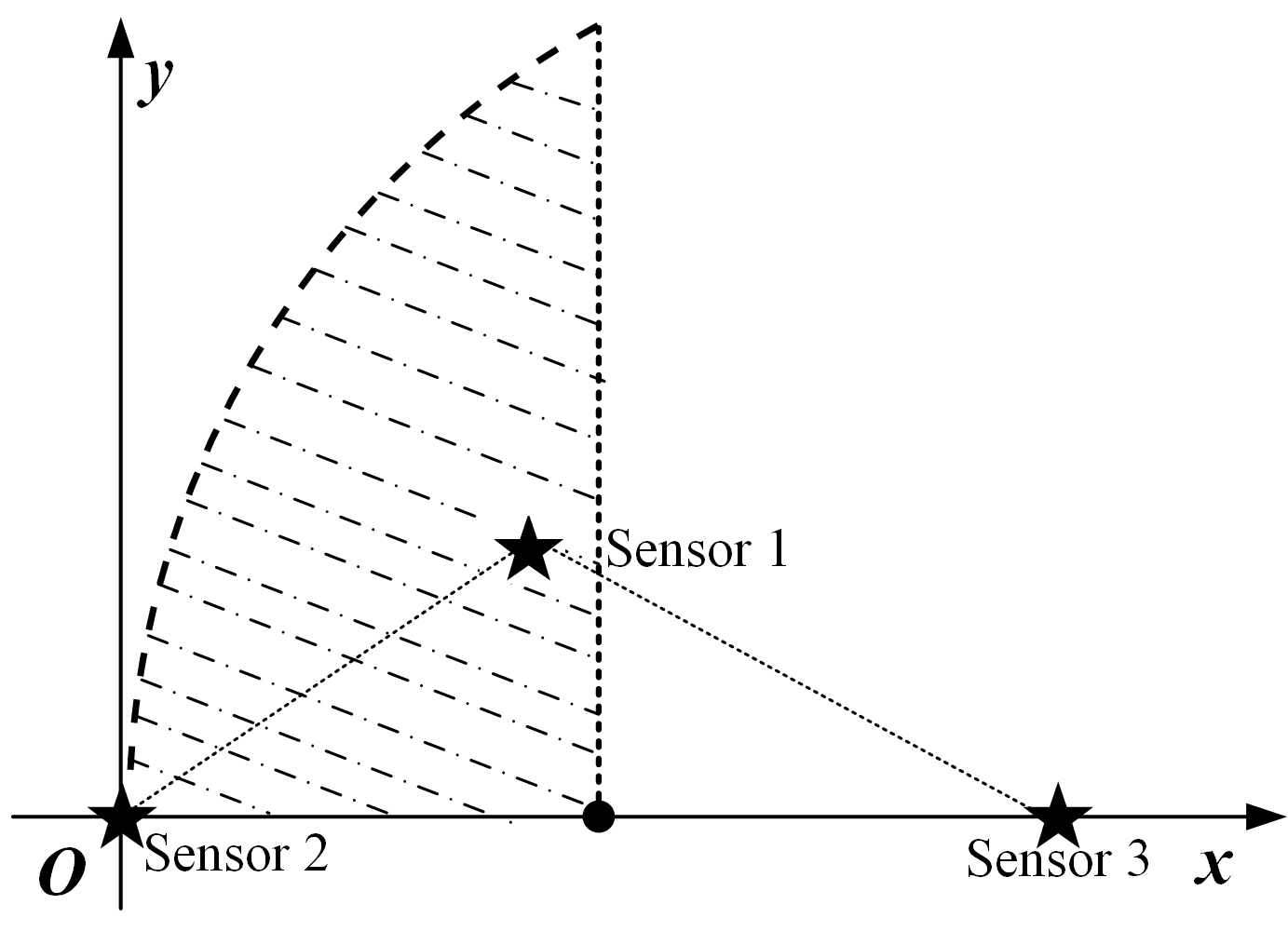} }&$23\ 23\cdots$& $inf{{o}_{[\inf,2,1]}}({{s}_{2}})>inf{{o}_{[\inf,2,1]}}({{s}_{1}})$ \\\cline{2-3}

  \multirow{4}{*}{}& boundary& $inf{{o}_{[\inf,2,1]}}({{s}_{2}})=inf{{o}_{[\inf,2,1]}}({{s}_{1}})$ \\ \cline{2-3}

  \multirow{4}{*}{} & $\vdots$& $\vdots$\\\cline{2-3}

  \multirow{4}{*}{}& $1232\ 1323\cdots$& \tabincell{c}{$inf{{o}_{[3,2,1]}}({{s}_{1}})<inf{{o}_{[3,2,1]}}({{s}_{2}})$ \\ 
  $inf{{o}_{[2,1,3]}}({{s}_{3}})>inf{{o}_{[2, 1, 3]}}({{s}_{1}})$  \\
  $inf{{o}_{[1,3,2]}}({{s}_{2}})>inf{{o}_{[1, 3, 2]}}({{s}_{3}})$\\
  $inf{{o}_{[4,2,1]}}({{s}_{1}})>inf{{o}_{[4,2,1]}}({{s}_{2}})$ } \\ \hline

  & boundary& $inf{{o}_{[3,2,1]}}({{s}_{1}})=inf{{o}_{[3,2,1]}}({{s}_{2}})$  \\ \hline

  \multirow{5}{*}{\includegraphics[width=1.2in]{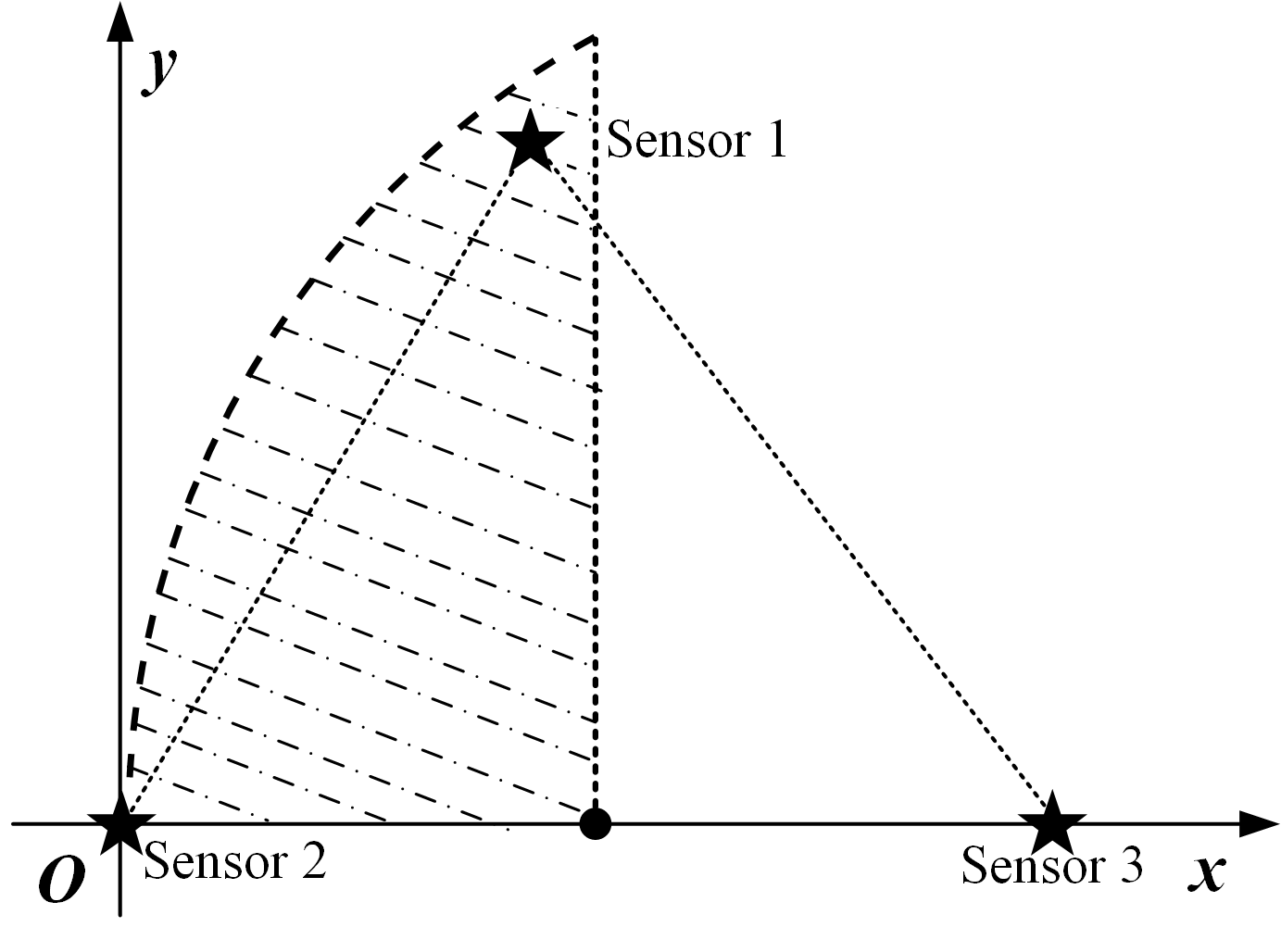}} & & \\
  \multirow{5}{*}{}& & \\
  \multirow{5}{*}{}& $123\ 123\cdots$& \tabincell{c}{$inf{{o}_{[3,2,1]}}({{s}_{1}})>inf{{o}_{[3,2,1]}}({{s}_{2}})$ \\ $inf{{o}_{[2,1,3]}}({{s}_{3}})>inf{{o}_{[2,1,3]}}({{s}_{1}})$  \\$inf{{o}_{[1,3,2]}}({{s}_{2}})>inf{{o}_{[1,3,2]}}({{s}_{3}})$} \\ 
  \multirow{5}{*}{}& & \\
  \multirow{5}{*}{}& & \\\hline

  & boundary& $inf{{o}_{[1,3,2]}}({{s}_{3}})=inf{{o}_{[1,3,2]}}({{s}_{2}}) $ \\ \hline

   \multirow{11}{*}{\includegraphics[width=1.5in]{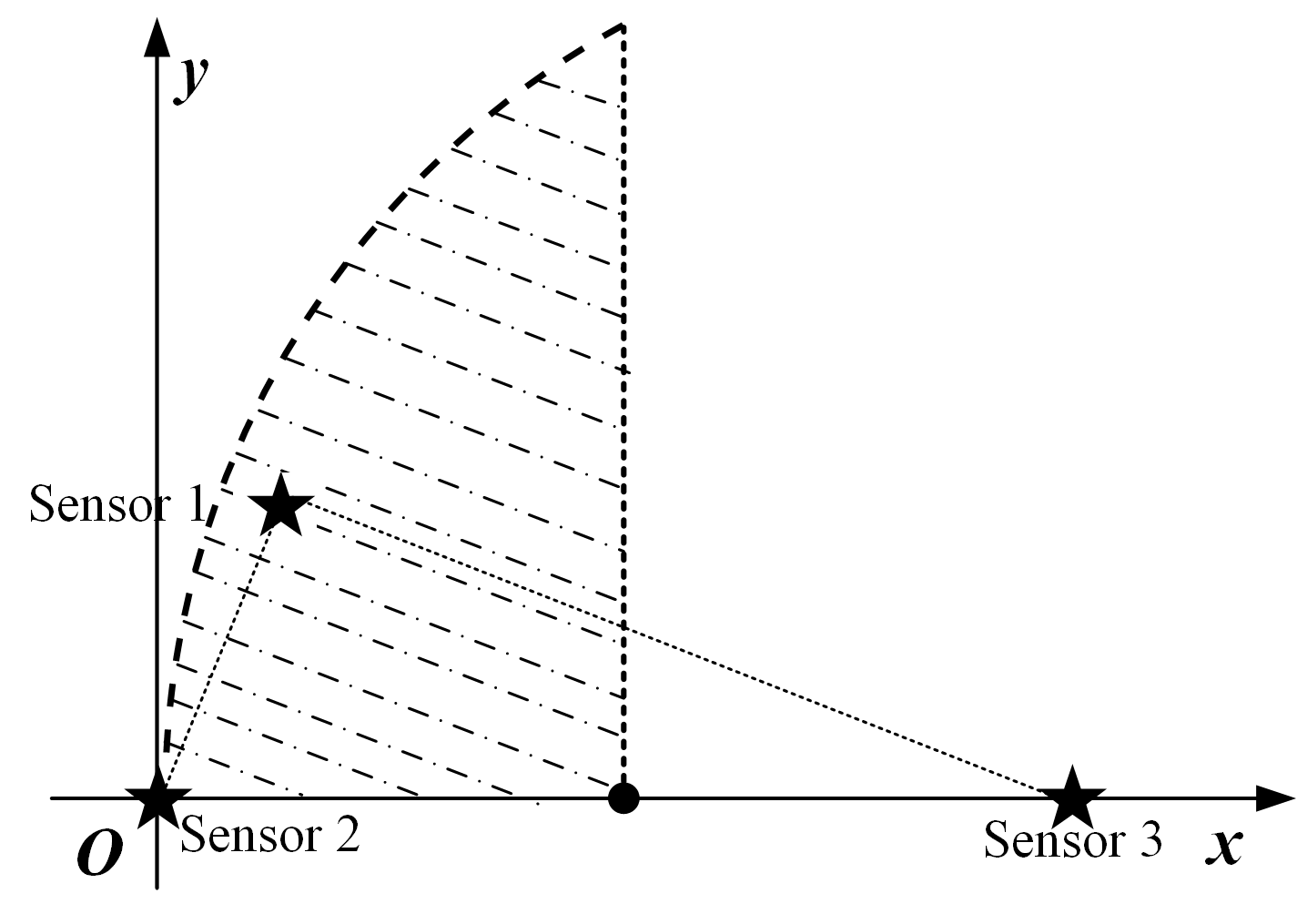}} & & \\  
  \multirow{11}{*}{} & & \\
  \multirow{11}{*}{}& & \\
  % \multirow{11}{*}{}& & \\
  \multirow{11}{*}{}& $3132\ 3132\cdots$ & \tabincell{c}{$inf{{o}_{[2,4,1]}}({{s}_{2}})>inf{{o}_{[2,4,1]}}({{s}_{1}})$\\ 
  $inf{{o}_{[1,3,2]}}({{s}_{3}})>inf{{o}_{[1,3,2]}}({{s}_{2}})$\\  %条件多余
  $inf{{o}_{[4,2,1]}}({{s}_{1}})>inf{{o}_{[4,2,1]}}({{s}_{2}})$\\
  
  $info_{[3,1,2]}({{s}_{3}})<info{_{[3,1,2]}}({{s}_{1}})$\\
  
  } \\  
  \multirow{11}{*}{}& & \\
  % \multirow{11}{*}{}& & \\
  \multirow{11}{*}{}& & \\\hline
\end{tabular}
\end{table*}
Therefore, the node positions can be traversed in the positive direction from the horizontal axis, and the vertical values can be traversed sequentially in the case of fixed values of each horizontal axis, then the numerical solutions of the following equations can be found respectively to provide the basis for judging the optimal scheduling of different regions as follows:
\begin{equation}
  \label{general equations}
  \left\{ \begin{split}
  & inf{{o}_{[\inf ,2,1]}}({{s}_{2}})=inf{{o}_{[\inf ,2,1]}}({{s}_{1}}) \\ 
 & inf{{o}_{[3,2,1]}}({{s}_{1}})=inf{{o}_{[3,2,1]}}({{s}_{2}}) \\ 
 & inf{{o}_{[1,3,2]}}({{s}_{3}})=inf{{o}_{[1,3,2]}}({{s}_{2}}) .\\ 
\end{split} \right.
\end{equation}

% 其中第一个方程代表了节点1长期是否会被激活的边界；第二个方程代表了三个节点交替激活和节点1的占比小于1/3的临界布局；第三个方程代表了节点3占比为1/2的临界，即周期调度为3132和三节点等比交替激活的边界布局。
% 且各区域的调度情况满足下列条件：
In Eq. (\ref{general equations}), the first equation represents the boundary of whether node $s_1$ will be activated in the long term; the second equation represents the critical case with three nodes alternately activated and the proportion of node $s_1$ less than $\frac{1}{3}$; the third equation represents the critical case with the proportion of node $s_3$ as $\frac{1}{2}$, i.e., the boundary layout with the periodic scheduling of $3132$ and three nodes alternately activated in equal proportion.
% Moreover, the scheduling results for each region meet the following conditions:
% \begin{equation}
%   \label{Scheduling judgment}
%   \left\{ \begin{split}
%     & 23\cdots :inf{{o}_{[\inf ,2,1]}}({{s}_{2}})>inf{{o}_{[\inf ,2,1]}}({{s}_{1}}) \\ 
%    & 123\cdots :\left\{ \begin{split}
%     & inf{{o}_{[3,2,1]}}({{s}_{1}})>inf{{o}_{[3,2,1]}}({{s}_{2}}) \\ 
%    & inf{{o}_{[2,1,3]}}({{s}_{3}})>inf{{o}_{[2,1,3]}}({{s}_{1}})\\
%    %inf{{o}_{[\inf ,2,1]}}({{s}_{2}})<inf{{o}_{[\inf ,2,1]}}({{s}_{1}}) \\ 
%    & inf{{o}_{[1,3,2]}}({{s}_{2}})>inf{{o}_{[1,3,2]}}({{s}_{3}}) \\ 
%   \end{split} \right. \\ 
%    & 3132\cdots :\left\{ \begin{split}
%     & inf{{o}_{[2,4,1]}}({{s}_{2}})>inf{{o}_{[2,4,1]}}({{s}_{1}}) \\ 
%    & inf{{o}_{[1,3,2]}}({{s}_{3}})>inf{{o}_{[1,3,2]}}({{s}_{2}}) \\ 
%    & inf{{o}_{[4,2,1]}}({{s}_{1}})>inf{{o}_{[4,2,1]}}({{s}_{2}}) \\
%   & inf{{o}_{[3,1,2]}}({{s}_{3}})>inf{{o}_{[3,1,2]}}({{s}_{1}}) .
%   \end{split} \right. 
%   \\ 
%   \end{split} \right.
% \end{equation}
% \hl{to be considered}
% %一般三角形布局时的积分表示方法与之前大体类似，但同样由于部分布局的特殊性可能会导致一些布局无法简单表示积分计算式，而且工作量较大。本文给出了一种利用simpson数值积分法的近似数值算法，能够通用的算出不同调度情形和不同布局的信息增量近似值, 从而为系统每次决策提供参考指标。

% 一般三角形布局时的积分表示方法与之前大体类似，但同样由于很难写出一个通用的信息增量计算式适用于所有布局情况，所以本文给出了一种利用simpson数值积分法的近似数值算法，能够通用的算出不同调度情形和不同布局的信息增量近似值。
% 从而为系统每次决策提供参考指标。
The integral information expressions in the general triangular layout are similar to the previous one, whose results can also be obtained by numerical calculation using Eq. (\ref{numerical calculation}) in order to obtain the scheduling results and the boundary distribution case. The details of the obtained results will be expanded in Section V.

\section{Long-term optimal mechanism}%长期平均信息获取最优机制
% 单步决策机制的核心是针对每个离散的时刻进行信息量预计算，激活能够在当前时刻获取最多信息增量的节点。但是从系统长期运行来看，因为单步决策机制每次决策只考虑了当前的收益，并没有考虑当前的决策可能对未来的信息收益造成的影响。所以本节主要分析若将系统长期运行信息获取均值作为优化目标，分析节点不同布局下的最优调度策略。
The core of the single-step decision mechanism is to precompute the amount of information and activate the node that can obtain the most information increment at the current moment. 
However, in terms of the long-term operation of the system, the single-step decision mechanism only concerns the current gain for each decision, and it does not consider the possible impact of the current decision on the future information gain. 
Therefore, this section focuses on analyzing the optimal scheduling strategy under different node layouts when the average information acquisition for the long-term operation is taken as the optimization goal.
% \hl{read}

% 本节主要采取的方案是将该过程建模为马尔科夫决策过程，将系统当前信息残余地图作为状态，当前时刻激活的节点获取的信息收益只与当前的信息残余地图有关，与先前所有时刻的信息残余地图无关。由于目前分析的系统包含三个节点，且定期激活某个节点，在这种情形下系统状态数是有限的，则理论上可以采用Q学习算法经过有限次训练最终收敛至最优调度结果。
The main scheme adopted in this section is to model the process as a Markov decision process, taking the current information residual map of the system as the state. The information gain obtained by the node activated at the current moment is only related to the current information residual map, independent of the residual information map at all previous moments. 
Since the currently analyzed system contains three nodes and activates a node periodically, the number of system states is finite. Then the Q-learning algorithm can be used to eventually converge to the optimal scheduling result after a finite number of training steps.
\subsection{States, Actions, and Rewards} %状态，动作与收益
\
\newline
\indent 
% 状态空间State则主要记录当前系统的信息残余地图，可直接取系统每个节点最新一次感知数据的AoI取值，即:
The state-space $State$ mainly records the current information residual map of the system and can directly take the AoI of the latest sensed data of each node as follows:
\begin{equation}
  State=AoI=\left[ AoI\left( {{s}_{1}} \right),AoI\left( {{s}_{2}} \right),AoI\left( {{s}_{3}} \right) \right].
  % State=\left[ AoI_ {{s}_{1}} ,\ AoI_{{s}_{2}} ,\ AoI_ {{s}_{3}} \right].
\end{equation}
% 因为决策时间间隔固定不变为$\Delta t$,所以AoI的取值可只简记为其关于$\Delta t$的系数。
% 由节点时空相关性以及之前的分析可知，系统连续采取相同动作信息收益不佳，所以可以默认规定系统不会连续两次及以上采取相同动作，此时能够减少系统总状态数，从而能够提高训练效率。
% 每个节点的数据信息存活时间是有限的，即经过一定时刻总会被临近节点的新数据所覆盖。所以每个节点数据的AoI有一个最大阈值，超过这个阈值即代表该数据不含有效信息，等效于AoI取值为inf。
Since the decision time interval is fixed as $\Delta t$, the AoI can be abbreviated only as its coefficient about $\Delta t$. 

From the node spatio-temporal correlation and the previous analysis, it is known that the system takes the same action continuously with poor information gain. Thus, it can be stipulated by default that the system will not take the same action twice and more consecutively, which can reduce the total number of states and improve the training efficiency.

The survival time of data information of each node is limited for the reason that it will always be covered by new data of adjacent nodes after a certain time. Thus the AoI of each node data has a maximum, greater than which the data contain no valuable information, and the AoI can be abbreviated as $inf$.
The maximum AoI of each node can be calculated %from the distance from this node to the other two nodes 
as Eq. (\ref{AoI maximum}), where 
\begin{equation}
  {{K}_{{{s}_{i}},{{s}_{j}}}}=\left\lfloor \frac{{{d}_{{{s}_{i}},{{s}_{j}}}}}{\frac{{{\lambda }_{t}}}{{{\lambda }_{d}}}\cdot \Delta t} \right\rfloor ,
\end{equation}
and $i,\ j,\ k\in \{1,2,3 \}$, $i\ne j\ne k$. %and.%$d$ represents the distance between nodes.
In addition, $s_j$ and $s_k$ represent the two nodes whose data AoI takes the values of 1 and 2, respectively.
% 另外，si和sj分别代表了数据AoI取值为1和2的两个节点。
% 在上述设定下系统状态值中AOI取值必然包含1和2，因为连续不采取相同动作，所以前一次激活的节点和前两次激活的节点必然不同，分别对应了AoI取值为1和2的情况，而另外一个节点的AoI值暂不确定，
% 最大AoI阈值可由该节点至另外两个节点距离进行计算得到，具体为：
% \hl{to be corrected}
% \begin{equation}
%   Ao{{I}_{{{S}_{i}},{{S}_{j}}}}{{({{S}_{k}})}_{th}}=\min \{\left\lfloor \frac{{{d}_{{{S}_{k}},{{S}_{i}}}}}{\frac{{{\lambda }_{t}}}{{{\lambda }_{d}}}\cdot \Delta t} \right\rfloor +1,\left\lfloor \frac{{{d}_{{{S}_{k}},{{S}_{j}}}}}{\frac{{{\lambda }_{t}}}{{{\lambda }_{d}}}\cdot \Delta t} \right\rfloor +2\}
% \end{equation}
\begin{figure*}
  \begin{equation}\label{AoI maximum}
    AoI{{({{s}_{i}})}_{\max }}=\max \left\{ \min \left( {{K}_{{{s}_{i}},{{s}_{j}}}}+AoI({{s}_{j}}),{{K}_{{{s}_{i}},{{s}_{k}}}}+AoI({{s}_{k}}) \right)\Big|AoI({{s}_{j}}),AoI({{s}_{k}})\in \{1,2\},AoI({{s}_{j}})\ne AoI({{s}_{k}}) \right\}
  \end{equation}
\end{figure*}
% 其中Si，Sj，Sk，分别为三个节点，且Si，Sj分别为AoI取值为1和2的节点。d代表节点之间的距离。
% Where $S_i$, $S_j$, $S_k$, are three nodes, and $S_i$ and $S_j$ are nodes with AoI taking values of 1 and 2, respectively. 
% 所以此时系统总状态数为(\ref{state nums})式所示：
The total number of states of the system is shown in Eq. (\ref{state nums}), 
% \begin{equation}
%   State\_nums=\sum\limits_{k=1}^{3}{\sum\limits_{i\ne j}^{i,j\in \{1,2,3\}/\{k\}}{[Ao{{I}_{{{S}_{i}},{{S}_{j}}}}{{({{S}_{k}})}_{th}}}-1]}
% \end{equation}
% \begin{figure*}
%   \begin{equation}\label{state nums}
%     State_{nums}=\sum\limits_{m=1}^{3}{\left\{ 1+\sum\limits_{i\ne j}^{i,j\in \{1,2,3\}/\{m\}}{\left[ u\left( {{K}_{{{s}_{m}},{{s}_{i}}}}-1 \right)+\max \left\{ 0,\min \left( {{K}_{{{s}_{m}},{{s}_{i}}}}+1,{{K}_{{{s}_{m}},{{s}_{j}}}}+2 \right)-2 \right\} \right]} \right\}}.
%   \end{equation}
% \end{figure*}
\begin{figure*}[ht]
\begin{equation}\label{state nums}
  \begin{split}
    &N_{State}=\sum\limits_{i=1}^{3}{N_{State}\Big|_{AoI_{max}=AoI(s_i)_{max}}}=\sum\limits_{i=1}^{3}\sum\limits_{m=1}^{AoI(s_i)_{max}}{N_{State}\Big|_{AoI_{max}=AoI(s_i)=m}}\\
  &=\sum\limits_{i=1}^{3}\left\{1+\sum\limits_{m=2}^{AoI(s_i)_{max}} \Bigg[u\Big(K_{s_i,s_j}-(m-1)\Big) \cdot u\Big(K_{s_i,s_k}-(m-2)\Big)+u\Big(K_{s_i,s_j}-(m-2)\Big) \cdot u\Big(K_{s_i,s_k}-(m-1)\Big)  \Bigg] \right\}
  \end{split}
\end{equation}
\end{figure*}
% 且其中${{K}_{{{S}_{i}},{{S}_{j}}}}=\left\lfloor \frac{{{d}_{{{S}_{i}},{{S}_{j}}}}}{\frac{{{\lambda }_{t}}}{{{\lambda }_{d}}}\cdot \Delta t} \right\rfloor $,u(.)为阶跃函数，当括号内取值大于等于0函数值为1，其余皆为0。
where %${{K}_{{{s}_{i}},{{s}_{j}}}}=\left\lfloor \frac{{{d}_{{{s}_{i}},{{s}_{j}}}}}{\frac{{{\lambda }_{t}}}{{{\lambda }_{d}}}\cdot \Delta t} \right\rfloor $, and 
$u(.)$ is a step function as follows:
\begin{equation}
  u(x)=\left\{
    \begin{aligned}
    1 \quad x\geq0\\
    0 \quad x<0\\
    \end{aligned}
  \right.
\end{equation}
% when the value inside the brackets is greater than or equal to 0 function value is 1, the rest are 0.
% 求和符号内分别代表了每个节点AoI取值最大分别为1,2,和大于等于3时的总状态数。
% \hl{whether improve}

% 动作集合为：
The set of actions is
\begin{equation}
  Action=\{s_1,s_2,s_3\},
\end{equation}
where $s_1$, $s_2$ and $s_3$ represent the activation of the corresponding nodes, respectively.
% 其中s1，s2和s3分别代表了激活对应节点。
% 状态转移过程如下，系统每次执行一个动作，激活相应节点，对应节点数据的AoI取值置为0，经过时间${\Delta t}$后，所有节点数据AoI取值加1。
% 然后判断是否有节点数据的AoI取值超过最大值，即判断该数据信息是否被相邻节点完全覆盖，若超过最大值即记为inf。

The state transfer process is as follows. Firstly, the system performs an action each time and activates the corresponding node, and the AoI value of the corresponding node data is set to 0. Secondly, the AoI value of all node data plus 1 elapsed time ${\Delta t}$. 
Then determine whether the AoI value of the node data exceeds the maximum value, and if it exceeds the maximum value, it is recorded as $inf$.
% 每种动作的即时奖励即为获取的信息增量。即

The immediate reward for each action is the incremental amount of information acquired, that is
\begin{equation}
  Reward=\frac{I_{gain}(State,s_i)}{I_{gain}([inf,\ inf,\ inf],s_1)},\ i\in \left\{1,2,3\right\},
\end{equation}
where the denominator is the information increment that the system can obtain by activating the node for the first time, whose role is to normalize the reward.
\subsection{Q learning algorithm} 

% Q学习中Q表的取值代表特定状态采取某个动作的价值（参考文献），算法最终输出结果为经过有限次训练基本完全收敛的Q表，依据该Q表取值系统可知在所有状态下最优的决策动作。训练过程主要采取贪婪策略，即前期主要采取随机动作探索环境，随着训练轮数的增多，逐渐增大贪婪系数，即系统趋向于选择Q值更大的动作。
% 等待Q表几乎完全收敛即可得到系统的长程最优调度。系统每次训练更新Q值过程如下（参考文献）：
% and the final output of the algorithm is a Q table that is almost after a finite number of training steps, 
The Q table in Q learning algorithm represents the quality of an action in a particular state \cite{ref26}.
The output of the algorithm is the Q table that tends to converge after finite training, based on which the system can know the optimal decision action in all states. 

The training process mainly adopts a greedy strategy, which means that the agent mainly takes random actions to explore the environment in the initial stage, and gradually increases the greedy coefficient as the number of training steps increases, i.e., the agent tends to choose the action with a larger Q value. 
The optimal long-term scheduling of the system is obtained by waiting for the almost complete convergence of the Q-table.
The process of updating the Q value for each training is as follows\cite{ref26}

\begin{equation}
  Q_{new}(s, a) \leftarrow Q(s, a)+\alpha\left(R\left(s^{\prime}\right)+\gamma \max _{a^{\prime}} Q\left(s^{\prime}, a^{\prime}\right)-Q(s, a)\right),
\end{equation}
% 其中s表示当前状态，a表示动作，Q（s，a）是先前的Q值，α是学习率，γ是discout factor，其取值决定了对未来所有时刻收益的不同重视程度，R是在新状态s中观察到的即时奖励。$max _{a^{\prime}} Q\left(s^{\prime}, a^{\prime}\right)$代表当时状态s采取动作a到达下一个状态获得的最优未来收益的最大估计。
where $s$ denotes the current state, $a$ denotes the action, and $Q(s, a)$ is the previous Q value. Meanwhile, $\alpha$ is the learning rate, and $\gamma$ is the discout factor whose value determines the different importance attached to the gain at all future moments. The $R$ is the immediate reward observed in the new state $s'$, and the $max _{a^{\prime}} Q\left(s^{\prime}, a^{\prime}\right)$ represents the estimate of the optimal future reward from the next state $s'$. %\hl{ref}
% 遍历不同布局得到的长程最优调度结果将在下一章节具体展开。
The long-term optimal scheduling results obtained by traversing different layouts are developed specifically in the next section.

\section{Numerical and simulation results} %数值与仿真结果
% \subsection{Single-step optimal decision mechanism scheduling results}
\subsection{Scheduling results with single-step optimal mechanism} %单步最优决策结果Single-step optimal decision mechanism 
% 本部分主要给出单步决策机制下的调度结果和部分理论分析出的数值近似边界。仿真运行中对于信息增量的计算方法采用先前的simpson数值积分法，为系统每次决策提供参考。
% 不失一般性，在本节中t的取值一律取一，时间参数的变化主要体现在lanmdat上。
This subsection gives the scheduling results under single-step decision mechanism and numerical approximation bounds from theoretical analysis. 
% The simpson numerical integration method is used for the calculation of information increment in the simulation to provide a reference for system decision.
Without loss of generality, the value of $\Delta t$ is  taken as $1$ in this section, and the variation of the time parameter is mainly reflected in $\lambda_t$.
\subsubsection{Three-node isosceles triangle layout} %三节点等腰三角布局结果和部分数值近似边界曲线
\begin{figure} [ht]
	\centering
	\subfloat[$\lambda_d=0.01$, $\lambda_t=0.3$.\label{fig:a}]{
		\includegraphics[width=3.5in]{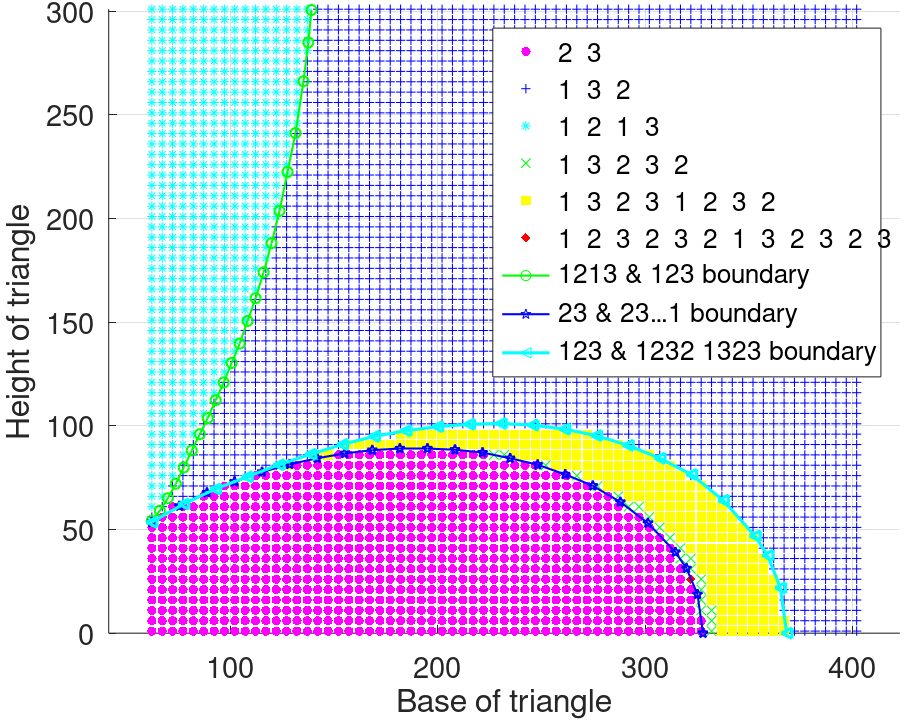}}
    \\
	\subfloat[$\lambda_d=0.025$, $\lambda_t=0.5$.\label{fig:b}]{
		\includegraphics[width=3.5in]{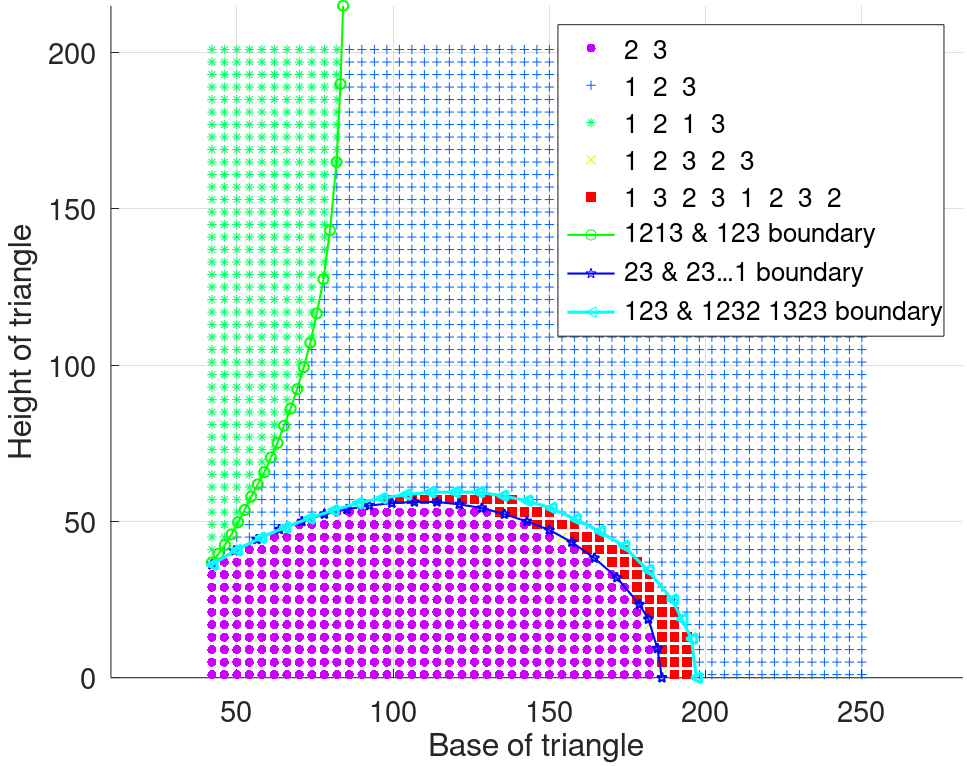}}
	\caption{Scheduling results of single step optimal decision mechanism with three nodes presenting an isosceles triangle layout. (a) Scheduling results with $\lambda_d=0.01$, $\lambda_t=0.3$. (b) Scheduling results with $\lambda_d=0.025$, $\lambda_t=0.5$.}
	\label{isosceles results} 
\end{figure}

\
\newline  
\indent 
% 当节点之间呈现等腰三角分布情形时，且当$\lambda_d$取0.01,$\lambda_t$取0.3时，主要调度结果如图\ref{isosceles results} 所示:
When three nodes present an isosceles triangle layout and traversal step is 5 with parameters $\lambda_d=0.01$ and $\lambda_t=0.3$, the main scheduling results are illustrated in Fig. \ref{isosceles results}(a), 
% \begin{figure}[ht]
%   \centering
%   \includegraphics[width=3.5in]{等腰调度结果V1.png}
%   \caption{isosceles one step }
%   % \label{fig_2}
% \end{figure}
% \thanks{This paper was produced liuyang}
% 其中横轴代表了等腰三角底边取值，纵轴代表了等腰三角高的取值，图中不同形状的散点代表了每一类调度情形，且其对应的周期调度序列分别如图中图例所示.
% 三条曲线为根据之前理论分析得到的关于临界调度情况的近似数值边界，即通过遍历d的取值，利用近似求解算法求得与每个d对应的临界h。
where the horizontal axis represents the values of the base of isosceles triangle, and the vertical axis represents the height of isosceles triangle. 
The scatter points of different shapes in the Fig. \ref{isosceles results}(a) represent different type of scheduling situation, and their corresponding periodic scheduling sequences are shown in the legend of the figure, respectively. 
The three curves are the approximate numerical bounds on the critical scheduling case obtained from the previous theoretical analysis, i.e., the critical $h$ corresponding to each $d$ is obtained by using the approximate solution algorithm. 

Specifically, the curve marked by circles represents the boundary node distribution for the cycle scheduling sequence of $1213$ and $123$. 
The curve marked by triangles represents the boundary node distribution for the cycle scheduling sequence of 123 and node $s_1$ activation percentage less than $\frac{1}{3}$, i.e., the scheduling result of $1232\ 1323 \cdots $.
In addition, the curve marked with pentagrams represents the boundary node distribution of whether node $s_1$ will be activated or not, in other words, it represents the critical distribution of the scheduling sequence $23$ and $23\cdots1$, and the interval between two node $s_1$ activations is not greater than 
$\left\lceil \frac{{{\lambda }_{d}\cdot{d}_{{{s}_{1}},{{s}_{2}}}}}{{{{{\lambda }_{t}}}}{{}}\cdot \Delta t} \right\rceil =\left\lceil \frac{{{\lambda }_{d}\cdot \sqrt{h_{th}^2+d_{th}^2/4}}}{{{{{\lambda }_{t}}}}{{}}\cdot \Delta t} \right\rceil$, where $h_{th}$ and $d_{th}$ correspond to the values taken on the boundary curve.

%$\left\lceil \frac{{{d}_{{{S}_{1}},{{S}_{other}}}}}{\frac{{{{\lambda }_{t}}}}{{{\lambda }_{d}}}\cdot \Delta t} \right\rceil $.
% 其中hth和dth对应该曲线上的取值
% 具体的，圆圈标记的曲线代表了周期调度序列分别为1213和123的边界节点分布关系，三角形标记的曲线代表了周期调度序列为123和节点1激活占比小于1/3即调度结果为1232 1323……的边界节点分布关系。五角星标记的曲线代表了节点1是否会被激活的临界分布，即代表了调度序列23和23...1的临界分布。且两次节点1激活间隔不大于
It can be observed that the theoretical numerical approximation bounds match the simulation, and the scheduling results of the simulation are consistent with the previous theoretical analysis. In addition, when traversal step is 4 with $\lambda_d=0.025$ and $\lambda_t=0.5$, the result is shown in the Fig. \ref{isosceles results}(b).
% \begin{figure}[ht]
%   \centering
%   \includegraphics[width=3.5in]{等腰调度结果0.025&0.5V1.png}
%   \caption{isosceles one step }
%   % \label{fig_2}
% \end{figure}

\subsubsection{Three-node general triangular layout} %一般布局结果和部分数值近似边界曲线
\begin{figure} [ht]
	\centering
	\subfloat[$d_{s_2,\ s_3}=220$.\label{fig:a}]{
		\includegraphics[width=1.7in]{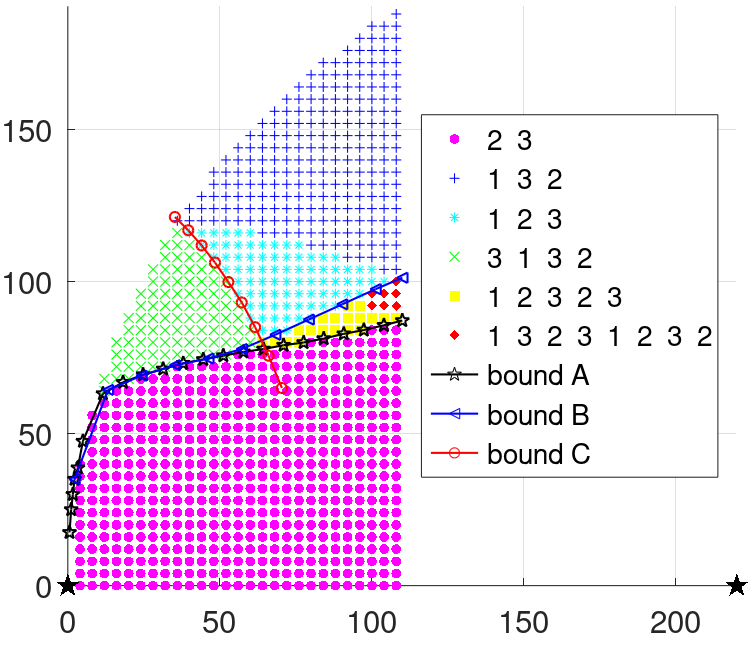}}
	\subfloat[$d_{s_2,\ s_3}=330$.\label{fig:b}]{
		\includegraphics[width=1.7in]{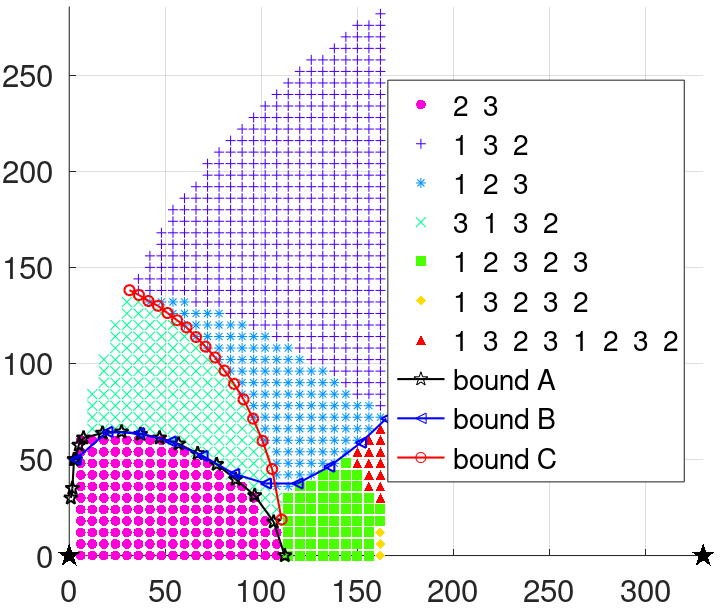}}
	\caption{Scheduling results of single step optimal decision mechanism with three nodes presenting a general triangle layout and parameters $\lambda_d=0.01$ and $\lambda_t=0.3$. (a) Scheduling results with the maximum distance between nodes $d_{s_2,\ s_3}=220$. (b) Scheduling results with the maximum distance $d_{s_2,\ s_3}=330$.}
	\label{general layout one step} 
\end{figure}

\
\newline  
\indent 
% 本小节主要给出当三个节点呈现一般三角形布局时的理论分析和仿真结果。
% 当$\lambda_d$取0.01,$\lambda_t$取0.3时，节点之间最大距离分别为280和450时的主要调度结果如图\ref{general layout one step} (a)和图\ref{general layout one step} (b)所示，且图中的曲线分别为利用理论分析数值近似求解得到的不同调度情形之间的边界情形。具体的，曲线bound A对应了方程组\ref{general equations}中的第1个方程，曲线bound B对应了第2个方程，曲线bound C对应了第3个方程。且根据不等式组\ref{Scheduling judgment}可判断在不同区域的调度情形，与仿真结果吻合。
% 由于篇幅的限制，更过不同参数取值对应的结果这里不再详细展示。
This subsection mainly presents the theoretical analysis and simulation results when the three nodes exhibit a general triangular layout. 
When $\lambda_d$ is 0.01 and $\lambda_t$ is 0.3, the main scheduling results with the maximum distance between nodes $d_{s_2,\ s_3}=280$ and $d_{s_2,\ s_3}=450$ are respectively illustrated in Fig. \ref{general layout one step}(a) and Fig. \ref{general layout one step}(b), where the curves are the boundary cases between different scheduling cases obtained by numerical approximation according to theoretical analysis. Specifically, curve bound A marked by pentagrams corresponds to the first equation in Eq. (\ref{general equations}), meanwhile, curve bound B marked by triangles corresponds to the second equation, and curve bound C marked by circles corresponds to the third equation. 
According to the discriminant conditions in Table \ref{table3}, the scheduling situation in different regions can be judged, which coincides with the simulation results.
% Due to the limited space, more detailed results corresponding to different parameter values are not shown here. \hl{modified or not}

% \begin{figure}[ht]
%   \centering
%   \includegraphics[width=3.0in]{d280_0.01&0.3_one_step.png}
%   \caption{}
%   \label{d280_0.01&0.3_one_step}
% \end{figure}

% \begin{figure}[ht]
%   \centering
%   \includegraphics[width=3.0in]{d450_0.01&0.3_one_step.png}
%   \caption{}
%   \label{d450_0.01&0.3_one_step}
% \end{figure}

%长程最优Q learning最终收敛结果
\subsection{Scheduling results with long-term optimal mechanism} 
% 本部分主要给出将系统调度建模为markov决策过程情形下得到的长程最优调度结果。为了方便与单步决策结果进行对比，所以在部分参数取值上的选择和之前单步决策相同。另外在Q学习算法中，学习率$\alpha$为0.1，折扣因子$\gamma$为0.9。
% 当$\lambda d$取0.01，$\lambda t$取0.3，且节点呈现等腰三角布局情形时，最终得到的调度结果如图\label{Q learning results} (a)所示，且每种布局的最优周期调度序列如图中图例所示。另外当$\lambda_d$取0.025,$\lambda_t$取0.5时的调度结果如图\label{Q learning results} (b)所示。
% % ，横轴代表底边取值，纵轴代表高。
% 与此同时，当三个节点呈现一般三角形分布情形时，图\ref{general Q learning results} (a)为$\lambda d$取0.01，$\lambda t$取0.3,最长节点距离d=280时的最终收敛调度结果;图\ref{general Q learning results} (b)为最长节点距离d=450时的最终收敛调度结果。
% 两种机制的调度结果的对比将在下一节中展开。

This subsection mainly presents the long-term optimal scheduling results obtained by modeling the system scheduling as a markov decision process. In order to facilitate the comparison with the single-step decision results, some parameter values is the same as the previous single-step decision results. 
In addition, the learning rate $\alpha$ is 0.1, and the discount factor $\gamma$ is 0.9 in the Q-learning algorithm. 

\begin{figure} [ht]
	\centering
	\subfloat[$\lambda_d=0.01$, $\lambda_t=0.3$.\label{fig:a}]{
		\includegraphics[width=3.5in]{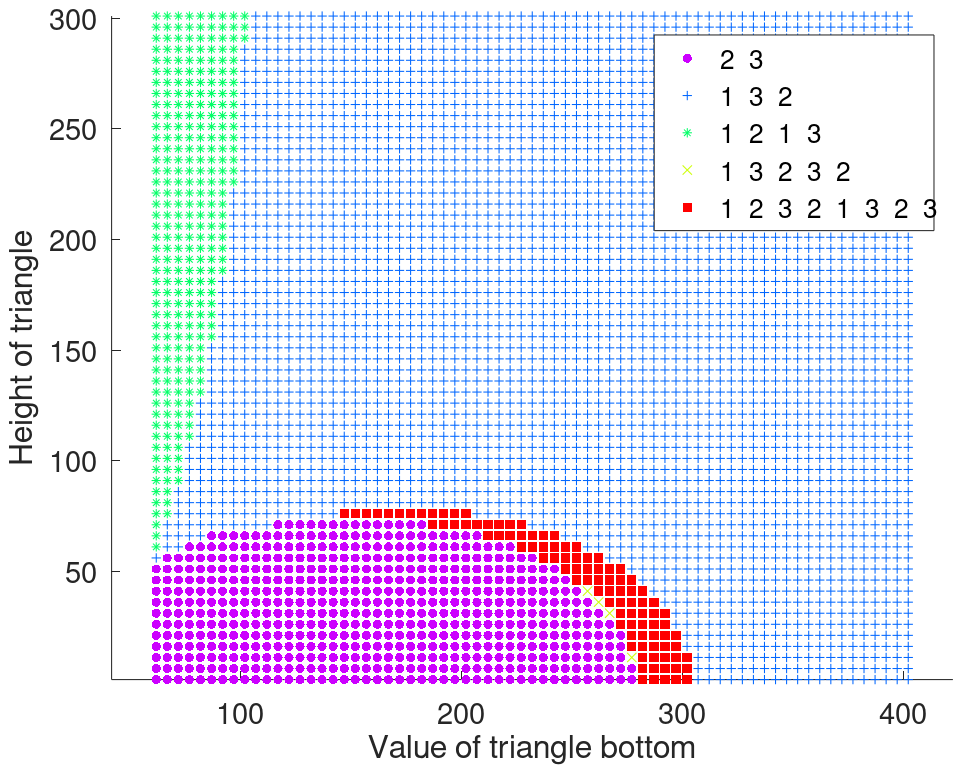}}
    \\
	\subfloat[$\lambda_d=0.025$, $\lambda_t=0.5$.\label{fig:b}]{
		\includegraphics[width=3.5in]{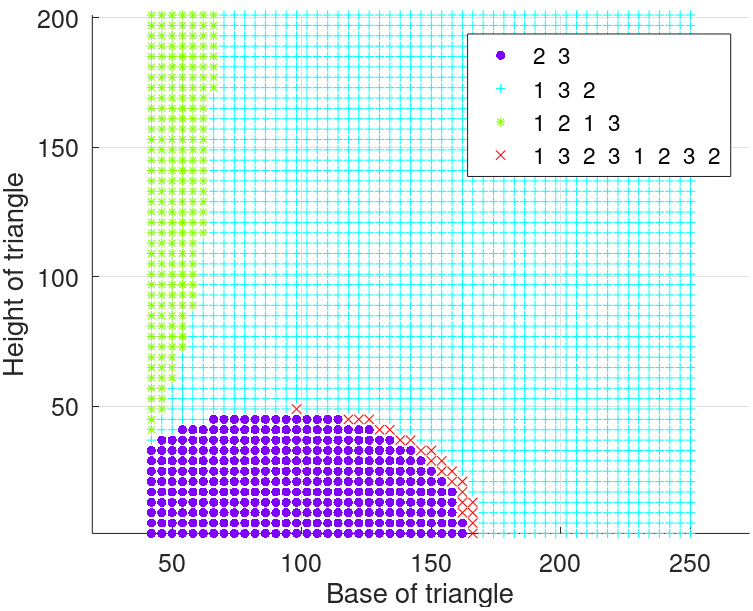}}

	\caption{Scheduling results of long-term average information acquisition optimal decision mechanism with three nodes presenting an isosceles triangle layout. (a) Scheduling results with $\lambda_d=0.01$, $\lambda_t=0.3$. (b) Scheduling results with $\lambda_d=0.025$, $\lambda_t=0.5$.}
	\label{Q learning results} 
\end{figure}

When three nodes present an isosceles triangular layout with parameters $\lambda_d=0.01$ and $\lambda_t=0.3$, the final obtained scheduling results are shown in Fig. \ref{Q learning results}(a), where the optimal cycle scheduling sequence for each layout is shown in the legend. 
Moreover, the scheduling results are shown in Fig. \ref{Q learning results}(b) with parameters $\lambda_d=0.025$, $\lambda_t=0.5$.

\begin{figure} [ht]
	\centering
	\subfloat[$d_{s_2,\ s_3}=220$.\label{fig:a}]{
		\includegraphics[width=1.7in]{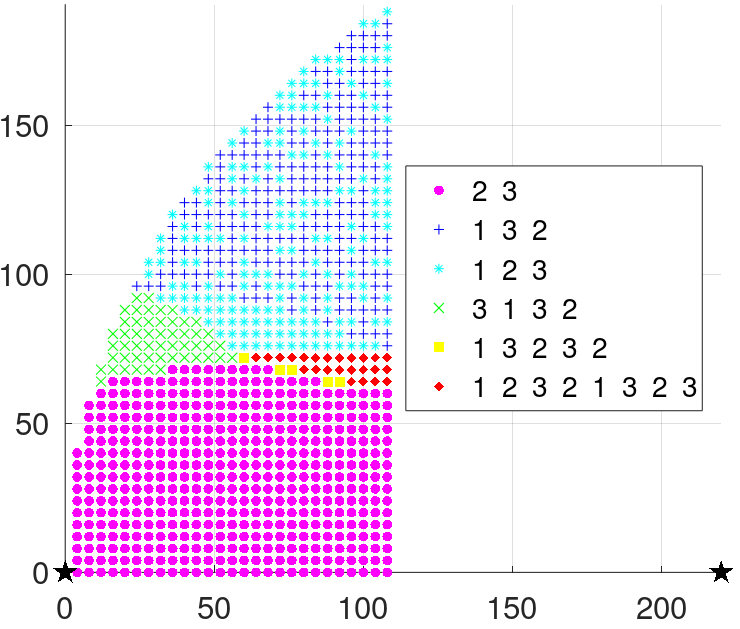}}
	\subfloat[$d_{s_2,\ s_3}=330$.\label{fig:b}]{
		\includegraphics[width=1.7in]{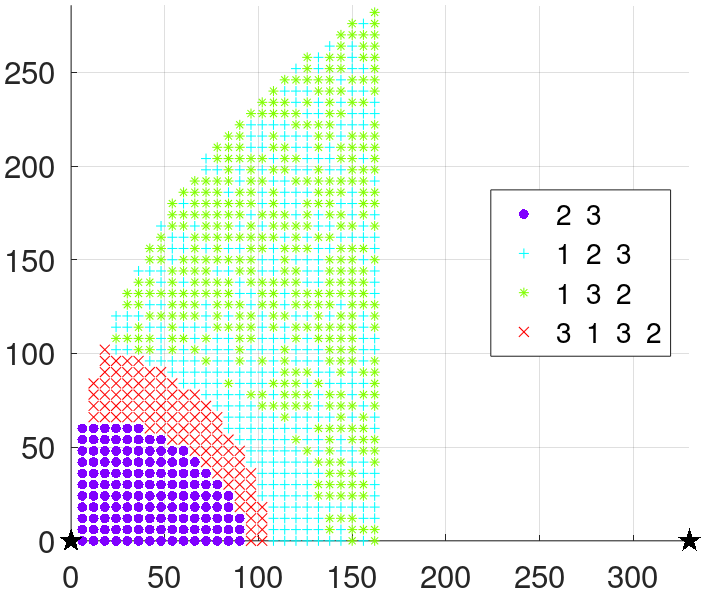}}
	\caption{Scheduling results of long-term optimal decision mechanism with three nodes presenting a general triangle layout and parameters $\lambda_d=0.01$, $\lambda_t=0.3$. (a) Scheduling results with the maximum distance $d_{s_2,\ s_3}=220$. (b) Scheduling results with the maximum distance $d_{s_2,\ s_3}=330$.}
	\label{general Q learning results} 
\end{figure}

Meanwhile, the scheduling results are shown in Fig. \ref{general Q learning results}(a) when three nodes present a general triangular layout with parameters $\lambda_d=0.01$, $\lambda_t=0.3$, and the longest node distance $d_{s_2,\ s_3}=280$. In addition, the the scheduling results are shown in Fig. \ref{general Q learning results}(b) with the longest node distance $d{s_2,\ s_3}=450$.
The comparison of the scheduling results of the two mechanisms is developed in the next subsection.
% \begin{figure}[ht]
%   \centering
%   \includegraphics[width=3.0in]{等腰Q学习0.01&0.3.png}
%   \caption{Q learning isosceles 0.01 0.3}
%   \label{等腰Q学习0.01&0.3}
% \end{figure}

% \begin{figure}[ht]
%   \centering
%   \includegraphics[width=3.0in]{等腰Q学习0.025&0.5.png}
%   \caption{Q learning isosceles 0.025 0.5}
%   % \label{fig_2}
% \end{figure}

% \begin{figure}[ht]
%   \centering
%   \includegraphics[width=3.0in]{d280_0.01&0.3_Q learning.png}
%   \caption{Q learning d280 0.01 0.3}
%   \label{Q learning d280_0.01_0.3}
% \end{figure}
% \begin{figure}[ht]
%   \centering
%   \includegraphics[width=3.0in]{d450_0.01&0.3_Q learning.png}
%   \caption{Q learning d450 0.01 0.3}
%   \label{Q learning d450_0.01_0.3}
% \end{figure}

\subsection{Performance comparison of two mechanisms} %两种机制调度结果性能对比

% 本部分主要分析两种机制最终得到的调度结果的不同性能表现。通过上一节给出的调度结果可以发现，单步决策机制的结果和长程优化最终收敛结果在不少区域内是相同的，即在这部分区域内，单步最优决策的结果同时也是满足长程均值最优的结果。% 具体可结合图7和图9，或图8和图10对比观察。
% 但是有小部分区域内两种机制的结果不相同，也就是在这部分区域单步最优决策的结果并非长程均值最优的结果，这就说明在这些区域内每次选取局部最优不能导致长程最优，所以在长期的决策过程中，需要选取特定时刻选取次优决策，而该次优决策可能能够带来更多的未来收益。
% 然而此时长期均值最优机制下获取信息增量的波动应该比单步最优决策下每次获取的信息增量略大，换句话说，虽然长程均值最优机制下信息增量均值比单步决策机制要大，但是长程均值最优机制下获取信息的不稳定度即方差也要比单步决策略大。
% 此外，长程最优求解算法的复杂度，消耗的资源也要高于单步决策。下面则随机选取两种机制结果不相同的几种布局情形进一步验证上述理论。

% 为了更具一般性，随机选取三种不同布局且三节点呈现一般三角布局下的几个调度结果进行对比。具体的，收益对比如图11所示
% 图\ref{performance comparison} (a)和图\ref{performance comparison} (b)为随机选取了三种不同布局下且两种机制调度结果不同的性能对比分析，且图\ref{performance comparison} (a)纵轴为每200轮长程信息均值收益归一化结果，图\ref{performance comparison} (b)纵轴为每200轮信息增量获取方差。
%显然如图中所示，长程最优机制下随着训练次数增多，长程收益最终趋于收敛，且高于单步决策机制下的长程收益；与此同时，随着训练次数的增多，长程最优机制下的信息增量获取方差逐渐降低，但最终收敛时仍然略高于单步决策机制，与之前理论思路相吻合。
% \hl{Same and different examples}

This subsection analyzes the different performance of the scheduling results obtained by two mechanisms.
The scheduling results given in the previous subsection show that the results with the single-step decision mechanism and the final convergence results with the long-term optimization are the same in quite a few regions, which can be observed specifically in combination with Fig. \ref{isosceles results} and Fig. \ref{Q learning results}, or Fig. \ref{general layout one step}  and Fig. \ref{general Q learning results}. 
In other words, the results of the single-step optimal decision are also the results that meet the long-term mean optimum in these regions. 
However, there are some regions where the results of the two mechanisms are not the same, that is, the results of the single-step optimal decision in these regions are not the results meet the long-term mean optimum, which indicates that each selection of the local optimum in these regions cannot lead to the long-term optimum. Therefore, in the long-term decision-making process, it is necessary to select a specific moment to select the second-best decision, which may be able to bring more future benefits.
Meanwhile, the fluctuation of the incremental information obtained under the long-term mean-optimal mechanism should be slightly larger than the incremental information obtained under the single-step optimal decision. In other words, although the mean value of the incremental information under the long-term mean-optimal mechanism is larger than that of the single-step decision mechanism, the instability of the information obtained under the long-term mean-optimal mechanism, i.e., the standard deviation, is also slightly larger than that of the single-step decision.

In the following, the above analysis is verified by the specific different simulation results obtained. When the three nodes show isosceles triangular distribution and traversal step is 5 with $\lambda_d=0.01$, $\lambda_t=0.3$, the highest mean value of information increment obtained with the long-term mechanism is about $2.5\%$ higher than that with the single-step mechanism. The mean value of information obtained under the long-term mechanism is about 0.8\% higher than that of the single-step mechanism in the term of node layouts with different results obtained by the two mechanisms. 
In addition, as for the inclusion of all layout cases, the average value of the long-term mechanism information is slightly higher by about 0.1\%.
The standard deviation of the incremental information acquisition with the long-term mechanism is also higher, up to about 300\% higher than that of the single-step, and it is on average about 60\% higher in the situation of node layouts for which the two mechanisms obtain different results.

Similarly, when the three nodes are distributed in a general triangle and traversal step length is 4 with $d_{s_2,\ s_3}=220,\lambda_d=0.01$, and $\lambda_t=0.3$, the mean value of information increment obtained by the long-term mechanism is up to 2.1\% higher than that of the single-step mechanism, which is about 0.6\% higher than that of the single-step for the situation of node layouts with different results obtained by the two mechanisms. As for the inclusion of all layout cases, the average value of the long-term mechanism information is slightly higher by about 0.1\%.
On the other hand, the standard deviation of information acquisition for the long-term mechanism is up to about 200\% higher than that of the single-step mechanism, and it is on average about 80\% higher in the node layout cases with different scheduling results for the two mechanisms. 
In summary, the information acquisition performance of the two scheduling results is consistent with the analysis that the information mean value of the long-range mechanism is slightly larger, but at the same time the standard deviation of information acquisition is also slightly larger.

\section{Experimental evaluation}
% 在本节中，中国陆地表面观测的齐次网格数据集其中的相对湿度数据被用来评估我们提出的调度机制。在实验中，假定网格数据与现实世界中对应的随机变量完全相同；换句话说，不考虑数据的误差，认为数据完全准确。
% 在本节的第一部分，分析了数据，以提取前面描述的协方差模型的缩放参数。在第二部分中，根据提取出的协方差缩放参数，结合时空范围信息模型，对四种调度方法进行了二维区域感知性能的比较。在第三部分中，进行了总结与讨论，阐述了模型性能和局限性。

In this section, the relative humidity data in the homogeneous grid dataset of China land surface observation\cite{dataset1}-\cite{dataset5} are used to evaluate our proposed scheduling mechanism. 
In the experiments, it is assumed that the grid data are identical to the corresponding random variables in the real world. In other words, the data are considered entirely accurate.
In the first part of this section, the data are analyzed to extract the scaling parameters of the covariance model described earlier. In the second part, four scheduling methods are compared in terms of their two-dimensional sense performance based on the extracted covariance scaling parameters in conjunction with the spatiotemporal scope information model. 
In the third part, a summary and discussion are presented, illustrating the model's performance and limitations.
 
\subsection{Data Analysis}

% 数据集为全中国近几十年来相对湿度的密集网格数据，时间分辨率为一个月。为了充分体现数据之间的随机变化，对网格数据进行步长为3的下采样操作，并选取了一个半径为10的圆形网格区域数据，以便于考量三个节点在不同布局下不同调度方法的全局感知性能。
% Pearson 相关系数公式被用来计算数据集中网格数据的时空相关性。
% 由于本文中前述建模采取了时空可分离的协方差函数，所以在实验提取时空相关参数可以分开提取。具体的，首先计算网格数据之间相同时刻的数据关于空间距离的相关系数，再拟合出距离相关参数lamda d。其次再计算相同网格数据不同时刻之间的时间相关性，拟合出时间相关参数lamda t。
% 然后再对数据之间的时空联合相关性进行验证，具体操作为当不同网格数据之间距离和时间差都不为0时，计算相关系数，与拟合结果进行比较，验证数据中的相关性是否符合可分离协方差函数的假设。若拟合结果偏离实际结果，则会影响后续调度的性能。从此也可以看出,参数拟合的是否准确，则是范围信息模型的可能局限性之一。

The data set is dense grid data of relative humidity for China in recent decades with a temporal resolution of one month. In order to fully reflect the stochastic variation among the data, a down-sampling operation with a step size of 3 is performed on the grid data, then a circular grid area data with a radius of 10 is selected to compare the global sensing performance of different scheduling methods for the three nodes in different layouts.

The Pearson correlation coefficient formula is used to calculate the spatio-temporal correlation of the grid data in the dataset. 
Since the modeling mentioned above in this paper adopts a spatio-temporal separable covariance function, the spatio-temporal correlation parameters can be extracted separately in the experiment. 
Specifically, firstly, the correlation coefficients of the data about the spatial distance between the same moments of grid data are calculated, and then the distance correlation parameter $\lambda_d$ can be fitted. 
Secondly, the time correlation between different moments of the same grid data is calculated, and the time correlation parameter $\lambda_t$ can be fitted. 
The joint spatio-temporal correlation between the data is then verified by calculating the correlation coefficient when both the distance and time difference between different grid data are not zero, and comparing it with the fitting results to verify whether the correlation in the data conforms to the assumption of separable covariance function. If the fitting results deviate from the actual results, it will affect the performance of the subsequent scheduling. It is also clear from this that the accuracy of the parameter fitting is one of the possible limitations of the scope information model.

% 下图即为拟合参数结果的示意图，图1a是相同时刻数据关于网格之间空间距离相关的拟合结果，图1b则是相同网格在不同时刻的数据关于时间差相关性的拟合结果，图1c和d则是验证分别单步拟合出的相关性参数lamda d 和lamda t是否符合可分离联合协方差的假设，其中c是选取了相对湿度数据集的拟合结果，可以看出拟合还是相对合理，符合时空协方差可分离联合分布的假设，而d则是选取了地表风速的拟合结果，其不太符合协方差可分离的假设，在这种情况下若直接代入参数，极大可能影响模型结果的性能。
% \begin{figure}[ht]
%   \centering
%   \includegraphics[width=3.8in]{拟合参数.png}
%   \caption{}
%   \label{}
% \end{figure}

% 以上差不多\hl{OK}
\subsection{Performance comparison}

% 实验具体实施步骤如下，在圆形区域网格数据内，选中一个大小合适的等边三角形，并将其中两个顶点设为节点所在位置，而另一节点的遍历区域，依旧如图4b所示。即确定节点布局最长边的位置，遍历第三个点的可行区域，比较四种不同调度方法的不同性能表现。
 
The specific implementation steps of the experiment are as follows: within the circular region mesh data, an equilateral triangle of suitable size is selected, and two of the vertices are set as the location of the node, while the traversal region of the other node, remains as shown in Fig. \ref{Nodes location distribution}(b). 
That is, the position of the longest edge of the node layout is determined, and the feasible region of the third point is traversed to compare the performance of the four different scheduling methods.
% 并将固定两个节点的位置，然后遍历另一节点的可行区域，这与先前仿真步骤类似，具体示意可参考下图，其中处于阴影区域内的网格则是节点遍历的集合。

% \begin{figure}[ht]
%   \centering
%   \includegraphics[width=2.5in]{实验节点遍历范围.png}
%   \caption{}
%   \label{}
% \end{figure}

% 实验具体衡量指标为整个二维圆形网格区域内的感知绝对误差。且每个节点的有效覆盖区域可由先前的分析，结合数据时空距离确定，在各自节点的覆盖区域内，以该点数据代表所有数据。四种调度方法包括本文先前提出的单步决策和长程决策方法，另外两种分别为理想调度方法和全局交替方法。理想调度方法是系统掌握实时所有网格数据，并实时选取能够带来最小范围总误差的调度决策；全局交替方法则是无论三个节点处于什么布局情形，一律采取等比交替激活节点的策略。
 
The specific metric of the experiment is the mean absolute error over the entire two-dimensional circular grid area. The effective coverage area of each node can be determined by the previous analysis, combined with the data spatiotemporal distance. Within the coverage area of the respective node, the data at the node position represent all the data. The four scheduling methods include the single-step and long-term mechanisms previously proposed in this paper, and the other two are the ideal scheduling and uniform alternation, respectively. The ideal scheduling method is a system that has all the grid data in real time and makes the scheduling decision that brings the minimum total scope error. The uniform alternation method is a strategy that activates nodes in equal proportional alternation regardless of the layout of the three nodes.

% 实验中，每次评估的时间跨度为9年，分别是1982-1990,1991-1999,2000-2008,2009-2017。由于每年有12条网格数据，则每次实验网格数据为108条。
% 模型中相关参数的提取主要是利用每次实验的前三年数据，提取出的参数用于整个时间段的实验。
In the experiments, the time span of each evaluation is nine years, which are 1982-1990,1991-1999,2000-2008,2009-2017, respectively. Moreover, the model parameters were extracted mainly using the data from the first three years used in each experiment. Then the extracted parameters are used for the entire period of the experiment.

\begin{figure}[ht]
  \centering
  \includegraphics[width=3.5in]{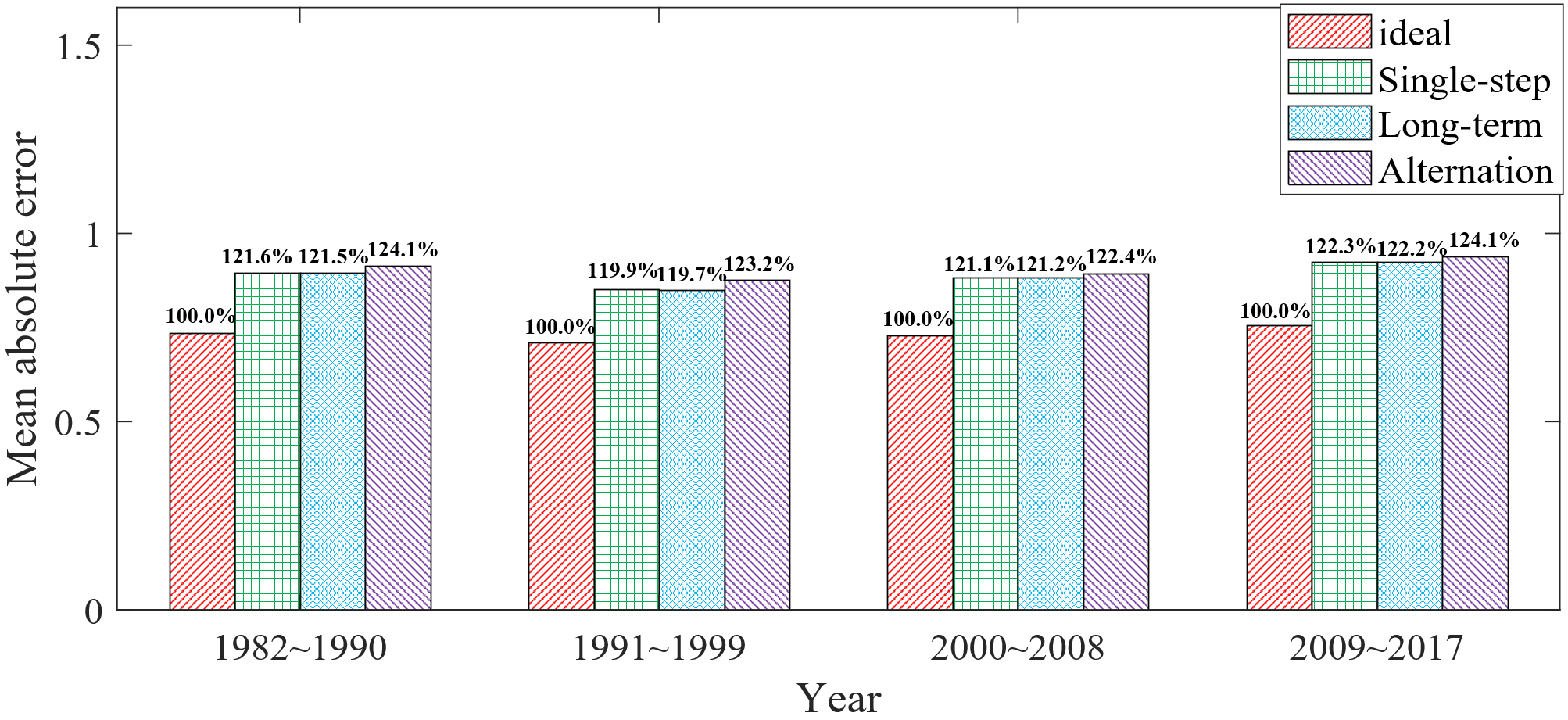}
  \caption{}
  \label{Global Results}
\end{figure}

The experimental results are shown in Fig. \ref{Global Results}. Obviously, the ideal scheduling method has the best performance with the minimum scope mean absolute error. Furthermore, the uniform alternation method has the maximum mean absolute error, about 23\% higher than the ideal scheduling method.
The performance of the single-step and long-term mechanisms lies between the above two, and the mean absolute error of both methods is about 21\% higher than that of the ideal scheduling method.
Since the single-step and long-term mechanisms get the same scheduling results in most node layout situations, and the global average difference in the information mean of the scheduling results obtained by the two mechanisms is less than 1\% in simulation, the global performance mean difference caused by the two methods in the experiment is also minimal, as shown in Fig. \ref{Global Results}. In other words, the mean absolute error of the long-term mechanism is only slightly higher than that of the single-step mechanism.

% 实验结果如图\ref{Global Results}所示。显然，理想调度方法的性能最好，平均范围绝对误差最小。而一律交替方法平均绝对误差最大，比理想调度方法高百分23左右。
% 而单步和长程机制性能位于上述两者之间，两种方法比理想调度方法平均绝对误差高百分之21左右。
% 由于长程决策和单步决策在大部分节点布局情形下得到的调度结果相同，且两种机制得到调度结果的信息均值差距全局平均不到1\%，所以在实验中两种方法造成的全局性能均值差距也微乎其微，具体表现如图所示。换句话说，长程机制的平均绝对误差只略高于单步机制。

% 若在遍历节点位置时，考虑节点在信息感知区域内布局的合理性，则统一考察圆形区域的范围总误差不是非常合理。因为若引入较多节点难以感知到有效信息的区域，则会带来更多范围总误差的波动。
% 简单来讲，例如当两个节点距离非常接近时，三个节点的有效覆盖范围可能趋近于一个椭圆形。
% 针对这个考虑，在实验中，在遍历不同节点布局时，分别以每个节点为圆心，选取合适大小的圆，并将三个圆的联合覆盖范围设为考察模型性能的数据范围。

If the reasonableness of node layout within the information sense region is considered when traversing node locations, it is not very reasonable to uniformly examine the total scope error in the circular region. 
Since if there are many areas where nodes have difficulty sensing effective information, it may lead to more fluctuations in the total scope error.
For example, when two nodes are very close to each other, the effective coverage of three nodes may converge to an ellipse.
For this consideration, in the experiments, when traversing different node layouts, circles of appropriate size are respectively selected with each node position as the center of the circle, then the joint coverage of the three circles is set as the data evaluation scope for examining the model performance.

\begin{figure}[ht]
  \centering
  \includegraphics[width=3.5in]{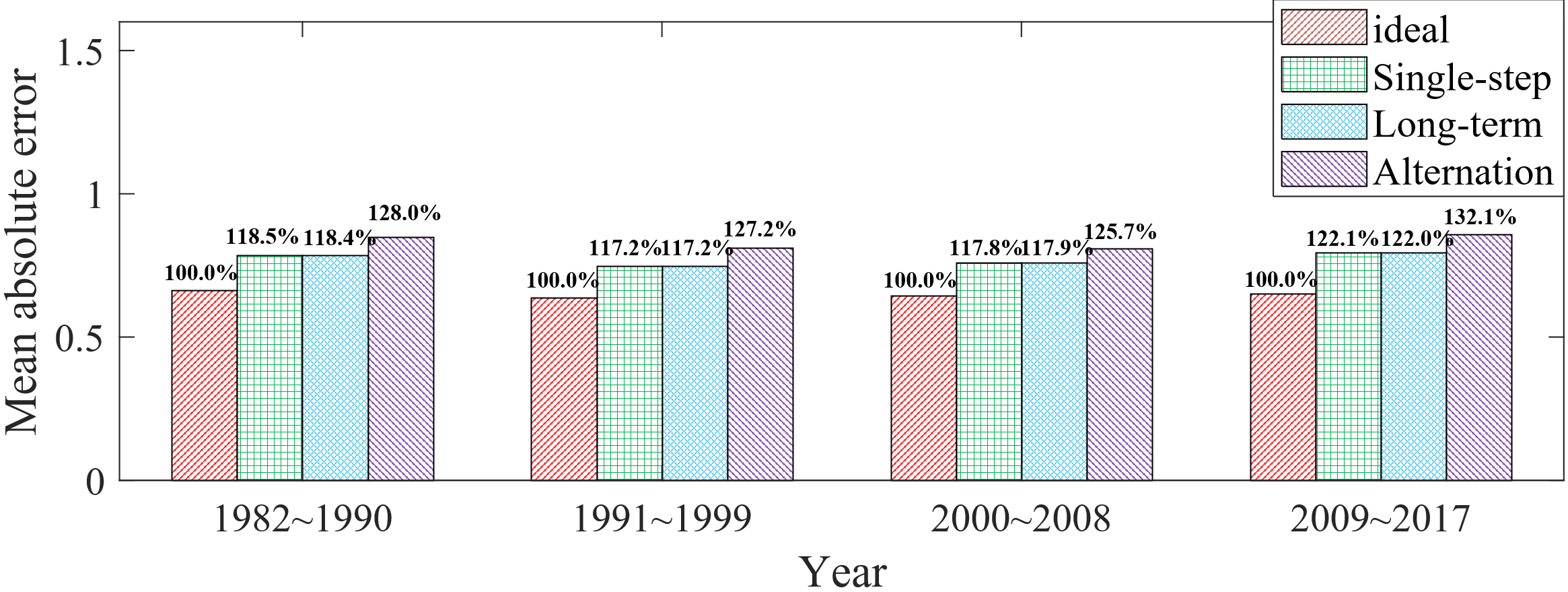}
  \caption{}
  \label{Adaptive coverage}
\end{figure}

Under the above experimental operation, the performance comparison of the four scheduling methods is obtained as shown in Fig. \ref{Adaptive coverage}. Samely, the ideal scheduling method has the minimum mean absolute error, the single-step decision and the long-term mechanism have about 19\% higher mean error than the ideal scheduling, and the performance difference between them is very little. The error of the uniform alternation scheduling method is about 28\% higher than that of the ideal scheduling method.

% 在上述实验操作下，得到四种调度方法的性能比较如下图\ref{Adaptive coverage}所示。同样理想调度方法平均绝对误差最低，单步决策和长程决策平均误差约比理想调度高百分之19左右，且两者性能差距微小。全局交替调度方法误差则比理想调度方法高百分之28左右。

\subsection{Summary and discussion}
% 由上述实验可以看出，提出的两种调度机制，单步决策和长程决策法基本符合预期，且全局平均性能长程决策比单步决策略高。
% 而单步决策的程序最大复杂度为O(3*n),n代表系统的AoI状态总数。而长程决策的算法的复杂度主要取决于收敛的速度，但总体来说，其复杂度比单步决策高一些。
% 由仿真和实验可知，单步决策和长程决策在大部分节点布局情况下得到的调度情况是相同的，只有在小部分节点布局的情况下，长程决策获取的信息性能比单步决策略高，仿真得到的信息获取性能一般比单步高一个百分点左右。

% 模型目前的局限性：
% 另外，经过实验可知，相关参数拟合的正确与否，是决定模型性能的首要因素。而对于随机过程来讲，其相关模型可能不是静态的，即相关系数可能是时变的。此外，采取的协方差函数模型是否符合数据集的分布，也是决定模型性能的重要因素。在未来的研究中，时变的相关模型，和不同协方差函数模型下的多节点合作感知情形可能是主要的研究兴趣。
From the above experiments, the two proposed scheduling mechanisms, single-step decision and long-term mechanisms, basically meet the expectations, and the global average performance of long-term mechanisms is slightly higher than that of single-step mechanisms.
The maximum program complexity of single-step mechanisms is $O(3n)$, where $n$ represents the total number of AoI states of the system. Moreover, the complexity of the algorithm for long-term mechanisms depends mainly on the speed of convergence, but in general, its complexity is higher than that of single-step mechanisms.

In addition, it is known experimentally that the correctness of the correlation parameter fit is the primary factor in determining the model performance. In the case of stochastic processes, the correlation model may not be stationary, i.e., the correlation coefficients may be time-varying. In addition, whether the adopted covariance function model fits the distribution of the data set is also an important factor in determining the model performance. In future research, time-varying correlation models and multi-node cooperative perception situations under different covariance function models may be major research interests.
% \begin{algorithm}[h]
%   \caption{单步决策算法}
%   \label{Alg1}
%   \begin{algorithmic}[1]
%   \REQUIRE $S\ \lambda_d,\ \lambda_t,\ N$.
%   \STATE Initialize $State=[inf\ inf\ inf],State\_Set,Info\_sense\_Matrix$ 
%   \FOR{$t_i=1$; $t_i\le N$; $t_i++$ }
%     \FOR{$si=1$; $si\le 3$; $si++$ }
%     \STATE $state\_temp=State;$%If the node si is currently active, the status is recorded as $State\_temp$%当前若激活节点si的状态记为state temp
%     \STATE $state\_temp(si)=0;$
%     \STATE %Find if the $State\_temp$ is already recorded in the $State\_Set$,with the result being recorded as $index$
%     $index=find(State\_Set==state\_temp);$%查找状态表中是否已经记录的该状态
%       % \IF{index==[]} 
%       % \EndIF
%       \IF{$index$ is not null} %$index~=[]$若已记录，查表读取信息增量值
%       \STATE $info(s_i)=info\_matrix(index,s_i);$
%       \ELSE%若无记录，将状态添加进状态集，并计算总信息增量，添加进信息获取矩阵中
%       \STATE $info(s_i)=info\_gain\_2D(state\_temp,si,S,\lambda_d,\ \lambda_t,);$   %//根据积分式计算二维总信息增量
%       \STATE $State\_Set\leftarrow[State\_Set;state\_temp];$
%       \STATE $info\_gain\_matrix\leftarrow[info\_gain\_matrix;(state\_temp,info\_sense(t_i))]$
%       \STATE Add the total amount of information obtained by the corresponding state and activation node into the information matrix%将对应状态和激活节点获取的信息总量添加进信息矩阵中
%       \ENDIF 
%     \ENDFOR
%     \STATE $info\_sense(t_i),sense\_order(t_i)=max(info);$
%     \STATE $State(sense\_order(t_i))=0;$
%     \STATE $State=State+1;$
%   \ENDFOR
%   \ENSURE $sense\_order,info\_sense$ 
%   \end{algorithmic}
% \end{algorithm}

\section{Conclusion} 
% 本文基于区域性信源的时空相关性质，建立了范围信息模型，用于量化感知数据实时有用信息。本文考虑一个包含三个感知节点（即传感器）的感知通信系统，定期激活一个节点获取感知数据。两种优化调度机制被提出，一种为单步信息获取最优决策机制，系统每次在离散决策时刻只激活能够在当前时刻获取最多信息增量的节点，并通过理论分析与数值计算求得出不同调度结果的近似数值边界，与仿真结果相吻合。另一种机制为长程平均信息获取最优机制，将该决策过程建模为markov决策过程，并利用Q学习算法遍历不同布局求出最优调度结果。分析比较两种机制优化结果的性能表现，分别指出其优缺点。
% \hl{to be modified}

In this paper, a Spatio-temporal Scope Information Model (SSIM) is developed to quantify the valuable information of sensor data, which decays with space and time. Periodically activating one node to obtain sensing data, a sensor monitoring system containing three sensor nodes is considered. For more efficient access to information, two optimal scheduling mechanisms are proposed. One is the single-step optimal decision mechanism, and the approximate numerical bounds for the node layout between partial scheduling results are obtained by theoretical analysis and numerical calculation, which coincide with the simulation results. 
The other one is the long-term optimal mechanism, which is modeled as a Markov decision process. The optimal scheduling results for long-term information with different node layouts are obtained using the Q-learning algorithm. 
Through simulation and experiments, the scheduling results of both mechanisms are the same in many node layout cases. In few node layouts, the mean value of the incremental information obtained by the long-term mechanism is higher than that of the single-step mechanism, but the standard deviation of the incremental information obtained by the long-term mechanism is also higher than that of the single-step mechanism. The average performance of the two mechanisms is similar under all node layouts, and the long-term mechanism is slightly higher.

In future work, we may focus on  spatio-temporal scope information models with time-varying and different covariance functions, and study the cooperative scheduling of multiple nodes to improve energy efficiency and extend lifetime.

% 通过仿真与实验，可知在许多节点布局情形下两种机制的调度结果相同。在少数节点布局情况下，长程机制下获取的信息增量均值要高于单步机制，但同时其获取信息增量的标准差也要高于单步机制。全部节点布局下两种机制的平均性能相差不多，长程机制略高。

% 在未来的工作中，我们可能会聚焦于时变的,不同协方差函数情况下的时空范围信息模型，研究多节点的合作调度，提高能效，延长寿命等问题。

% \section*{Acknowledgments}
% This should be a simple paragraph before the References to thank those individuals and institutions who have supported your work on this article.

\end{document}